\documentclass[aps,,twocolumn,nofootinbib,preprintnumbers,superscriptaddress]{revtex4-1}

\usepackage{psfrag}
\usepackage{mathrsfs}

\usepackage{amsfonts}
\usepackage{amsmath}
\usepackage{empheq}		
\usepackage{cancel}     
\usepackage[toc,page]{appendix}

\usepackage{amssymb}
\usepackage{graphicx}
\usepackage{epsfig}
\usepackage{float}
\usepackage{color,array,subfigure}
\usepackage{epstopdf}
\usepackage{esint}

\usepackage{geometry}
\RequirePackage{CJKutf8}

\usepackage[colorlinks=true,  
            citecolor=red,
            urlcolor= blue,                     %
            filecolor=black,
            linktocpage=true,  
            linkcolor=blue,
             ]{hyperref}

\newcommand{\beq}{\begin{equation}}
\newcommand{\eeq}{\end{equation}}
\newcommand{\bq}{\begin{equation}}
\newcommand{\eq}{\end{equation}}
\newcommand{\ba}{\begin{array}}
\newcommand{\ea}{\end{array}}
\newcommand{\beqa}{\begin{eqnarray}}
\newcommand{\eeqa}{\end{eqnarray}}
\newcommand{\bea}{\begin{eqnarray}}
\newcommand{\eea}{\end{eqnarray}}
\newcommand{\bit}{\begin{itemize}}
\newcommand{\eit}{\end{itemize}}
\newcommand{\bsel}{{\begin{subequations}\begin{empheq}}}
\newcommand{\bse}{{\begin{subequations}\begin{empheq}[left={\ii}\empheqlbrace]{align}}}
\newcommand{\ese}{{\end{empheq}\end{subequations}}}
\def\bc{\begin{center}}
\def\ec{\end{center}}
\def\bnum{\begin{enumerate} }
\def\enum{\end{enumerate}}
\def\nn{\nonumber}
\def\ii{\!\!\!\!\!\!}  
\def\3i{\!\!\!}
\def\2i{\!\!}

\def\ea{{e_a}}

\def\ec{{e_c}}

\def\nn{\nonumber}

\def\[{\left[}
\def\]{\right]}
\def\({\left(}
\def\){\right)}

\def\gev{\rm GeV}
\def\tev{\rm TeV}

\def\abs#1{\left|#1\right|}
\def\vev#1{\left\langle#1\right\rangle}
\def\>{\rightarrow}
\def\bf#1{\textbf #1}

\def\Diracslash#1{\not{\hbox{\kern-4pt $#1$}}}

\def\Dslash{\not{\hbox{\kern-4pt $D$}}}
\def\pslash{\not{\hbox{\kern-4pt $p$}}}
\def\qslash{\not{\hbox{\kern-4pt $q$}}}

\def\lv{\not{\hbox{\kern-4pt $L$}}}
\def\lsim{\mathrel{\raise.3ex\hbox{$<$\kern-.75em\lower1ex\hbox{$\sim$}}}}
\def\gsim{\mathrel{\raise.3ex\hbox{$>$\kern-.75em\lower1ex\hbox{$\sim$}}}}
\def\ifmath#1{\relax\ifmmode #1\else $#1$\fi}

\evensidemargin=\oddsidemargin

\numberwithin{equation}{section}

\begin{document}
\begin{CJK}{UTF8}{gbsn}

\begin{titlepage}
\begin{flushright}
\end{flushright}

\begin{center}
 \vspace*{10mm}

{\LARGE 
Effective Field Theory Perspective on Next-to-Minimal Composite Higgs 
}\\
\medskip
\bigskip\vspace{0.6cm}
{ 
{Yong-Hui Qi$^{\dag,\star,\pounds}$,}  ~ 
 {Jiang-Hao Yu$^{\ast,\ddag}$,}  
\,{Shou-Hua Zhu$^{\S,\dag}$} 
 }
\\[7mm]
{\it
$^{\dag}$Center for High Energy Physics, Peking University, Beijing, 100871, Peoples Republic of China\\
$^{\star}$Asia Pacific Center for Theoretical Physics, Pohang, Gyeongbuk, 37673, Korea \\
$^{\pounds}$Department of Physics, Pohang University of Science and Technology, Pohang, Gyeongbuk, 37673, Korea \\
$^{\ast}$Institute of Theoretical Physics, Chinese Academy of Sciences, Beijing 100190, P. R. China \\
$^{\ddag}$School of Physical Sciences, University of Chinese Academy of Sciences, Beijing 100049, P.R. China\\
$^{\S}$Department of Physics and State Key Laboratory of Nuclear Physics and Technology, Peking University, Beijing 100871, P.R. China \\
}

\vspace*{0.3cm}
 {\tt  yonghui.qi@apctp.org, 
 ~ jhyu@itp.ac.cn,  
 ~ shzhu@pku.edu.cn
 }
\bigskip\bigskip\bigskip

{
\centerline{\large  Abstract}
\begin{quote}
We study both the CP-even and CP-odd effective chiral Lagrangians of next minimal composite Higgs model {with symmetry breaking pattern depicted by the coset $SO(6)/SO(5)$} through the sigma/omega decomposition, in which the Goldstone matrix of the coset is decomposed in terms of the standard model Higgs doublet and an additional scalar singlet $s$ at the electroweak scale. The effective Lagrangian is described by the electroweak chiral Lagrangian up to $p^4$ order, with a function dependence on the Higgs boson and new scalar $s$, named Higgs function. This function in the effective Lagrangian incorporates the Higgs non-linearity or vacuum misalignment effects in the next minimal composite Higgs model, due to the Riemann curvature on the Nambu-Goldstone bosons scalar manifold, leads to various Higgs couplings deviated from the standard model ones, and also indicates the relations among different Higgs couplings in the low energy. Matching to the Higgs effective field theory below the electroweak scale,  we obtain various low energy observables such as the electroweak oblique parameters, anomalous triple and quartic gauge couplings, anomalous couplings of Higgs to gauge bosons, in which the Higgs non-linearity effects are encoded in the ratio $\xi\equiv v^2/f^2$ of the electroweak scale and the new physics scale. 
\end{quote}}
\end{center}
\end{titlepage}


\tableofcontents


\section{Introduction}

The standard model (SM) of elementary particle physics provides a framework for all of the visible matters in nature among the three known fundamental interactions except the gravity~\cite{Weinberg:1967tq}. The model is described by the gauge theory~\cite{Yang:1954ek,Abers:1973qs,Weinberg:1996kr} with symmetry $SU(3)_c\times SU(2)_L\times U(1)_Y$. 

Theoretically the electroweak (EW) sector of the SM still owns problems such as vacuum stability, too many free parameters and naturalness or hierarchy problem. Thus, it is commonly believed that the SM is not a UV complete theory at energy above trillion electron volt ($\tev$) scale, at where underlying dynamics is still unclear. To solve these problems suffered by the SM, many models beyond the SM are proposed, with different motivations emphasized. Among them, the composite Higgs model (CHM) provides a scenery~\cite{Kaplan:1983fs,Kaplan:1983sm,Georgi:1984af}, that might be related to a strong dynamics in the high energy, solve the electroweak (EW) hierarchy problem. In the scenario, the Higgs-like particle arise as a pseudo Nambu-Goldstone Boson (NGB or GB)~\cite{Weinberg:1972fn} of a global symmetry breaking at a higher energy scale $f$ higher than EW scale $v=2m_W/g$, and is related with strong dynamical scale $\Lambda_s \simeq 4\pi f$ for an unbroken symmetry as the subgroup of the global symmetry~\cite{Weinberg:1975gm}.

There are many kinds of composite Higgs models in literature, for review, see Refs.~\cite{Contino:2010rs,Bellazzini:2014yua,Panico:2015jxa}. We could classify different composite Higgs models based on the numbers and EW representations of GBs in the setup.  
The most economical and popular composite Higgs model is the $SO(5)/SO(4)$ minimal composite Higgs model (MCHM)~\cite{Agashe:2004rs,Csaki:2008zd,Contino:2011np,vonGersdorff:2015fta,Panico:2015jxa}. 
The symmetric coset $SO(5)/SO(4)$ provides only four GBs: three GBs eaten by the longitudinal components of the W and Z bosons, and the Higgs boson as the pseudo-Nambu-GB. 
The next to minimal model is the $ SU(4)/Sp(4)\simeq SO(6)/SO(5) $ next-to-minimal composite Higgs model (NMCHM)~\cite{Gripaios:2009pe,Frigerio:2012uc,Redi:2012ha,Marzocca:2014msa}, in which five GBs are presented: four SM Higgs components and additional singlet scalar. 
For a heavy singlet scalar GB, this model recovers to the MCHM. Thus the NMCHM contains the MCHM as the limiting case. 
On the other hand, if this singlet scalar is not so heavy, it provides new low energy phenomenology, such as dark matter candidate~\cite{Frigerio:2012uc,Marzocca:2014msa}, electroweak phase transition, etc.  

Further classification of the CHMs could be justified using the custodial symmetry. The symmetry is invariant for the three gauge bosons in the fundamental representation of $SO(3)$ or adjoint representation of $SU(2)_c$~\cite{Georgi:1984af}, guarantees that the $\rho$ parameter, the ratio defined with both mass of gauge bosons and EW mixing angle as below,
\beqa
\rho \equiv \frac{m_W^2}{m_Z^2c_\theta^2} = 1, \label{rho_SM_0}
\eeqa
at the tree level, although there could have small correction at the loop-level. 
This custodial protection mechanism in the SM, are valid both before and after electroweak symmetry breaking (EWSB), which entails a strong constraint for model buildings beyond the SM, namely, whatever underlying dynamics as a possible UV completion beyond $\tev$ scale, its symmetry breaking pattern down to the EW scale, always entails $\rho=1$ at the tree level. 
The electroweak precision tests indicate that we should consider CHMs with the custodial symmetry imposed. 
Both the MCHM and NMCHM contains the $SO(3)$ group as the subgroup and thus the custodial symmetry is guaranteed. 

On the other hand, given the experimental results at the LHC, the lack of the evidence of new physics and the precision measurement of the Higgs property have already push the new physics scale up to the TeV scale, unless the new particles do not carry the electroweak charge. 
In CHMs, the composite particles related to the strong dynamics scale $f$ are typically heavy~\cite{Contino:2010rs,Bellazzini:2014yua}.
If the additional GBs rather than the pseudo Goldstone Higgs exist, they should not be lighter than the electroweak scale unless the GB is the electroweak singlet.  Although there are many kinds of composite Higgs models with different symmetry breaking pattern depicted by the coset $G/H$ \cite{Bellazzini:2014yua}, after integrating out heavy composite states, and heavy GBs, what is left should at least be the SM contents: the matter fields, the SM gauge fields, and the Higgs boson as the pseudo GB. There could have additional pseudo GB being the electroweak singlet. 
Thus the effective Lagrangian description of the composite Higgs models should contain at least $SO(5)/SO(4)$ symmetry~\cite{Contino:2011np}. 
In this work, we consider more general case in which additional light GB exist at the electroweak scale. 
The typical example is the $SO(6)/SO(5)$ symmetry at the electroweak scale, containing one light scalar singlet GB.
If the scalar singlet GB becomes heavy, it recovers to the $SO(5)/SO(4)$ symmetry at the electroweak scale. 

From the bottom-up perspective, the low energy effective field theory (EFT) approaches ~\cite{Appelquist:1980vg,Longhitano:1980tm,Buchmuller:1985jz,Appelquist:1993ka,Giudice:2007fh,Grzadkowski:2010es,Buchalla:2012qq,Alonso:2012px,Buchalla:2013rka,Brivio:2013pma,Hierro:2015nna,Brivio:2016fzo,Alonso:2016oah,Pich:2016lew} with the general principle of quantum field theory, such as Lorentz invariance, unitary, causality etc,  have provides a universal model independent description for new physics beyond the SM. There are usually two ways to describe the low energy effective field theory: the SM EFT in the symmetric phase~\cite{Buchmuller:1985jz,Grzadkowski:2010es,Giudice:2007fh}, and the EW chiral Lagrangian in the broken phase~\cite{Appelquist:1980vg,Longhitano:1980tm,Appelquist:1993ka}.  
For the case that the Higgs boson as the pseudo-Nambu-GBs, the EW chiral Lagrangian is usually adopted due to the Higgs nonlinearity in the scalar manifold of the GB fields. 
In the EW chiral Lagrangian, the Higgs field is non-linearly realized, e.g., in the coset construction of Callan-Coleman-Wess-Zumino (CCWZ) approach~\cite{Coleman:1969sm,Callan:1969sn}. 
On the other hand, the standard model effective field theory (SMEFT) up to dimensional-$6$ operators cannot fully characterize high energy effective Lagrangian up to the $p^4$ order, since the operators at this order can also obtain contributions from dimensional-$8$ operators of SMEFT.
Moreover, in the EW chiral Lagrangian, the global $SO(4)$ Higgs singlet $h$ supplies as a natural embedding of the custodial symmetry $SU(2)_C\sim SO(3) \subset SO(4) $ after the EW symmetry is spontaneously broken 
\beqa
|h|^2 
\to 
|\vec\varphi|^2 + (v+h)^2,
\eeqa 
where $\vec{h}=(h_1,h_2,h_3,h_4)$, and $\vec\varphi=(\varphi_1,\varphi_2,\varphi_3)$ consists of the three would-be GBs for the EW symmetry. 
In the CHM scenery, again we will specifically focus on those with the $ {\mathcal G}/SO(4)$ symmetry breaking pattern. We will take the custodial symmetric CHM as an example to study the non-linearity of Higgs as well as its possible mixing with a light scalar. 

In this work, we take the $ SU(4)/Sp(4)\simeq SO(6)/SO(5) $ next-minimal composite Higgs model (NMCHM)~\cite{Gripaios:2009pe,Frigerio:2012uc,Redi:2012ha,Marzocca:2014msa}, as an extension of
MCHM.
In the NMCHM, after the explicit breaking of the global symmetry $f$, the gauge symmetry and the Yukawa terms induce the radiative potential for the SM Higgs, which acquires dynamically a non-vanishing 
vacuum expectation value (vev)
\beqa
&&   v = f\sin\left(\frac{v_{\phi}}{f}\right)  \cos\left(\frac{v_\psi}{f}\right),   \label{xi_psi-phi_0} 
\eeqa
where $\phi$ denote the quantum fluctuation field around the vacuum with expectation value $v_\phi$ and $v_\psi$ denoted as the vacuum phase of an extra singlet $s$ relative to the Higgs $h$.  
In the $f\gg v$  limit, the nonlinear NMCHM should recover the SM limit, in which the EWSB is linearly realized and the Lagrangian is written in terms of the SM Higgs as an $SU(2)_L$ doublet. 

We connect the low energy EW chiral Lagrangian to the UV CHM valid up to an energy scale $\Lambda_s\sim 4\pi f$. 
In the CHMs, according to the CCWZ formalism, the kinetic term of the Higgs boson should origin from~\cite{Alonso:2015fsp,Alonso:2016btr,Alonso:2016oah,Nagai:2019tgi}
\beqa
	{\mathcal L}_2 = f^2 \textrm{Tr} d_\mu d^\mu \supset g_{a b}(h) \partial^\mu \pi^a \partial_\mu \pi^a,
\eeqa
where the metric $g_{a b}(h)$ parametrize the curvature in the scalar manifold. Given the non-flat metric, 
the degree of non-linearity of the theory can be quantified by the non-linear parameter
\beqa
\xi \equiv \frac{v^2}{f^2},  \label{xi}
\eeqa
which recovers the flat metric if $f\to \infty$.

On the other hand, the Wilson coefficients of the effective operators in the EW chiral Lagrangian
can be described by the Higgs functions (or radial functions) ${\mathcal F}_H$~\cite{Alonso:2014wta,Hierro:2015nna,Alonso:2016oah},  as
\beqa
{\mathcal L}_h &=&   c_H {\mathcal F}_H(h,s) \mathcal{L}_{H}  + \ldots,  \qquad  \label{L_h-LEFT} 
\eeqa
where $\mathcal{L}_{H}$ is defined in Eq.(\ref{L_H}), 
$\ldots$ denote the higher order terms, e.g., at the $p^4$ order, include the CP-even ${\mathcal L}_{C,T}$ or CP-odd ${\mathcal L}_{\widetilde{T}}$ defined in Eqs.(\ref{L_C-T}) and (\ref{L_Ct-Tt})  in Appendix.~\ref{app:LCEFT}. The Higgs functions encode all the information of the Higgs nonlinearity due to the non-flat metric of pseudo-Nambu-GBs manifold of the CHM.

The Higgs functions provide the connection between the low energy EW EFT and the chiral Lagrangian for the CHM. 
The effective Lagrangian of the model are matching with the EW chiral effective Lagrangian up to the $p^4$ order.
We extract the Higgs functions of the model from the low energy EW chiral Lagrangian, which contains information of not only the Higgs itself, but also that of an additional light scalar $s$. In a series expansion of $h/v$ in the Higgs effective field theory (HEFT) with the parameter $\xi$  fixed,  we find the effective Wilson coefficients associated with the high energy effective chiral Lagrangian in the NMCHM. Observables such as the EW oblique parameters, anomalous triple and quartic gauge couplings, anomalous couplings of Higgs to gauge bosons are obtained.

The structure of the paper is organized as below:  
In Sec.\ref{sec:NMCHM}, we study the building blocks for both Sigma and Omega representation in the NMCHM and discuss its symmetries. We also illuminate the significance of Higgs nonlinearity.
In Sec.\ref{sec:EWCL}, we study high-energy chiral effective Lagrangian of NMCHH up to the $p^4$ order.
In Sec.\ref{sec:HEEFT}, we match the chiral effective Lagrangian of the NMCHM with low energy effective EW chiral Lagrangian (EWCL), and extract the Higgs functions. 
In Sec.\ref{sec:HEFT-NMCHM}, we study the Higgs function at EW scale. 
In Sec.\ref{sec:HEFT-obs}, we study the connection of effective field theory to the corresponding physical observables.
Conclusions are made in Sec.\ref{sec:dis-con}.

\section{Next minimal composite Higgs model}
\label{sec:NMCHM}

In generic composite Higgs (CH) scenery, a global symmetry group ${\mathcal G}$ is spontaneously broken by some strong dynamics mechanism at the scale $f$ down to a subgroup ${\mathcal H}$. The coset ${\mathcal G}/{\mathcal H}$ is assumed symmetric and entails that $\text{dim}({\mathcal G}/{\mathcal H})\ge 4$, e.g. the minimal version in terms of minimal composite Higgs model (MCHM)~\cite{Agashe:2004rs}. Consequently, there are (at least) four GBs arise from the non-linear symmetry breaking mechanism of the global symmetry ${\mathcal G}$, one is identified with the light Higgs-like scalar field $h$ and three are identified with the longitudinal components of the SM gauge bosons. For the next minimal composite Higgs model (NMCHM) with the coset $SO(6)/SO(5)$, there is an additional new singlet GB $s$, comparing with those in the MCHM.
Following CCWZ in Appendix.~\ref{sec:CCWZ}, we can introduce the Goldstone field matrix $\Omega$, which transform nonlinearly under the group ${\mathcal G}$. For a symmetric coset, equivalently the $ \boldsymbol\Sigma$ field could be introduced, transforming linearly under the group ${\mathcal G}$.

\subsection{Goldstone boson field matrix}

The GBs of the coset $SO(6)/SO(5)$ in the fundamental representation can be parameterized by\footnote{We have introduced $\Phi_0$ to be dimensionless, so that the kinetic terms should be defined as ${\mathcal L}_{\text{kin}}=f^2(\partial_\mu\Phi)^T(\partial^\mu \Phi)/2$. Alternatively, one may defined $\Phi_0=(0_{5},f)^T$ with a dimensional scale $\vev{f}$, in this case, the kinetic terms should be defined as in Eq.(\ref{L_C-phi-psi}). }
\beqa
\Phi\left(h_{\hat{a}}\right)=\Omega \Phi_{0}, 
\quad \Phi_{0}=\left(\begin{array}{c}{0_{5}} \\ {1}\end{array}\right), \label{Sigma_0}
\eeqa
where $\Phi_0$ is the vacuum expectation value in the fundamental representation of $SO(6)$ as a $6$-dimensional unit vector.

\subsubsection{Omega parametrization}

In the symmetric coset $SO(6)/SO(5)$, it is equivalent to use the $\Omega$ or $ \boldsymbol\Sigma$ to describe the Goldstone degree of freedom (d.o.f). as
\beqa
\Omega 
= e^{   i \frac{ \sqrt{2}}{f}   T^{\hat{a}} h_{\hat{a}}  } \equiv  e^{i\Pi}, \quad \boldsymbol\Sigma = \Omega^2 = e^{2i\Pi},  \label{Omega-Sigma}
\eeqa
where $T^{\hat{a}}$ are the generators in the coset $SO(6)/SO(5)$ as given in
Eq.(\ref{T_ahat}) in Appendix.~\ref{app:rep},  where $\hat{a}=(\hat\alpha,5)$ with $\hat\alpha=1,2,3,4$. $T^{\hat\alpha}$ are the unbroken generators of $SO(4)$ and $T^{\hat{5}}$ is that of $SO(2)$.
The first four index span a $4$-parameter coset space and $h_{\hat{\alpha}}$ is a $SO(4)$ vector.
Denoting the GBs in an array
\beqa
\hat{\phi}\equiv (\hat{h}_\alpha,\hat{h}_5)^T, \quad \textrm{with} \quad \phi = \sum_a \sqrt{h_a h_a} 
\eeqa
where $\hat{h}_\alpha \equiv(\hat{h}_1,\hat{h}_2,\hat{h}_3,\hat{h}_4)^T$ and we have $\hat{h}_{\hat{a}} \equiv {h_{\hat{a}}}/{\phi}$.
The Goldstone boson field matrix is expressed as
\beqa
\Pi =  \frac{\phi}{ f} \Xi, \quad \textrm{with} \quad
&& \Xi = \sqrt{2} T^{\hat{a}} \hat{h}_{\hat{a}} = -i \left(
\begin{array}{c|c}
\textbf{0}_{5\times 5}&  \hat{\phi} \\
 \hline
- \hat{\phi}^{T} &0 \\
\end{array}
\right). 
\eeqa
It is convenient to define the mixing angle $\phi$ at strong dynamics breaking scale $f$ as 
\beqa
s_\phi \equiv \sin{\left(\frac{\phi}{f}\right)}, \quad c_\phi \equiv \cos{\left(\frac{\phi}{f}\right)}. 
\label{s_c-phi}
\eeqa
Thus, one can express the Goldstone matrix $\Omega$ as 
\beqa
 \Omega 
&=&
\left(
\begin{array}{c|c}
 1-(1-c_\phi) \hat{\phi} \hat{\phi}^{T} & s_\phi \hat{\phi} \\
 \hline
-s_\phi \hat{\phi}^{T} & c_\phi    \\
\end{array}
\right)  
,
\label{Sigma-Omega_h}
\eeqa
where \beqa
 \hat{\phi}  \hat{\phi}^{T} = \left(
\begin{array}{cc}
 \hat{h} \hat{h}^{T} & \hat{h}_5  \hat{h} \\
 \hat{h}_5\hat{h}^T & \hat{h}_5^2 \\
\end{array}
\right),  
\eeqa
and $\hat{\phi}^{T}  \hat{\phi}    =\text{Tr}( \hat{h} \hat{h}^{T} )+  \hat{h}_5^2 = 1 $. 

After the electroweak symmetry is broken, in the unitary gauge, i.e., $\hat{h}_1=\hat{h}_2=\hat{h}_3=0$, it is convenient to define another angle as~\cite{Redi:2012ha}
\beqa  
&&  c_\psi \equiv \cos\left(\frac{\psi}{f}\right) = \frac{h_4}{\sqrt{h_4^2+h_5^2}}  , \nn\\   
&&  s_\psi \equiv \sin\left(\frac{\psi}{f}\right) = \frac{h_5}{\sqrt{h_4^2+h_5^2}}  ,   \label{c_psi-s_psi}  \label{psi_h4-h5} 
\eeqa
where $\psi=\arctan{(h_5/h_4)}$ is pseudo-scalar under CP symmetry due to the relative phase between the $h_5$ and $h_4$. In the unitary gauge, the amplitude $\phi$ reduces to 
$\phi = \sqrt{h_4^2+h_5^2} $ and 
%
 the $(h_4,h_5)$ can be expressed in terms of $\phi$ and $\psi$ as
\beqa
h_4 = \phi c_\psi, \quad h_5 = \phi s_\psi,  \label{h4-h5_phi-psi} 
\eeqa
Thus, the GB matrix $\Omega$ in Eq.(\ref{Sigma-Omega_h}) can be re-expressed as
\beqa
\Omega = \left(
\begin{array}{cccc}
 \textbf{1}_3 & 0 & 0 & 0 \\
 0 & c_{\phi } c_{\psi }^2+s_{\psi }^2 & \left(c_{\phi }-1\right) c_{\psi } s_{\psi } & c_{\psi } s_{\phi } \\
 0 & \left(c_{\phi }-1\right) c_{\psi } s_{\psi } & c_{\psi }^2+c_{\phi } s_{\psi }^2 & s_{\psi } s_{\phi } \\
0 & -c_{\psi } s_{\phi } & -s_{\phi } s_{\psi } & c_{\phi } \\
\end{array}
\right) . \qquad \qquad \label{Sigma-Omega_phi-psi}
\eeqa
In the original Cartesian $(h_4,h_5)$ basis, one has
\beqa
\Omega = \left(
\begin{array}{cccc}
 \textbf{1}_3 & 0 & 0 &  \\
0 & \frac{h_4^2 c_\phi+h_5^2}{h_4^2+h_5^2} & \frac{h_4 h_5 \left(c_\phi-1\right)}{h_4^2+h_5^2} & \frac{h_4 s_\phi}{\sqrt{h_4^2+h_5^2}} \\
0 & \frac{h_4 h_5 \left(c_\phi-1\right)}{h_4^2+h_5^2} & \frac{h_5^2 c_\phi+h_4^2}{h_4^2+h_5^2} & \frac{h_5 s_\phi}{\sqrt{h_4^2+h_5^2}} \\
0 & -\frac{h_4 s_\phi}{\sqrt{h_4^2+h_5^2}} & -\frac{h_5 s_\phi}{\sqrt{h_4^2+h_5^2}} & c_\phi \\
\end{array}
\right).
\eeqa

For the latter convenience of relating the $h_{4,5}$ to the SM Higgs $h$, we introduce a new basis $(h,s)$ as
\beqa
&& 
 \frac{h}{f} \equiv \frac{h_4}{\sqrt{h_4^2+h_5^2}} \sin{\left(\frac{\sqrt{h_4^2+h_5^2}}{f}\right)}  \overset{f\to \infty}{=} \frac{h_4}{f}  , \nn\\
&& 
 \frac{s}{f} \equiv \frac{h_5}{\sqrt{h_4^2+h_5^2}} \sin{\left(\frac{\sqrt{h_4^2+h_5^2}}{f}\right)}   \overset{f\to \infty}{=} \frac{h_5}{f}  . \label{h-s_h4-h5}
\eeqa
in the decoupling limit, i.e., $f\to \infty$, $(h,s) $ just recover $(h_4,h_5)$, respectively. 
One can check that
\beqa
\sin\left( \frac{h_4^2+h_5^2}{f^2} \right) = \frac{h^2 + s^2}{f^2}, \quad \frac{h_4}{h_5} = \frac{h}{s}, 
\eeqa
and the phase $\psi$ also represent the relative phase between the singlet $s$ and Higgs $h$ as
\beqa
t_\psi = \tan\left(\frac{s}{h}\right). \label{psi_s-h}
\eeqa
With Eqs.(\ref{s_c-phi}) and (\ref{c_psi-s_psi}), the $(h,s)$ fields can be expressed in terms of $(\phi,\psi)$ as
\beqa
&& h  =  f c_\psi s_\phi , \quad 
s =  f s_\psi s_\phi , \label{h-s_phi-psi}  
\eeqa
The amplitude $\phi=\sqrt{h^2+s^2}$ is a scalar consists of $h$ and $s$. In the $(h,s)$ basis, the GBs matrix in Eq.(\ref{Sigma-Omega_phi-psi}) can be expressed as
{ 
\small
\beqa
&& \Omega = \left(
\begin{array}{cccc}
  \textbf{1}_3 & 0 & 0 & 0 \\
 0 & \frac{h^2 \sqrt{1-\frac{h^2+s^2}{f^2}} + s^2 }{h^2+s^2} & \frac{h s \left(\sqrt{1-\frac{h^2+s^2}{f^2}}-1\right)}{h^2+s^2} & \frac{h}{f} \\
 0 & \frac{h s \left(\sqrt{1-\frac{h^2+s^2}{f^2}}-1\right)}{h^2+s^2} & \frac{s^2 \sqrt{1-\frac{h^2+s^2}{f^2}} +h^2 }{h^2+s^2} & \frac{s}{f} \\
 0 & -\frac{h}{f} & -\frac{s}{f} & \sqrt{1-\frac{h^2+s^2}{f^2}} \\
\end{array}
\right).  \nn \\
&& \label{Sigma-Omega_h-s}
\eeqa
}

Finally let us consider the situation in the absence of CP-odd singlet, in which the NMCHM just recovers that of MCHM, with 
\beqa
 \frac{h}{f} \equiv   \sin{\left(\frac{h_4}{f}\right)}  \overset{f\to \infty}{=} \frac{h_4}{f} , \quad s = 0.
\eeqa
 In the $(\phi,\psi)$ basis, with $\psi=0$, the GB matrix in Eq.(\ref{Sigma-Omega_phi-psi}) recovers that for MCHM in $SO(5)\subset SO(6)$ as
\beqa
\Omega &=& \left(
\begin{array}{cccc}
  \textbf{1}_3 & 0 & 0 & 0  \\
 0 & c_\phi & 0 & s_\phi \\
 0 & 0 & 1 & 0 \\
 0 & -s_\phi   & 0 & c_\phi \\
\end{array}
\right).
\eeqa
Equivalently, in the original Cartesian $(h_4,h_5)$ basis with $h_4\equiv h$, $h_5=0$, it is just recovers
\beqa
\Omega = \left(
\begin{array}{cccc}
 1 & 0 & 0 & 0  \\
0 & \cos \left(\frac{h}{f}\right) & 0 & \sin \left(\frac{h}{f}\right) \\
0 & 0 & 1 & 0 \\
0 & -\sin \left(\frac{h}{f}\right) & 0 & \cos \left(\frac{h}{f}\right) \\
\end{array}
\right).
\eeqa
While also equivalently, in the $(h,s)$ basis, with $s=0$, the GBs matrix becomes
\beqa
\Omega = \left( 
\begin{array}{cccc}
  \textbf{1}_3 & 0 & 0 & 0  \\
 0 & \sqrt{1-\frac{h^2}{f^2}}  & 0 & \frac{h}{f} \\
 0 & 0 & 1 & 0 \\
 0 & -\frac{h}{f} & 0 & \sqrt{1-\frac{h^2 }{f^2}} \\
\end{array}
\right) ,
\eeqa
which gives the exponential non-linear parametrization of $SO(5)$ transformation on GBs in terms of unconstrained coordinates $(h_1,h_2,h_3,h_4)\in SO(4)$.

\begin{table*}[ht]
\centering
\caption{Fields coordinate transformation among $(\phi,\psi)$, $(h_4,h_5)$ and $(h,s)$ for $\Omega$ parametrization.}\label{tab_phi-psi_h4-h5_h-s}
\begin{tabular}{|c|c|c|c|}
\hline
  & $(\phi,\psi)$ & $(h_4,h_5)$  & $(h,s)$   \\
\hline
$(\phi,\psi)$  & $-$  & $ \sqrt{h_4^2+h_5^2}$, $ f \arctan\left({h_5}/{h_4}\right)$   &  $ f \arcsin\left( \sqrt{h^2+s^2}/f \right)$, $ f \arctan\left({s}/{h}\right)$  \\
\hline
$(h_4,h_5)$  & $ \phi c_\psi$, $\phi s_\psi$ &  $-$  & $ \frac{h f \arccos\left(\sqrt{1-{(h^2+s^2)}/{f^2}}\right)}{\sqrt{h^2+s^2}} $, $ \frac{s f \arccos\left(\sqrt{1-{(h^2+s^2)}/{f^2}}\right)}{\sqrt{h^2+s^2}}  $   \\
\hline
$(h,s)$ & $  f c_\psi s_\phi$, $ f s_\psi s_\phi$  & $  \frac{h_4  f \sin{\left({\sqrt{h_4^2+h_5^2}}/{f}\right)}}{\sqrt{h_4^2+h_5^2}}  $, $  \frac{h_5 f\sin{\left({\sqrt{h_4^2+h_5^2}}/{f}\right)}}{\sqrt{h_4^2+h_5^2}}  $  &  $-$  \\
\hline
\end{tabular}
\end{table*}

\subsubsection{From Omega to Sigma parametrization}

For symmetric coset, the GB matrix can be also parameterized as 
\beqa
\boldsymbol\Sigma = \Omega^2,
\eeqa
which can be obtained from $\Omega$ in Eqs.(\ref{Sigma-Omega_phi-psi}), 
by making the replacement as
\beqa
f\to \frac{f}{2} , \quad \psi \to \frac{1}{2}\psi, \label{Omega-to-Sigma}
\eeqa
so that one need to making the replacement as
\beqa
&& c_{\phi} \to c_{2\phi}, \quad s_{\phi} \to s_{2\phi}, \nn\\
&& c_\psi \to c_\psi, \quad s_\psi \to s_\psi,
\eeqa
and in the $(h,s)$ basis, equivalent to making the replacement 
\beqa
 f \to \frac{f}{2}, \quad h \to 2h, \quad s \to 2s .
\eeqa
In this case, according to Eq.(\ref{c_psi-s_psi}), the singlets can be re-expressed as
\beqa
&& (h,s) = \frac{\sin\left(\frac{n}{f}\phi\right)}{\left(\frac{n}{f}\right)}  (c_\psi, s_\psi ) ,
\overset{f\to\infty}{=} (h_4,h_5)  , \qquad \label{h-s_phi-psi-n}
\eeqa
from which, one can re-express the phase $\phi$ and $\psi$, respectively, in terms of singlets $h$ and $s$ as
\beqa
&& \phi = \frac{f}{n} \arcsin\left( \frac{n}{f}\sqrt{h^2+s^2} \right) = \sqrt{h_4^2+h_5^2} , \nn\\
&& \psi = f \arctan\left(\frac{s}{h}\right) = f \arctan\left(\frac{h_5}{h_4}\right) , \qquad \label{phi-psi_h-s}
\eeqa
where $n=1$ for $\Omega$ parametrization and $n=2$ for $\Sigma$ parametrization, respectively. This implies that for $\mathbf\Sigma\equiv\Omega^n$, one just makes the replace 
\beqa
f\to \frac{f}{n}, \quad \psi \to \frac{\psi}{n},
\eeqa
so that the ratio $s/h$ are unchanged. For the convenience of later usage, we summarize the fields coordinate transformation among $(\phi,\psi)$, $(h_4,h_5)$, $(h,s)$ and list in Table.\ref{tab_phi-psi_h4-h5_h-s}.

\subsection{Symmetries}

\subsubsection{Rotational symmetry}

One may rotate the generators of unbroken $SO(4)\subset SO(5)$ in Eqs.(\ref{T_L-R}) as well as the coset generator $SO(5)/SO(4)$ in Eq.(\ref{T_ahat}) by an angle $\theta$ in the $SO(5)$ inner space, 
\beqa
&& T^\alpha \equiv (T^a_L,T^a_R,T^{\hat{a}}) \to T^{\alpha \prime} \equiv R(\theta) T^\alpha R^{-1}(\theta) , 
\eeqa
by a rotation matrix $R(\theta)$
\beqa
&& R(\theta) =\left(\begin{array}{cccc}{\textbf{1}_3} & {} & {} &\\ 
{} & {\cos\theta} & & {\sin\theta}\\ 
{} &  & 1 & \\
& {-\sin\theta}& & {\cos\theta}
\end{array}\right),  
\label{R_theta}
\eeqa
where $\textbf{1} \equiv \textbf{1}_{3\times 3}$. The angle $\theta$ parameterizes the misalignment of $SO(6)$ vacuum as we will discuss in the following section.
It turns out that the generators rotated becomes
\beqa
&& T_{L }^{a}(\theta) = \frac{1+c_\theta}{2}T_{L}^{a} + \frac{1-c_\theta}{2} T_{R}^{a} - \frac{s_\theta}{\sqrt{2}} T^{\hat{a}},  \nn\\
&& T_{R }^{a}(\theta) = \frac{1-c_\theta}{2}T_{L}^{a} + \frac{1+c_\theta}{2} T_{R}^{a} + \frac{s_\theta}{\sqrt{2}} T^{\hat{a}},  \nn\\
&& T^{\hat{a}}(\theta)  =  \frac{s_\theta}{\sqrt{2}}T_{L}^{a} - \frac{s_\theta}{\sqrt{2}} T_{R}^{a} + c_\theta T^{\hat{a}}, \quad \hat{a}=1,2,3, \nn\\
&& T^{\hat{4}}(\theta)  =   T^{\hat{4}}, \quad T^{\hat{5}}(\theta)  =   c_\theta T^{\hat{5}} - s_\theta T^{4},
\label{T_theta}
\eeqa
where $s_\theta \equiv \sin\theta$, $c_\theta \equiv \cos\theta$, $T^{\hat{5}}(\theta)$ is the generator of $SO(2)$ as defined in Eq.(\ref{T_alpha-T_5}) in Appendix.~\ref{app:rep}.

It can be check that all the generators are also normalized as
\beqa
&& \text{Tr}[T_L^a(\theta)T_L^b(\theta)] = \delta^{ab}, \quad \text{Tr}[T_R^a(\theta)T_R^b(\theta)] = \delta^{ab}, \nn\\
&& \text{Tr}[T^{\hat{a}}(\theta)T^{\hat{b}}(\theta)] = \delta^{\hat{a}\hat{b}} = 2 (T^{\hat{a}}T^{\hat{b}})_{66},
\eeqa
and one can also check that $ (T^{a}_{L,R}T^{\hat{b}})_{66} =  (T^{\hat{a}}T^{b}_{L,R})_{66} = (T^a_{L,R} T^b_{L,R})_{66} = (T^a_{L,R} T^b_{R,L})_{66} = 0 $.
From the generator with rotation angle $\theta$, one can read the components of the external gauge fields are given by
\beqa
&& A_\mu^a(\theta): A_\mu^{a L}(\theta) = \frac{1+c_\theta}{2}W_{\mu}^{a} + \frac{1-c_\theta}{2} B_{\mu}\delta^{a3}, \nn\\
&& A_\mu^{a R}(\theta) = \frac{1-c_\theta}{2}W_{\mu}^{a} + \frac{1+c_\theta}{2} B_{\mu}\delta^{a3},  \nn\\
&& A_\mu^{\hat{a}}(\theta) : A_\mu^{\hat{a}}(\theta) = \frac{s_\theta}{\sqrt{2}}(W_\mu^{\hat{a}}-\delta^{\hat{a}3}B_\mu), \quad A_\mu^{\hat{4}} = 0, \qquad \label{A_theta}
\eeqa
where $W_\mu^a$, $B_\mu$ are the EW $SU(2)_L\times U(1)_Y$ vector bosons. 

\subsubsection{Automorphism symmetry}

Since the quotient space $SO(6)/SO(5)$ is symmetric, the symmetric coset has a automorphism or ``grading'' symmetry that acts upon the generators of ${\mathcal G}$ as defined in Eq.(\ref{R_automorphism}) or (\ref{R_auto_2}) both lead to the same representation as
\beqa
\mathcal{R} :
\left\{\begin{array}{l}{T_{a}(\theta) \rightarrow+T_{a}(\theta) } \\ 
{T_{\hat{a}}(\theta) \rightarrow-T_{\hat{a}}(\theta) }.
\end{array}\right.   \label{R_auto}
\eeqa
It is a linear transformation among the generators that preserves the algebra. 
There is other automorphism given by~\cite{Gripaios:2009pe}
\beqa
\mathcal{R}_2 :
\left\{\begin{array}{l}{T_{a}(\theta) \rightarrow - T_{a}^T(\theta) } \\ 
{T_{\hat{a}}(\theta) \rightarrow T_{\hat{a}}^T(\theta) }.
\end{array}\right.  \label{R_auto_2}
\eeqa
The two linear transformation preserve the Lie algebra following from the fact that the $SO(6)/SO(5)$ is symmetric space. Both these two automorphism lead to the same representation as Eq.(\ref{RR_theta}). 

With the generators in Eq.(\ref{T_theta}), the automorphism symmetry of $SO(6)/SO(5)$ in Eq.(\ref{R_auto}) is
\beqa
\mathcal{R}(\theta) &=& \left(
\begin{array}{cccc}
 \textbf{1}_{3\times 3} & 0 & 0 & 0 \\
0 & \cos (2\theta ) & 0 & -\sin (2\theta ) \\
0 & 0 & 1 & 0 \\
0 & -\sin (2\theta ) & 0 & -\cos (2\theta ) \\
\end{array}
\right) \nn\\
& \overset{\theta=0,\pi}{=} & \text{diag}(1,1,1,1,1,-1), 
\label{RR_theta}
\eeqa
which satisfies
\beqa
 \mathcal{R}(\theta) T_{L,R}^a(\theta)  \mathcal{R}^{-1}(\theta) &=&  T_{L,R}^a(\theta) , \nn\\
 \mathcal{R}(\theta) T_{\hat{a}}(\theta)  \mathcal{R}^{-1}(\theta) &=&  -T_{\hat{a}}(\theta)  ,
\eeqa
where $a=1,2,3$ and $\hat{a}=1,2,3,4,5$.

When the $SU(2)_L\times U(1)_Y$ symmetry is turned on, the "grading" symmetry ${\mathcal R}$ is explicitly breaking, to the discrete one as
\beqa
\theta=0,\pi. 
\eeqa
Since ${\mathcal R}$ is an element of the unbroken $SO(4)$ symmetry, i.e, it is an internal automorphism of the algebra, it will ben an exact symmetry of the low energy Lagrangian up to any order in the absence of  gauging the EW symmetry.

It is also interesting that when $\theta=\pi/2$, the automorphism corresponds to the Higgs parity as~\cite{Hosotani:2010hx}
\beqa
\mathcal{R}(\theta) \overset{\theta=\pi/2}{=}& \text{diag}(1,1,1,-1,1,1) \equiv P_H,
\eeqa
which transforms the generators as
\beqa
T^A = \{  T^a_L, T^a_R, T^{\hat{a}}, T^{\hat{4}} \} \to T'^A = \{  T^a_R, T^a_L, T^{\hat{a}}, -T^{\hat{4}} \} , \qquad\quad
\eeqa
implying that the operator $P_H$ flips the direction of $T^{\hat{4}}$ through
\beqa
T'^{A} = P_H T^A P_H^{-1}  . 
\eeqa

\subsubsection{Left-right parity symmetry}

There is also a left-right parity symmetry $P_{LR}$ in the NMCHM, 
\beqa
P_{LR} = \text{diag} (1,1,1,-1,1,-1), \label{P_LR}
\eeqa
which exchanges the generators of $SU(2)_L$ and $SU(2)_R$ subgroup of $SO(4)$, and also changes the sign of the first three broken generators $T^{\hat{a}}$ as
\beqa
&& P_{LR} T_{L,R}^a P_{LR}^{-1} = T_{R,L}^a, \nn\\
&& P_{LR} T^{\hat{a}} P_{LR}^{-1} = - T^{\hat{a}}, \nn\\ 
&& P_{LR} T^{\hat{4}} P_{LR}^{-1} =  T^{\hat{4}}, \label{P_L-R}
\eeqa
with $a=1,2,3$ and $\hat{a}=1,2,3,5$. For the broken generators, one should rewritten as $P_{LR} T^{\hat{a}} P_{LR}^{-1} = - \eta^{\hat{a}} T^{\hat{a}}$ with $\eta^{\hat{a}}=(1,1,1,-1)^T$. Since it is not element of $SO(4)$, one would expect it broken at ${ }(p^4)$ order, although it is an accidental symmetry of GBs Lagrangian at ${ }(p^2)$. When the SM gauge symmetry $SU(2)_L\times U(1)$ is turned on, $P_{LR}$ is explicitly broken for a generic value of $\theta$.

\subsubsection{CP symmetry}

The CP symmetry is a symmetry of the sigma model of the Higgs sector in the $SO(6)/SO(5)$ model~\cite{Gripaios:2009pe}. The first automorphism symmetry ${\mathcal R}$ in Eq.(\ref{R_auto}) makes the vielbein $\overline{V}_\mu= \left( D_\mu \boldsymbol\Sigma  \right) \boldsymbol\Sigma^{-1}$ change the sign, so the Wess-Zumino-Witten (WZW) term
\beqa
{\mathcal L}_{WZW} =\epsilon^{\mu\nu\rho\sigma\tau} \text{Tr}\left[\overline{V}_\mu \overline{V}_\nu \overline{V}_\rho \overline{V}_\sigma \overline{V}_\tau\right],
\eeqa
changes the sign. While the second automorphism symmetry ${\mathcal R}_2$ in Eq.(\ref{R_auto_2}) does not make the vielbein $\overline{V}_\mu= (D_\mu \mathcal\Sigma) \mathcal\Sigma^{-1}$ change the sign, so that it is a symmetry of the WZW term.

The WZW term is unchanged under two condition 
\beqa
{\mathcal R}P_0, \quad {\mathcal R}_2,
\eeqa
where
\beqa
P_0 : x\to -x ,
\eeqa
is the spacetime parity. In the NMCHM, ${\mathcal R}{\mathcal R}_2P_0$ corresponds to 
\beqa
h \to h, \quad s \to -s,
\eeqa 
which defines the CP symmetry of the Higgs sector, including the WZW term. In this case, one can write down a gauge-invariant Lagrangian of the form
\beqa
{\mathcal L} = \frac{s}{(4\pi)^2} (n_B B_{\mu\nu}\widetilde{B}_{\mu\nu} + n_W W_{\mu\nu}\widetilde{W}_{\mu\nu} + n_G G_{\mu\nu}\widetilde{G}_{\mu\nu} ), \qquad \quad
\eeqa
where $\widetilde{B}^{\mu\nu}=\epsilon^{\mu\nu\rho\sigma}B_{\rho\sigma}/2$ and $n_{B,W,G}$ are integers that measure the strengths of the gauge anomalies, which are fixed by the fermion content in the UV.

\subsection{Vacuum misalignment}

The electroweak symmetry should be broken at the electroweak scale. It has been shown that the electroweak symmetry breaking can be viewed a due to the misalignment angel $\theta$ with respect to the vacuum of $SO(6)$. 
Even assuming there is no misalignment at the tree level, a non-vanishing $\theta=\vev{h}/f$ is generated at the loop level after the GBs obtain a vev $\vev{h}=v$.  
There are two energy scales $f$ and $v$, the EWSB can be described as a two-step process: at first, $SO(6)$ is spontaneously breaking down to $SO(5)$ at the scale $f$, giving rise to an $SU(2)_L$ doublet of GBs; secondly, the EW symmetry is spontaneously breaking from $SO(4)$ down to $SO(3)$ at the EW scale, which is defined as
~\footnote{In fact, this definition comes from the mass terms of the gauge boson $W$ and $Z$: 
\beqa
\frac{v}{f} = \sqrt\xi = \frac{v_{4}}{\sqrt{v_{4}^2+v_{5}^2}}  \sin
   \left(\frac{\sqrt{v_{4}^2+v_{5}^2}}{f}\right),  \qquad \label{xi_h4-h5}
\eeqa
where $v_{4} \equiv \vev{h_4} $ and $ v_{5} \equiv \vev{h_5} $ are vevs of $h_4$ and $h_5$, respectively. }
\beqa
v=f\sin\left(\frac{\vev{h}}{f} \right) \equiv f\sin\theta, 
\eeqa
leaving an approximate custodial symmetry, where in the last equality, we have used Eq.(\ref{theta_hv}). Therefore, the vacuum misalignment parameter $\theta$ is related to the non-linear parameter as
\beqa
 \sqrt{\xi} = \sin\theta \overset{\vev{h} \gg f}{\approx} \theta. 
\eeqa

\subsubsection{Embedding of $SO(4)$ symmetry}

Here we study the correspondence between the $SO(6)/SO(5)=S^5$ PNGBs in the NMCHM and the SM Higgs. 
It it convenient to re-parametrize the first three components $h_a$ as the three massless SM GBs $\varphi_a$ with $a=1,2,3$, 
\beqa
 \hat{h}_a \equiv  \frac{h}{v}\varphi_a ,
\eeqa
and define the mixing angle $\varphi$ as
\beqa
s_\varphi \equiv \sin\left(\frac{|\varphi|}{v}\right), \quad c_\varphi \equiv \cos\left(\frac{|\varphi|}{v}\right), \label{s_varphi-c_varphi} 
\eeqa
with absolute value $|\varphi|=\sqrt{\varphi^a \varphi^a}$. 

In the following, the first four elements of the GB scalar with $SO(4)$ symmetry are embedded into a $6$-dimensional fundamental scalar in $SO(6)$. 
We can rewritten the four components as a $4$-vector times with a phase factor ${\mathbb U}$ as embedding of ${\mathbf U}$ in Eq.(\ref{U_varphi}) into $SO(6)$ as
\beqa
{\mathbb U}  & \equiv & \exp{\left(i \frac{\varphi^{a} }{v} t_L^{a}\right)} = c_\varphi + i \hat\varphi^a t_L^a s_\varphi, 
\eeqa
where $t_L^a=2T_L^a$ with $T_L^{a}$ are $ SO(6)$ embedding generators of custodial symmetry $ SU(2)_L \simeq SO(3) \subset SO(4)$. One may define the GBs as
\beqa
\pi^a \equiv v s_\varphi \hat\varphi^a, \label{pi_varphi}
\eeqa
so that the unitary matrix can be re-expressed in analogy to Eq.(\ref{square root-parametrization}) as
\beqa
{\mathbb U}  
= \sqrt{1 - \frac{\pi^2}{v^2} } + i  \frac{\pi^a}{v} t_L^a, 
\eeqa
which defines the coordinate change.





In terms of the phase factor ${\mathbb U}$, the GBs matrix of NMCHM can also be re-expressed 
\beqa
\Omega=e^{i \frac{|h|}{2 f} \Xi }, \quad \boldsymbol{\Sigma}(x)=e^{i \frac{|h|}{f} \Xi},
\eeqa 
where $|h|$ is the scalar singlet field and $\Xi$ is the would-be GBs non-linear field given by
\beqa
\Xi = \sqrt{2} T^{\hat{a}} \hat{h}_{\hat{a}} = - i \sqrt{2} \operatorname{Tr}\left(\mathbf{U} \sigma_{\hat{a}}\right) T^{\hat{a}}, 
\eeqa
where $ \hat{a}=1, \ldots, 5$, and we have matching it to Eq.(\ref{Omega-Sigma}), by using Eq.(\ref{ha-h4_U})
and $\sigma^{\hat{a}} \equiv (  \sigma^{1}, \sigma^{2}, \sigma^{3}, i \textbf{1}_{2} , i \textbf{1}_{2}   )$ with $\psi$ defined in Eq.(\ref{psi_s-h}) and the definition of $\hat{h}_a$ in the unitary gauge as
\beqa
\hat{h}_{a}=0, \quad \hat{h}_4 = c_\psi, \quad \hat{h}_5 = s_\psi.
\eeqa
In this case, $\Xi$ reduces to $\sqrt{2} (T^{\hat{4}} c_\psi + T^{\hat{5}}h_5 s_\psi )$. 

In general case, we can parameterize the $6$-dimensional fundamental scalar in $SO(6)$ as Eq.(\ref{Sigma_0}),  embedding with the $4$-component Higgs vector in the fundamental representation of $SO(4)$ as Eq.(12) in Ref.~\cite{Redi:2012ha} or Eq.(A.5) in Ref.~\cite{Marzocca:2014msa} as
\beqa
\Phi_{\hat{a}} &=& \left(\begin{array}{c}
s_\phi \hat{h}_{{\hat\alpha}}  \\
s_\phi \hat{h}_{{5}} \\ 
c_\phi
\end{array}\right)  = \left(\begin{array}{c}  
 s_\phi \frac{1}{2}\text{Tr}[\mathbf{U} \sigma_{\hat{\alpha}}]  \\ 
{s_\phi} \hat{h}_5 \\
{c_\phi}
\end{array}\right) ,
\eeqa
where for $h_{\hat{\alpha}}$ with $\hat{\alpha}=1,2,3,4$ and in the second equality, we have used Eq.(\ref{h_a-varphi_a}). Thus, by using the definition of Eq.(\ref{h4-h5_phi-psi}), one can parameterize the $SO(6)$ fundamental scalar with SM higgs embedded as
\beqa
\Phi_{\hat{a}} &\equiv& \left(\begin{array}{c}  
s_\phi  c_\psi (s_\varphi \hat\varphi_a) \\ 
{s_\phi}  c_\psi (c_\varphi) \\
 {s_\phi} s_\psi \\
{c_\phi}
\end{array}\right) \overset{v  \gg 1 }{=}  
\left(\begin{array}{c}
{s_\phi}{c_\psi} \frac{\varphi_a}{v} \\
{s_\phi}{c_\psi}  \\
{s_\phi}{s_\psi} \\ 
{c_\phi}
\end{array}\right)  . \label{Omega-Psi_h-s-vaphi-SO(6)} 
\eeqa

It is convenient to re-express the fundamental scalar in the unitary gauge
\beqa
\Phi_{\hat{a}}
& \overset{}{=} &
\left(\begin{array}{c}
\text{0}_3 \\
{c_\psi}{s_\phi}\\
{s_\psi}{s_\phi}\\ 
{c_\phi}
\end{array}\right) = \left(\begin{array}{c}
\text{0}_3 \\
\frac{h_4 s_\phi}{\sqrt{h_4^2+h_5^2}} \\
\frac{h_5 s_\phi}{\sqrt{h_4^2+h_5^2}} \\ 
{c_\phi}
\end{array}\right) =  \left(\begin{array}{c}
\text{0}_3 \\
\frac{h}{f}\\
\frac{s}{f} \\
\sqrt{1- \frac{h^2+s^2}{f^2}}
\end{array}\right)  ,
\qquad \quad 
 \label{Omega-Psi_h-s-SO(6)} 
\eeqa
where in the second equality, we change into $(h,s)$ basis by using definitions in Eq.(\ref{h-s_phi-psi}). This gives the square root non-linear parametrization of $SO(6)$ in analogy to that of $SO(4)$ in Eq.(\ref{pi_nonlinear}).
In the absence of the singlet $s$,  the fundamental representation of the scalar recovers that of MCHM in $SO(5)/SO(4)$ as~\cite{Agashe:2004rs,Contino:2010rs}
\beqa
\Phi_{\hat{a}}
= \left(\begin{array}{c}  
s_h  s_\varphi \hat\varphi_a\\ 
{s_h} c_\varphi \\
0 \\
{c_h}
\end{array}\right) \overset{\varphi\to 0}{=} \left(\begin{array}{c}  
0\\ 
{s_h}\\
0 \\
{c_h}
\end{array}\right).  \qquad\quad \label{Omega-Psi_h-SO(6)} 
\eeqa

\subsubsection{Embedding with rotation $\theta$}

The vacuum can be associated with a angle $\theta$, as the rotation angle in a $S^5$ unit sphere.  
The vacuum in the fundamental representation in Eq.(\ref{Sigma_0}), i.e., a $5$-dimensional unit vector $\Sigma_0$, under the rotation $R(\theta)$ in Eq.(\ref{R_theta}), 
becomes
\beqa
 \Phi_{0}(\theta) &=& R(\theta)\Phi_0 
=\left(\begin{array}{c}
{ 0}_{3\times1} \\ 
{s_\theta}\\
{0} \\
{c_\theta}\\
\end{array}\right)
\eeqa
where $0_{3\times 1}= (0,0,0)^T$, and the angle $\theta$ parameterize the misalignment of the vacuum with respect to the original vacuum. 


The pseudo-Nambu goldstone bosons (PNGBs) in the fundamental representation of $SO(6)$ in Eq.(\ref{Omega-Psi_h-s-vaphi-SO(6)}) becomes
\beqa
\Phi_{\hat{a}}
&=& \left(\begin{array}{c}
 \left(c_{\theta } s_{\phi }-2 \hat{h}_4 s_{\theta } s_{\frac{\phi }{2}}^2\right) \hat{\vec{h}}^T  \\
\hat{h}_4 c_{\theta } s_{\phi }-2\hat{h}_4^2 s_{\theta } s_{\frac{\phi }{2}}^2+s_{\theta } \\
\hat{h}_5 \left(c_{\theta } s_{\phi }-2 \hat{h}_4 s_{\theta } s_{\frac{\phi }{2}}^2\right) \\
c_{\theta } c_{\phi }-\hat{h}_4 s_{\theta } s_{\phi }
\end{array}\right) 
 \overset{\theta = 2n\pi}{=} \left(\begin{array}{c}
\hat{h}_1 s_{\phi }\\
\hat{h}_2 s_{\phi }\\
\hat{h}_3 s_{\phi }\\
\hat{h}_4 s_{\phi }\\
\hat{h}_5 s_{\phi }\\
c_{\phi } \\
 \end{array}\right)  , \qquad \quad
\eeqa
where $\hat{\vec{h}} = (\hat{h}_1,\hat{h}_2,\hat{h}_3)$ with $\hat{h}_a\equiv h_a/\phi$, $a=1,2,3$. The SM Higgs doublet can be defined as through the first four elements of the fundamental $SO(6)$ scalar as Eq.(\ref{Phi-H}).
Having the $\hat{h}_5 $ d.o.f decoupled, it reduces to 
\beqa
\Phi_{\hat{a}}
\overset{\hat{h}_5 \sim 0 }{=} \left(\begin{array}{c}
{ s_\phi} \hat{\vec{h}}^T \\ 
{\hat{h}_4 c_{\theta } s_\phi+c_\phi s_{\theta }}\\
0 \\
{- \hat{h}_4 s_\phi s_{\theta } + c_\phi c_{\theta }}\\
\end{array}\right), \label{Phi_h4=1}
\eeqa
where $\hat{\vec{h}}=(\hat{h}_1,\hat{h}_2,\hat{h}_3)$. In the GB-less limit, i.e., $\varphi\to 0$ by redefining $\hat{h}_{1,2,3}=\varphi_{1,2,3}/v$, $\hat{h}_4=c_\psi$ and $\hat{h}_5=s_\psi$, one has
\beqa
\Phi_{\hat{a}} \left(\theta\right) = \Omega(h) \Phi_0(\theta)   
   \overset{\theta=0}{=} 
   \left(\begin{array}{c}
   \frac{\varphi_1 }{v}c_{\psi } s_{\phi } \\
   \frac{\varphi_2 }{v}c_{\psi } s_{\phi } \\
   \frac{\varphi_3 }{v}c_{\psi } s_{\phi } \\
   c_{\psi } s_{\phi } \\
   s_{\psi } s_{\phi } \\
   c_{\phi } \\
   \end{array}\right)  \overset{f\to \infty}{=} \Phi_0(\theta),.\qquad \quad \label{Omega-Psi_h-s-theta-SO(6)}
\eeqa


When $\theta=0$, the SM electroweak group is unbroken, and are embedded into the global $SO(4)$ symmetry, and the $3+1=4$ EW GBs forms a complex doublet of $SU(2)_L$. 
When $\psi=0$ with a small $\theta$, then
\beqa
\Phi\left(\theta\right)  
& \overset{\theta\sim 0}{=}& \left(\begin{array}{c}
 \sin  \left(\frac{h}{ f}+\theta \right) \frac{\vev\varphi^T }{v} \\
\sin\left(\frac{h}{f}+\theta \right) \\
0 \\
\cos \left(\frac{h }{f}+\theta \right)\\
\end{array}\right). \label{Phi_theta}
\eeqa

When $\theta\ne 0$, the SM vector bosons gauge the $SO(6)/SO(5)$ broken generators $T^{\hat{a}}$.  It triggers the spontaneous symmetry breaking so that three EW GBs are eaten to give mass to the $W^\pm$ and $Z$, respectively, while a fourth one is identified as the Higgs boson. Therefore, the EWSB is due to the misalignment $\theta$, which can be generated at the loop level as long as the GB $4$-vector acquires a vev $\vev{h}\ne 0$, so that 
\beqa
\theta\equiv \frac{\vev{h}}{f}. \label{theta_hv}
\eeqa

It can be checked that under the automorphism symmetry in Eq.(\ref{RR_theta}), the vacuum is inverse to itself
\beqa
{\mathcal R}(\theta) \Phi(\theta) &=&  \left(\begin{array}{c}
\frac{2 }{v}\sin \left(\frac{h}{2 f}\right) \cos \left(\frac{h}{2 f}+\theta \right) \vev\varphi^T \\
\sin\left(\frac{h}{f}-\theta \right) \\
0 \\
-\cos \left(\frac{h }{f}-\theta \right)\\
\end{array}\right) \nn\\
&\overset{h=0}{=} & - \Phi(\theta)  .
\eeqa
Thus, given a generic $\theta$, the action of ${\mathcal R}$ is linear on the $SO(6)/SO(5)$ GBs at scale $f$, while is non-linear on the Higgs field $h$ as well as the $SU(2)_L$ GBs at scale $v$.


Thus, according to Eq.(\ref{Phi_theta}), the fundamental scalar in $SO(6)$ becomes 
\beqa
\Phi\left(h_{\hat{a}}\right) 
&=& \left(\begin{array}{c}
\sin\left(\frac{h}{f}+\theta\right) \left(\begin{array}{c}
{\hat{\varphi}^1 s_\varphi} \\ 
{\hat{\varphi}^2 s_\varphi}\\
{\hat{\varphi}^3 s_\varphi}\\
c_\varphi \\
\end{array}\right)\\
0\\
\cos\left(\frac{h}{f}+\theta\right)\\
\end{array}\right)   ,  \label{Sigma_pi-theta}
\eeqa
where $h$ parameterize the $SO(4)$ invariant quantum fluctuation around the vacuum $ \vev{h}  = f \theta$. 


Under the left-right parity symmetry in Eq.(\ref{P_L-R}), the GBs changes sign, while the Higgs is invariant,
\beqa
\varphi^{\hat{a}} \to -\varphi^{\hat{a}} , \quad h \to h.
\eeqa

From the two definitions in Eqs.(\ref{Phi_h4=1}) and (\ref{Sigma_pi-theta}), one can identify the relations between $\hat{h}$ and $\hat{\phi}$ as
\beqa
&& \hat{h}^a s_\phi  = \sin\left(\frac{h}{f}+\theta\right) \frac{\pi^a}{v} ,\nn\\
&& \hat{h}^4 s_\phi = c_\theta \sin\left(\frac{h}{f}+\theta\right)  \sqrt{1-\frac{\pi^2}{v^2}} - s_\theta  \cos\left(\frac{h}{f}+\theta\right) , \nn\\
&& c_\phi = c_\theta \cos\left(\frac{h}{f}+\theta\right) + s_\theta \sin\left(\frac{h}{f}+\theta\right) \sqrt{1-\frac{\pi^2}{v^2}}, \qquad
\eeqa
which leads to a re-parametrization between $\hat{h}_a$ and $(\hat\phi_a,\hat{h}^4)$ with $a=1,2,3$ in $SO(4)$. In above, we rewrite the linearly realized $\varphi$ in terms of non-linearly realized $\pi$ by using Eq.(\ref{pi_varphi}).




\section{High energy effective chiral Lagrangian}
\label{sec:EWCL}

\subsection{Non-linear chiral Lagrangian for Sigma parametrization}

To construct the nonlinear chiral Lagrangian, different building blocks in the last section are used.
The building blocks for the Sigma parametrization in NMCHM are
\beqa
\overline{\mathrm{W}}_{\mu\nu}, \quad  \overline{\mathrm{B}}_{\mu\nu}, \quad \mathbf{\Sigma}, \quad \overline{\mathrm{V}}_{\mu}, \quad \overline{\mathbf{T}} ,  \label{BL_SM-Sigma}
\eeqa
where the corresponding gauge fields $\overline{\mathrm{W}}_{\mu}$ and $\overline{\mathrm{B}}_{\mu}$ are defined in Eq.(\ref{W_L-B}).
$Q_{L}^{a}$ with $a = 1,2,3$ and $Q_Y$ are the $SU(2)_L\times U(1)_Y$ generators embedded in $SO(6)$ 
\beqa
Q_L^a \equiv T_L^a, \quad Q_Y \equiv T_R^3,
\qquad \quad \label{Q_L-Y}
\eeqa
where we have identified $SU(2)_L$ charge operator as $T_L^a$ and the hypercharge operator as $T_R^3$ as defined in Eq.(\ref{T_L-R}). When fermions are taken into account, the realistic hypercharge operator is defined as $Y=T_R^3+X$, where $X$ is a new non-vanishing charge under an additional $U(1)_X$, in order to reproduce the correct hypercharges of fermions.

\subsubsection{CP-even sector}
For the CP-even case, one has thirteen independent operators as up to ${\mathcal O}(p^4)$,
\beqa
\mathscr{L}_{\text { high }}=\mathscr{L}_{\text { high }}^{p^{2}}+\mathscr{L}_{\text { high }}^{p^{4}} ,  \label{L_high_CP-even}
\eeqa
where the high energy effective chiral Lagrangian $\mathscr{L}_{\text { high }}$ becomes
\beqa
&& \mathscr{L}_{\text { high }}^{p^{2}} = \overline{\mathcal{L}}_{C}, \nn\\
&& \mathscr{L}_{\mathrm{high}}^{p^{4}} = \overline{\mathcal{L}}_{B}+\overline{\mathcal{L}}_{W}+\overline{c}_{B \Sigma} \overline{\mathcal{L}}_{B \Sigma}+\overline{c}_{W \Sigma} \overline{\mathcal{L}}_{W \Sigma}  + \sum_{i=1}^8 c_i \overline{\mathcal{L}}_{i} , \qquad \quad
\eeqa
with the coefficients $\overline{c}_i$ are expected to be all of the same order of magnitude as
\beqa
\begin{array}{lll} 
\overline{\mathcal{L}}_{C} &=&-\frac{f^{2}}{4} \operatorname{Tr}\left(\overline{\mathrm{V}}_{\mu} \overline{\mathrm{V}}^{\mu}\right)       \\
\overline{\mathcal{L}}_{B} &=&-\frac{1}{4}  g^{\prime 2}\operatorname{Tr}\left(\overline{\mathrm{B}}_{\mu \nu} \overline{\mathrm{B}}^{\mu \nu}\right)     \\
\overline{\mathcal{L}}_{W} &=&-\frac{1}{4} g^2 \operatorname{Tr}\left(\overline{\mathrm{W}}_{\mu \nu} \overline{\mathrm{W}}^{\mu \nu}\right)      \\
\overline{\mathcal{L}}_{B \Sigma} &=&g^{\prime 2} \operatorname{Tr}\left(\boldsymbol{\Sigma} \overline{\mathrm{B}}_{\mu \nu} \boldsymbol{\Sigma}^{-1} \overline{\mathrm{B}}^{\mu \nu}\right)     \\
\overline{\mathcal{L}}_{W \Sigma} &=&g^{2} \operatorname{Tr}\left(\boldsymbol{\Sigma} \overline{\mathrm{W}}_{\mu \nu} \mathrm{\boldsymbol{\Sigma}}^{-1} \overline{\mathrm{W}}^{\mu \nu}\right)       \\
\overline{\mathcal{L}}_{1} &=&g g^{\prime} \operatorname{Tr}\left(\boldsymbol{\Sigma} \overline{\mathrm{B}}_{\mu \nu} \boldsymbol{\Sigma}^{-1} \overline{\mathrm{W}}^{\mu \nu}\right)     \\
 \overline{\mathcal{L}}_{2} &=&i g^{\prime} \operatorname{Tr}\left(\overline{\mathrm{B}}_{\mu \nu}[\overline{\mathrm{V}}^{\mu}, \overline{\mathrm{V}}^{\nu}]\right)   \\
 \overline{\mathcal{L}}_{3} &=&i g \operatorname{Tr}\left(\overline{\mathrm{W}}_{\mu \nu}\left[\overline{\mathrm{V}}^{\mu}, \overline{\mathrm{V}}^{\nu}\right]\right)   \\
  \overline{\mathcal{L}}_{4} &=&\operatorname{Tr}\left(\overline{\mathrm{V}}_{\mu} \overline{\mathrm{V}}^{\mu}\right) \operatorname{Tr}\left(\overline{\mathrm{V}}_{\nu} \overline{\mathrm{V}}^{\nu}\right)  \\
  \overline{\mathcal{L}}_{5} &=&\operatorname{Tr}\left(\overline{\mathrm{V}}_{\mu} \overline{\mathrm{V}}_{\nu}\right) \operatorname{Tr}\left(\overline{\mathrm{V}}^{\mu} \overline{\mathrm{V}}^{\nu}\right)  \\
  \overline{\mathcal{L}}_{6} &=&\operatorname{Tr}\left( (\mathrm{D}_{\mu} \overline{\mathrm{V}}^{\mu})^{2}\right)  \\
  \overline{\mathcal{L}}_{7} &=&\operatorname{Tr}\left(\overline{\mathrm{V}}_{\mu} \overline{\mathrm{V}}^{\mu} \overline{\mathrm{V}}_{\nu} \overline{\mathrm{V}}^{\nu}\right) \\
  \overline{\mathcal{L}}_{8} &=&\operatorname{Tr}\left(\overline{\mathrm{V}}_{\mu} \overline{\mathrm{V}}_{\nu} \overline{\mathrm{V}}^{\mu} \overline{\mathrm{V}}^{\nu}\right),  \\
\end{array}   \label{Lbar_CP-even}
\eeqa
where $f$ is strong dynamics scale, that can be related to the EW scale through Eq.(\ref{xi}). $\overline{\mathrm{V}}_\mu$ is defined in Eq.(\ref{Vbar}). The corresponding covariant derivative is defined in Eq.(\ref{DVbar}) as
\beqa
\mathrm{D}_{\mu} \overline{\mathrm{V}}^{\mu} =\partial_{\mu} \overline{\mathrm{V}}^{\mu}+i g_{A} [ \overline{\mathrm{A}}_\mu, \overline{\mathrm{V}}^{\mu} ] . \label{DVbar}
\eeqa
For the SM, it becomes
\beqa
\mathrm{D}_{\mu} \overline{\mathrm{V}}^{\mu}=\partial_{\mu} \overline{\mathrm{V}}^{\mu}+i g\left[\overline{\mathrm{W}}_{\mu}, \overline{\mathrm{V}}^{\mu}\right]+i g^{\prime}\left[\overline{\mathrm{B}}_{\mu}, \overline{\mathrm{V}}^{\mu}\right] .
\eeqa
The gauging of the SM symmetry represents an explicit breaking of global symmetry ${\mathcal G}$. For SM, the explicit expression $g_A\overline{A}_\mu = g\overline{W}_\mu+g' \overline{B}_\mu$ is given in Eq.(\ref{Wb-Bb}). The gauging of the SM symmetry represents an explicit breaking of global symmetry $SO(6)$.

\subsubsection{CP-odd sector}
For the CP-odd case, the EW high-energy chiral Lagrangian describing the bosonic interactions, up to the fourth derivatives, one has six independent operators as
\beqa
\widetilde{\mathscr{L}}_{\text { high }}=\widetilde{\mathscr{L}}_{\text { high }}^{p^{2}}+\widetilde{\mathscr{L}}_{\text { high }}^{p^{4}} ,  \label{L_high_CP-odd}
\eeqa
where the leading o
\beqa
 \widetilde{\mathscr{L}}_{\text { high }}^{p^{2}} &=& \overline{c}_{\widetilde{W}} \overline{\mathcal{L}}_{ \widetilde{W}}  , \nn\\
 \widetilde{\mathscr{L}}_{\text { high }}^{p^{4}} &=&  \overline{c}_{\widetilde{B} \Sigma} \overline{\mathcal{L}}_{\widetilde{B} \Sigma}+ \overline{c}_{\widetilde{W} \Sigma} \overline{\mathcal{L}}_{\widetilde{W} \Sigma} + \overline{c}_{\widetilde{1}} \overline{\mathcal{L}}_{\widetilde{1}} +\overline{c}_{\widetilde{2}} \overline{\mathcal{L}}_{\widetilde{2}}+\overline{c}_{\widetilde{3}} \overline{\mathcal{L}}_{\widetilde{3}}, \qquad \quad
\eeqa
with $c_{\widetilde{i}}$ as the coefficients corresponding to the operators as
\beqa
\begin{array}{lll}
\overline{\mathcal{L}}_{ \widetilde{W}} &=&-\frac{g^{2}}{4} \operatorname{Tr}\left(\widetilde{\overline{\mathrm{W}}}_{\mu \nu} \overline{\mathrm{W}}^{\mu \nu}\right)   \\ 
\overline{\mathcal{L}}_{\widetilde{B} \Sigma} &=&g^{\prime 2} \operatorname{Tr}\left(\widetilde{\overline{\mathrm{B}}}_{\mu \nu} \mathbf{\Sigma} \overline{\mathrm{B}}^{\mu \nu} \mathbf{\Sigma}^{-1}\right)     \\ 
\overline{\mathcal{L}}_{\widetilde{W} \Sigma} &=&g^{2} \operatorname{Tr}\left(\widetilde{\overline{\mathrm{W}}}_{\mu \nu} \mathbf{\Sigma} \overline{\mathrm{W}}^{\mu \nu} \mathbf{\Sigma}^{-1}\right)     \\
 \overline{\mathcal{L}}_{\widetilde{1}} &=&g g^{\prime} \operatorname{Tr}\left(\widetilde{\overline{\mathrm{W}}}_{\mu \nu} \mathbf{\Sigma} \overline{\mathrm{B}}^{\mu\nu} \mathbf{\Sigma}^{-1}\right)  \\
 \overline{\mathcal{L}}_{\widetilde{2}} &=&\epsilon_{\mu \nu \rho \sigma} \operatorname{Tr}\left(\overline{\mathbf{T}}\left[\overline{\mathrm{V}}^{\mu}, \overline{\mathrm{V}}^{\nu}\right]\right) \operatorname{Tr}\left(\overline{\mathbf{T}}\left[\overline{\mathrm{V}}^{\rho}, \overline{\mathrm{V}}^{\sigma}\right]\right) \\
 \overline{\mathcal{L}}_{\widetilde{3}} &=&\epsilon_{\mu \nu \rho \sigma} \operatorname{Tr}\left(\overline{\mathrm{V}}^{\mu} \overline{\mathrm{V}}^{\nu} \overline{\mathrm{V}}^{\rho} \overline{\mathrm{V}}^{\sigma}\right),  \\
\end{array} \label{Lbar_CP-odd}
\eeqa
where $\overline{\mathbf{T}}$ is defined in Eq.(\ref{Tbar}).

By substituting the explicit expressions of the building blocks for the Sigma parametrization in Eq.(\ref{BL_SM-Sigma}), into the CP-even or the CP-odd operators in the high-energy effective Lagrangian basis $\overline{\mathcal{L}}_{\text{high}}$ of Eqs.(\ref{L_high_CP-even}) or (\ref{L_high_CP-odd}), respectively, one can produce the low energy effective chiral Lagrangian basis ${\mathcal L}_{\text{low}}$ in Eqs.(\ref{L_CP-even}) or (\ref{L_CP-odd}) for the NMCHM with $SO(6)/SO(5)$ symmetry breaking pattern, as functions of the SM gauge bosons, the SM would-be GBs and the CP-even scalar field $h$ as well as the CP-odd scalar singlet $s$.

\subsection{Non-linear chiral Lagrangian in the Omega parametrization}

To construct the nonlinear chiral Lagrangian, different building blocks in the last section are utilized.
The building blocks for Omega in NMCHM are
\beqa
d_\mu, \quad  e_{\mu\nu}, \quad f_{\mu\nu} .\label{BL_SM-Omega}
\eeqa
With the building blocks shown above, we can obtain all of the low energy effective Lagrangian up to $p^4$ order.





In CCWZ approach, according to Eq.(\ref{omega_d-E}), by projecting upon the broken generators of the Maurer-Cartan form, one obtains the elements of $d_\mu$ and $e_\mu$ fields, along broken and unbroken generators, respectively, as expressed more explicitly as

\beqa
&& d_{\mu}^{\hat{a}} =  \text{Tr}\left( \omega_\mu T^{\hat{a}}\right)  =  - i  \text{Tr} \left[ \left( \Omega^\dag \partial_\mu \Omega \right) T^{\hat{a}} \right]  
 \nn\\
&& e_{\mu}^{a}  = \text{Tr} \left( \omega_\mu T^{a} \right) =  - i  \text{Tr} \left[ \left( \Omega^\dag \partial_\mu \Omega \right) T^{a} \right]  , 
\eeqa
where $\Omega$ is defined in Eq.(\ref{Sigma-Omega_h}), $\Omega^\dag = \Omega^{-1}$ and $\omega_\mu \equiv -i \Omega^{-1}\partial_\mu \Omega $ is defined as the Maurer-Cartan one form $d\omega$ as introduced in Eq.(\ref{omega_d-E}) in the vector formalism as
\begin{widetext}
\beqa
\omega_\mu 
&=& -i \left(
\begin{array}{c|c}
- \frac{1}{2}s_\phi^2 [ (\partial_\mu  \hat{\phi} ) \hat{\phi}^{T}  -  \hat{\phi} (\partial_\mu   \hat{\phi}^{T}) ] & \frac{1 }{f} (\partial_\mu |\phi|) \hat{\phi} + s_\phi  (\partial_\mu \hat{\phi})      -  (1-c_\phi)  s_\phi  \hat{\phi} \hat{\phi}^{T}   (\partial_\mu \hat{\phi}) \\
 \hline
-   \frac{1}{f} (\partial_\mu |\phi|)   \hat{\phi}^{T}     +  s_\phi (1-c_\phi)  (\partial_\mu \hat{\phi}^{T}) ( \hat{\phi} \hat{\phi}^{T}) - s_\phi  (\partial_\mu  \hat{\phi}^{T})  & s_\phi  (\partial_\mu \hat{\phi}) \\
\end{array}
\right), \qquad \quad \label{omega_m}
\eeqa
where $\hat\phi  \hat\phi^T = \textbf{1}$ so that $\hat{\phi} \hat{\phi}^{T} =  (\hat{\phi} \hat{\phi}^{T})^2$ and we also used $\partial_\mu(\hat\phi^T \hat\phi)=0$.
\end{widetext}

In the NMCHM, the broken generators $T^{\hat{a}}$ are given in Eqs.(\ref{SO6-Tc_1-3}) and (\ref{SO6-Tc_4-5}), and can be expressed as
\beqa
\quad T^{\hat{a}}= -i \frac{1}{\sqrt{2}}\left(
\begin{array}{c|c}
\textbf{0}_4&  \text{1}_{\hat{a}} \\
 \hline
- \text{1}_{\hat{a}}^{T} &0 \\
\end{array}
\right),
\eeqa
where $\textbf{0}_4=\text{diag}(0,0,0,0)$, $\text{1}_{\hat{a}}$ is a $4$-vector with only non-vanishing component $1$ along the $\hat{a}$-th direction. The factor in front is due to that we have used the fact of normalization of generators in Eq.(\ref{T_norm}).

With the explicit expression of $\Omega_\mu$ given in Eq.(\ref{omega_m}), one can read
\beqa
d_\mu^{\hat{a}}  
&=& \sqrt{2}\left(   \frac{1}{f} (\partial_\mu |\phi|)   \hat{\phi}^{\hat{a}} + s_\phi c_\phi  (\partial_\mu \hat{\phi}^{\hat{a}})  \right) . 
\eeqa
By using the identities as below
\beqa
&& \partial_\mu ( \hat{\phi} \hat{\phi}^{T}) =\frac{ \partial_\mu ( \phi \phi^{T})}{|\phi|^2}  -2  \frac{\partial_\mu |\phi|}{|\phi|^3} (\phi \phi^{T})   ,\nn\\
&& \partial_\mu \hat\phi = \frac{1}{|\phi|}\partial_\mu \phi - \frac{\partial_\mu |\phi|}{|\phi|^2} \phi ,
\eeqa
$d_\mu$ can be re-expressed as 
\beqa
d_\mu^{\hat{a}}  
= \sqrt{2}\bigg[ \frac{1}{2}\left( \frac{2s_\phi c_\phi}{|\phi|}  - \frac{1}{f} \right) \frac{\phi^{\hat{a}}}{|\phi|^2} \partial_\mu |\phi|^2 - \frac{2s_\phi c_\phi}{|\phi|} \partial_\mu \phi^{\hat{a}}  \bigg]. \qquad \quad
\eeqa
The kinetic derivative $\partial_\mu$ above can be generalized to be the covariant derivative as that in Eq.(\ref{Dm_Sigma-Aa-Aat})
\beqa
\partial_\mu \to D_\mu = \partial_\mu + i [A_\mu^a T^a , \Sigma] + i \{ A_\mu^{\hat{a}}T^{\hat{a}}, \Sigma \}.
\eeqa
For the SM, $\{W_\mu^a, B_\mu\}\in A_\mu^a$, we has
\beqa
D_\mu = \partial_\mu + i [W_\mu^a Q_L^a , \Sigma] + i [B_\mu Q_Y, \Sigma]. 
\eeqa
where the gauge couplings $g,g'$ are absorbed into the $W_\mu^a$ and $B_\mu$, respectively. As stated before, the gauging of the SM symmetry represents explicit breaking of the global symmetry $SO(6)$. The explicit expression of the covariant derivatives terms as well as the Omega parametrization building blocks in EW sector of NMCHM are shown in Appendix.~\ref{app:cov_deri}.

\subsubsection{$p^2$ order in Omega parametrization}

At the $p^2$ order where at most two derivative, there is only one CP-even high energy effective Lagrangian as
\beqa
{\mathcal L}^{(2)} = \frac{f^2}{4}\text{Tr}\left(d_\mu d^\mu\right).
\eeqa

\subsubsection{$p^4$ order in Omega parametrization}

At the $p^4$ order where at most four derivative, there is a complete basis for high-energy effective Lagrangian with $11$ operators
\beqa
{\mathcal L}^{(4)} = \sum_i c_i {\mathcal O}_i,
\eeqa
where $i=1,\cdots, 11$, and to be explicitly, are

The next-leading order CP-even and odd operators can be constructed from the building blocks as below~\cite{Alonso:2014wta,Buchalla:2014eca},
\beqa
\begin{array}{lll}
 {\mathcal O }_1 &=& \left[ \text{Tr}\left(d_\mu d^\mu\right) \right]^2 ,\\
 {\mathcal O}_2 &=& \text{Tr}\left( d_\mu d_\nu \right) \text{Tr}\left( d^\mu d^\nu \right). ,\\
 {\mathcal O}_3 &=&  \text{Tr}\left[ (e_{\mu\nu}^L)^2 - (e_{\mu\nu}^R )^2 \right], \\
 {\mathcal O}_4^\pm &=& \text{Tr}\left[ (f_{\mu\nu}^L \pm f_{\mu\nu}^R ) i[ d^\mu, d^\nu] \right] ,\\
 {\mathcal O}_5^+ &=& \text{Tr}\left[ (f_{\mu\nu}^-)^2 \right],  \\
 {\mathcal O}_5^- &=& \text{Tr}\left[ (f_{\mu\nu}^L)^2- (f_{\mu\nu}^R)^2 \right] , \\
\end{array} \label{O_1-7}
\eeqa
and~\cite{Contino:2011np} 
\beqa
\begin{array}{lll}
O_{6}^{\pm} &= \operatorname{Tr}\left[\left(\widetilde{f}_{\mu \nu}^{L}\pm \widetilde{f}_{\mu \nu}^{R}\right) i\left[d^{\mu}, d^{\nu}\right]\right] \\
O_{7}^{+} &= \operatorname{Tr}\left(\widetilde{f}_{\mu \nu}^{-} f^{-\mu\nu}\right) \\ 
O_{7}^{-} &= \operatorname{Tr}\left(\widetilde{f}^{L\mu \nu} f_{\mu\nu}^{L}-\widetilde{f}^{R\mu \nu} f_{\mu \nu}^{R}\right), 
\end{array} \label{O_8-11}
\eeqa
where $\widetilde{f}_{\mu\nu}^\pm = \epsilon_{\mu\nu\rho\sigma}f^{\rho\sigma \pm}$ is a dual antisymmetry tensor.

\section{Matching to EW chiral Lagrangian}
\label{sec:HEEFT}

\subsection{CP-even sector}

\subsubsection{The ${ }p^2$ order: kinetics and mass terms}

The leading order low energy effective Lagrangian up to the two derivatives is
\beqa
\overline{\mathcal{L}}_{C} 
&=&  \frac{1}{2} (\partial_\mu\phi)^2 + 2 \sin^2\left( \frac{\phi}{2f} \right)  (\partial_\mu\psi)^2 \nn\\
&& + \frac{4}{\xi}  \cos^2\left(\frac{\psi}{f}\right) \sin ^{2}\left(\frac{\phi}{2 f}\right)   \mathcal{L}_{C},  \qquad \quad \label{LC_Omega}
\eeqa
where  we have used Eq.(\ref{Vbar}) in the $\Omega$ parametrization. 
The first line are canonically normalized kinetic terms for the field $\phi$ and the phase $\psi$, respectively, in analogy to $\mathcal{L}_{H}$ defined in Eq.(\ref{L_H}).
$\xi$ 
quantifying the Higgs non-linearity due to the strong dynamics breaking scale $f$.
The leading order operator $\mathcal{L}_{C}$ is associated with the two gauge bosons as defined in Eq.(\ref{L_C-T}) or more explicitly shown in Eq.(\ref{LCEFT_CP-even}) in Appendix.~\ref{app:LCEFT}.  
 

In the absence of the singlet, i.e., $\psi=0$ (or $s=0$), the leading order high-energy effective Lagrangian in the NMCHM just leads to those in the MCHM as~\cite{Agashe:2004rs,Contino:2010rs}
\beqa
\overline{\mathcal{L}}_{C} &= &  \mathcal{L}_{H} + \frac{4}{\xi} \sin ^{2}\left(\frac{h}{2 f}\right) \mathcal{L}_{C},
\eeqa
where the first two terms describe the low energy projection of the custodial preserving two derivative operators. The first term $\mathcal{L}_{H}=(\partial_\mu h)^2/2$ give the correctly normalized kinetic term of $h$.

The identities above are model dependent relations relating the d.o.f of high energy sector to that of low energy sector. 
These identities are still valid even after EWSB in which Higgs obtains vev at EW scale $v$ as will be discussed more in Eq.(\ref{xi_psi-phi}).

\subsubsection{Vacuum expectation values}

After EWSB, one can derive the mass terms of the gauge boson $W$ and $Z$, from which, one can read the EW vev as
\beqa
\frac{v}{f} = \sqrt\xi =  \cos\left(\frac{v_\psi}{f}\right) \sin\left(\frac{v_\phi}{f}\right) ,  \qquad \label{xi_psi-phi}
\eeqa
where $v_\phi \equiv \vev\phi $ and $ v_\psi \equiv \vev\psi $ are vacuum expectation values (vev) of the $\phi$ and $\psi$, respectively.
In the $(h,s)$ basis, the EW vev $\vev{h} = v_h$ and the singlet vev $\vev{s}=v_s$ can be expresses, respectively, as
\beqa
&&  \tan\left(\frac{v_\psi}{f}\right) = \frac{v_s}{v_h}, \qquad \quad  \label{v_psi-v_s-v_h}
\eeqa
which recovers that given in Eq.(\ref{xi_psi-phi_0}). This supplies as a model dependent structure for the NMCHM, that related to the EW vev $v$, Higgs vevs $v_\phi$, vacuum phase $v_\psi$, and the strong dynamics breaking scale $f$.

One can transfer from the $(\phi,\psi)$ basis to the $(h_4,h_5)$ basis, 
\beqa
\frac{v}{f} = \sqrt\xi = \frac{v_{4}}{\sqrt{v_{4}^2+v_{5}^2}}  \sin
   \left(\frac{\sqrt{v_{4}^2+v_{5}^2}}{f}\right),  \qquad \label{xi_h4-h5}
\eeqa
where $v_{4} \equiv \vev{h_4} $ and $ v_{5} \equiv \vev{h_5} $ are vevs of $h_4$ and $h_5$, respectively. 
In addition one could transfer from the $(h_4,h_5)$ basis to the $(h,s)$ basis, 
\beqa
\frac{v}{f} = \sqrt\xi  = \frac{v_h}{f}, \label{v-vh}
\eeqa
where $v_h\equiv \vev{h}$ is the vev of $h$, which is exactly equal to the EW vev.

From Eqs.(\ref{v_psi-v_s-v_h}) and (\ref{xi_psi-phi}), the vevs $(v_\phi, v_\psi)$ can be expressed in terms of $(v_h, v_s)$ as
\beqa
&& v_\psi = f \arctan\left(\frac{v_s}{v_h}\right),  \nn\\  
&& v_\phi 
= f \arcsin{\left(\frac{\sqrt{v_h^2+ v_s^2}}{f} \right)} ,    \label{v_phi-v_psi}
\eeqa
where in the last equality, we have used the identification of $v_h=v$ in Eq.(\ref{v-vh}).
In the EWSB vacuum identified as $ v_\psi=0$ (or $v_s=0$), one has
which leads to 
\beqa
&& v_\phi = f \arcsin\sqrt\xi .  
\eeqa
In this case, the non-linearity exactly recovers that in the MCHM as
\beqa
\xi = \sin\left({v_\phi}/{f}\right). 
\eeqa

\subsubsection{${ }p^4$  order: Higgs couplings to gauge bosons}

In the NMCHM, the operators in high energy effective chiral Lagrangian in Eq.(\ref{Lbar_CP-even}), are related to the low energy chiral EW Lagrangian in Eq.(\ref{L_CP-even})
\beqa
\begin{array}{lll} 
\overline{\mathcal{L}}_{B} &=& \mathcal{L}_{B} ,  \\
\overline{\mathcal{L}}_{W} &=& \mathcal{L}_{W}  ,  \\
\overline{\mathcal{L}}_{B \Sigma} &=& -4  \left[  1-\cos ^2\left(\frac{\psi }{f}\right) \sin ^2\left(\frac{\phi }{2 f}\right)\right]  \mathcal{L}_{B}   , \\
\overline{\mathcal{L}}_{W \Sigma} &=& -4  \left[  1-\cos ^2\left(\frac{\psi }{f}\right) \sin ^2\left(\frac{\phi }{2 f}\right)\right]  {\mathcal L}_W  , \\
\overline{\mathcal{L}}_{1}  &=& \cos ^2\left(\frac{\psi }{f}\right) \sin ^2\left(\frac{\phi }{2 f}\right)\mathcal{L}_{1}  ,  \\ 
 \overline{\mathcal{L}}_{2} &=& \cos ^2\left(\frac{\psi }{f}\right) \sin ^2\left(\frac{\phi }{2 f}\right) {\mathcal L}_2 \\
 && + \sqrt{\xi } \cos ^2\left(\frac{\psi }{f}\right) \sin \left(\frac{\phi }{f}\right) {\mathcal L}_4^{(\phi)}   \\
   && - 2 \sqrt{\xi } \sin\left(\frac{2\psi }{f}\right) \sin^2\left(\frac{\phi }{2f}\right) {\mathcal L}_4^{(\psi)} , \\
\overline{\mathcal{L}}_{3} &=& 2 \cos ^2\left(\frac{\psi }{f}\right) \sin ^2\left(\frac{\phi }{2 f}\right) {\mathcal L}_3  \\
&& -2 \sqrt{\xi } \cos ^2\left(\frac{\psi }{f}\right) \sin \left(\frac{\phi }{f}\right) {\mathcal L}_5^{(\phi)} \\
&& + 4 \sqrt{\xi } \sin \left(\frac{2\psi }{f}\right) \sin ^2\left(\frac{\phi }{2 f}\right) {\mathcal L}_{5}^{(\psi)} ,   \\
\overline{\mathcal{L}}_{4} &=& 4 \xi^2 {\mathcal L}_{D \phi} + 16 \cos ^4\left(\frac{\psi }{f}\right) \sin ^4\left(\frac{\phi }{2 f}\right) {\mathcal L}_6  \\
&& -16 \xi  \cos ^2\left(\frac{\psi }{f}\right) \sin ^2\left(\frac{\phi }{2 f}\right){\mathcal L}_{20}^{(\phi)}  \\
&& +  64 \xi^2 \sin^4\left(\frac{\phi}{2f} \right) {\mathcal L}_{D \psi} \\
&& -64 \xi  \cos ^2\left(\frac{\psi }{f}\right) \sin ^4\left(\frac{\phi }{2 f}\right){\mathcal L}_{20}^{(\psi)}, \\
 \overline{\mathcal{L}}_{5}&=&  4\xi^2 {\mathcal L}_{D \phi} + 16 \cos ^4\left(\frac{\psi }{f}\right) \sin ^4\left(\frac{\phi }{2 f}\right) {\mathcal L}_{11}  \\
 && -16 \xi  \cos ^2\left(\frac{\psi }{f}\right) \sin ^2\left(\frac{\phi }{2 f}\right) {\mathcal L}_8^{(\phi)} \\
 && + 64 \xi^2 \sin^4\left( \frac{\phi}{2f} \right) {\mathcal L}_{D \psi}  \\
 && -64 \xi  \cos ^2\left(\frac{\psi }{f}\right) \sin ^4\left(\frac{\phi }{2 f}\right)  {\mathcal L}_8^{(\psi)} , \\
 \overline{\mathcal{L}}_{6} &=& - 2 \cos ^2\left(\frac{\psi }{f}\right) \sin ^2\left(\frac{\phi }{2 f}\right) \left[ 1 - \cos ^2\left(\frac{\psi }{f}\right) \sin ^2\left(\frac{\phi }{2 f}\right) \right]  {\mathcal L}_6 \\
 && + 4 \cos ^2\left(\frac{\psi }{f}\right) \sin ^2\left(\frac{\phi }{2 f}\right) {\mathcal L}_9 \\
 && -2 \xi {\mathcal L}_{\square \phi}  + 4 \xi  \cos ^2\left(\frac{\psi }{f}\right) \cos ^2\left(\frac{\phi }{2 f}\right){\mathcal L}_{8}^{(\phi)}  \\
 && - 8 \xi \sin^2\left(\frac{\phi}{2f}\right) {\mathcal L}_{\square \psi}  + 16 \xi  \sin ^2\left(\frac{\psi }{f}\right) \sin ^2\left(\frac{\phi }{2 f}\right) {\mathcal L}_{8}^{(\psi)}   \\
 &&-2 \sqrt{\xi } \cos ^2\left(\frac{\psi }{f}\right) \sin\left(\frac{\phi }{f}\right) ({\mathcal L}_{7}^{(\phi)} - 2{\mathcal L}_{10}^{(\phi)}  )  \\
 && + 4 \sqrt{\xi } \sin \left(\frac{2 \psi }{f}\right) \sin^2\left(\frac{\phi }{2 f}\right) ({\mathcal L}_{7}^{(\psi)} -2{\mathcal L}_{10}^{(\psi)}) , \\
 \overline{\mathcal{L}}_{7}&=& \frac{1}{4}\left(\overline{\mathcal{L}}_{4}+\overline{\mathcal{L}}_{5}\right),  \\
 \overline{\mathcal{L}}_{8}&=& \frac{1}{2} \overline{\mathcal{L}}_{5}, \\
\end{array} \label{Lb-L_CP-even}
\eeqa
where ${\mathcal L}_n^{(\phi)},{\mathcal L}_n^{(\psi)}$ are defined by ${\mathcal L}_n$ with $h$ replaced with terms of $\phi$ or $\psi$, respectively, and similarly for ${\mathcal L}_{D\phi}$ and ${\mathcal L}_{\box \phi}$ etc.  The $\overline{\mathcal{L}}_{7,8}$ do not give independent contributions since they can be expressed in linear combinations of other operators. The traces of four $\overline{\mathrm{V}}_{\mu}$ can be expressed as products of traces of two $\overline{\mathrm{V}}_{\mu}$. Therefore, in a whole, there are $2+6=8$ independent operators. The explicit expression of the low energy chiral Lagrangian in NMCHM are list in Eq.(\ref{LCEFT_CP-even}) in Appendix.~\ref{app:LCEFT}. In the absent of the CP-odd singlet, $\psi=0$ ($s=0$, or $h_5=0$), the result just recover Eq.(4.18) in Ref.~\cite{Alonso:2014wta} in the MCHM. 

\subsection{CP-odd sector}

\subsubsection{${ }p^2$  order: topological $\theta$ phase}

\beqa
\begin{array}{lll} 
\overline{\mathcal{L}}_{ \widetilde{B}} &=& 0  ,   \\
\overline{\mathcal{L}}_{ \widetilde{W}} &=& \mathcal{L}_{\widetilde{W}} ,    \nn\\
\end{array}
\eeqa
The leading order operator with at most two derivative in terms of ${ }(p^2)$ order, is a topological $\theta$ phase term for non-Abelian gauge symmetry in analogous to the QCD $\theta$ term. Namely, in principle, one has EW $\theta$ term as a free parameter in the SM.

\subsubsection{${ }p^4$ order: Higgs couplings to gauge bosons}

For NMCHM, the dimension four operators in high-energy effective Lagrangian in Eq.(\ref{Lbar_CP-odd}), are related to the low energy EW chiral Lagrangian in Eq.(\ref{L_CP-odd}) as
\beqa
\begin{array}{lll} 
\overline{\mathcal{L}}_{\widetilde{B} \Sigma} &=& -4 \left[1-\cos ^2\left(\frac{\psi }{f}\right) \sin ^2\left(\frac{\phi }{2 f}\right)\right]  \mathcal{L}_{\widetilde{B}}  ,  \\
\overline{\mathcal{L}}_{\widetilde{W} \Sigma} &=& - 4 \left[ 1 - \cos ^2\left(\frac{\psi }{f}\right) \sin ^2\left(\frac{\phi }{2 f}\right) \right] \mathcal{L}_{\widetilde{W}}  , \\
\overline{\mathcal{L}}_{\widetilde{1}} &=& \frac{1}{2} \cos ^2\left(\frac{\psi }{f}\right) \sin ^2\left(\frac{\phi }{2 f}\right) \mathcal{L}_{\widetilde{1}} ,\\
\overline{\mathcal{L}}_{\widetilde{2}} &=&   4 \cos ^{4}\left(\frac{\psi}{ f}\right)\sin ^{4}\left(\frac{\phi}{2 f}\right) \left(\mathcal{L}_{\widetilde{B}}-\mathcal{L}_{ \widetilde{W}}\right)  \\
&& +2 \sqrt{\xi} \cos^4 \left(\frac{\psi}{f}\right) \cos \left(\frac{\phi}{2 f}\right)   \sin ^{3}\left(\frac{\phi}{2 f}\right)\left(\mathcal{L}_{\widetilde{2}}+2 \mathcal{L}_{\widetilde{3}}\right) , \\
 \overline{\mathcal{L}}_{\widetilde{3}}  & =& 0,
\end{array} \label{Lb-L_CP-odd}
\eeqa
where we list the explicit expressions of CP-odd low energy chiral effective Lagrangian in Eqs.(\ref{LCEFT_CP-odd}). In the last equality, we have used that $ \overline{\mathcal{L}}_{\widetilde{3}}  = \epsilon^{\mu\nu\rho\sigma} \operatorname{Tr}\left(\overline{\mathrm{V}}_{\mu} \overline{\mathrm{V}}_{\nu}\right) \operatorname{Tr}\left(\overline{\mathrm{V}}_{\rho} \overline{\mathrm{V}}_{\sigma}\right) $ for $SO(6)/SO(5)$. It is interesting to observe that, at ${ }(p^4)$ order, which can be due to contributions from one loops, the low energy chiral Lagrangian of the $\overline{\mathcal{L}}_{\widetilde{B}, \widetilde{W} \Sigma}$ would lead to non-vanishing topological $\theta$ phase. In the absent of the CP odd singlet, i.e, $\psi=0$ or $s=0$, the results just recover Eq.(4.6) in Ref.\cite{Hierro:2015nna} for MCHM. 


\subsection{High-energy effective Lagrangian for Omega}

\subsubsection{$p^2$ order in Omega representation}

In the CCWZ formalism~\cite{Coleman:1969sm,Callan:1969sn}, the leading order CP-even low energy effective Lagrangian with at most two derivatives is
\beqa
{\mathcal L}_{\text{eff}}^{(2)} = \frac{f^2}{4}\text{Tr}\left( d_\mu d^\mu \right). \label{L0-kin}
\eeqa
In the radial coordinate $(\varphi,\psi)$, 
it is straightforward to check that the leading order Lagrangian just gives~\cite{Marzocca:2014msa}
\beqa
{\mathcal L}_{\text{eff}}^{(2)} 
&=& {\mathcal L}_\phi +  \sin^2\left(\frac{\phi}{f}\right) {\mathcal L}_\psi \nn\\
&& + \frac{1}{\xi} \cos^2\left(\frac{\psi}{f}\right) \sin^2\left(\frac{\phi}{f}\right)  {\mathcal L}_{C}, \qquad \quad \label{L2_dd-phi-psi}
\eeqa
where $\mathcal{L}_{C}$ is the custodial preserving two derivative operators as defined in Eqs.(\ref{L_C-T}) with explicit expression given in Eq.(\ref{LCEFT_CP-even}). $\mathcal{L}_{\phi,\psi}$ is defined in analogy to $\mathcal{L}_{H}$ in Eq.(\ref{L_H}), which are canonically normalized kinetic terms for the field $\phi$ and the phase $\psi$. To be more explicitly, the kinetic term is
\beqa
{\mathcal L}_{kin}^{(\phi,\psi)} = \frac{1}{2} (\partial_\mu \phi)^2 + \frac{1}{2} \sin^2\left(\frac{\phi}{f}\right) (\partial_\mu \psi)^2.  \label{Lkin_phi-psi}
\eeqa
The results just recovers Eq.(\ref{L_C-h-s}) in the Cartesian $(h,s)$ basis. By combing with Eq.(\ref{LC_Omega}), this result is consistent with that in Eq.(A.34) for MCHM in Ref.~\cite{Alonso:2014wta}.




\subsubsection{$p^4$ order in Omega parametrization}

The first seven $p^4$ operator in the $\Omega$ parametrization defined in Eq.(\ref{O_1-7}) can be expressed in EW chiral Lagrangian as
\beqa
\begin{array}{lll}
\mathcal{O}_{k} 
&=& -4 {\mathcal L}_W - 4 {\mathcal L}_B, \\
{\mathcal O}_1 &=& \cos^4 \left(\frac{\psi }{f}\right) \sin^4 \left(\frac{\phi }{f}\right)  {\mathcal L}_6   \\
&& - 4 \xi  \cos ^2\left(\frac{\psi }{f}\right) \sin ^2\left(\frac{\phi }{f}\right) {\mathcal L}_{20}^{(\phi)} + 4\xi^2 {\mathcal L}_{D\phi} \\
&& - 4 \xi  \cos ^2\left(\frac{\psi }{f}\right) \sin ^4\left(\frac{\phi }{f}\right) {\mathcal L}_{20}^{(\psi)} + 4\xi^2 \sin ^4\left(\frac{\phi }{f}\right) {\mathcal L}_{D\psi} ,\\
{\mathcal O}_2 &=& \cos^4 \left(\frac{\psi }{f}\right) \sin^4 \left(\frac{\phi }{f}\right)  {\mathcal L}_{11} \\
&& -4 \xi  \cos ^2\left(\frac{\psi }{f}\right) \sin ^2\left(\frac{\phi }{f}\right) {\mathcal L}_8^{(\phi)} + 4\xi^2 {\mathcal L}_{D\phi} \\
&& -4 \xi  \cos ^2\left(\frac{\psi }{f}\right) \sin ^4\left(\frac{\phi }{f}\right) {\mathcal L}_8^{(\psi)} + 4\xi^2 \sin ^4\left(\frac{\phi }{f}\right) {\mathcal L}_{D\psi} ,\\
{\mathcal O}_4^+&=& -\frac{1}{4} \cos ^2\left(\frac{\psi }{f}\right) \sin ^2\left(\frac{\phi }{f}\right) ( {\mathcal L}_2 + 2{\mathcal L}_3 )  \\
&& +  \sqrt{\xi } \cos ^2\left(\frac{\psi }{f}\right) \sin \left(\frac{\phi }{f}\right)\\
&& \times \left(1 - 2 \cos ^2\left(\frac{\psi }{f}\right) \sin ^2\left(\frac{\phi }{2
   f}\right)\right) ({\mathcal L}_4^{(\phi)} - 2{\mathcal L}_5^{(\phi)}) \\
   && - \sqrt{\xi } \sin \left(\frac{\psi }{f}\right) \cos \left(\frac{\psi }{f}\right) \sin ^2\left(\frac{\phi }{f}\right) \\
   && \times \left(1 - 2 \cos ^2\left(\frac{\psi }{f}\right) \sin
   ^2\left(\frac{\phi }{2 f}\right)\right)  ({\mathcal L}_4^{(\psi)} - 2{\mathcal L}_5^{(\psi)})  ,  \\
{\mathcal O}_4^-&=& \frac{1}{4} \cos ^2\left(\frac{\psi }{f}\right) \sin ^2\left(\frac{\phi }{f}\right) \left(1 - 2 \cos ^2\left(\frac{\psi }{f}\right) \sin ^2\left(\frac{\phi }{2
   f}\right)\right) \\
   && \times ( {\mathcal L}_2 - 2{\mathcal L}_3) - \sqrt{\xi } \cos ^2\left(\frac{\psi }{f}\right) \sin \left(\frac{\phi }{f}\right)({\mathcal L}_4^{(\phi)} + 2{\mathcal L}_5^{(\phi)}) \\
   && + \frac{1}{2}\sqrt{\xi } \sin \left(\frac{2 \psi }{f}\right) \sin ^2\left(\frac{\phi }{f}\right) ({\mathcal L}_4^{(\psi)} + 2{\mathcal L}_5^{(\psi)}), \\
{\mathcal O}_5^+ &=& -\cos ^2\left(\frac{\psi }{f}\right) \sin ^2\left(\frac{\phi }{f}\right)({\mathcal L}_1+2{\mathcal L}_B+2{\mathcal L}_W) , \\
{\mathcal O}_5^- &=& 4\left(1-2 \cos ^2\left(\frac{\psi }{f}\right) \sin ^2\left(\frac{\phi }{2 f}\right) \right)({\mathcal L}_B- {\mathcal L}_W), \\
\end{array}  \label{O_1-7-Omega}
\eeqa
where the low energy effective Lagrangian ${\mathcal L}_i$ is defined in Eq.(\ref{L_CP-even}). The results are consistent with Eq.(A.34) in Ref.~\cite{Alonso:2014wta}, after combining the known matching results in Eq.(\ref{Lb-L_CP-even}). We have also calculated ${\mathcal O}_3$ and find the exact relation as below
\beqa
{\mathcal O}_3={\mathcal O}_5^- - 2{\mathcal O}_4^-.
\eeqa 
Thus, ${\mathcal O}_3$ is not a linearly independent operator. As a results, it can be dropped as already neglected in Eq.(\ref{O_1-7-Omega}). 


It is also worthing of noticing that the first two operators can also be re-xpressed without expanding the square as
\beqa
{\mathcal O}_1 
&=& \frac{16}{f^4}\bigg[ {\mathcal L}_\phi  + \sin^2\left( \frac{\phi}{f} \right) {\mathcal L}_\psi +  \frac{1}{\xi} \cos^2\left(\frac{\psi}{f}\right) \sin^2\left(\frac{\phi}{f}\right)  {\mathcal L}_{C}  \bigg]^2, \nn\\
{\mathcal O}_2 
&=& \frac{16}{f^4} \bigg[  \frac{1}{2} \partial_\mu \phi \partial_\nu \phi + \sin^2\left( \frac{\phi}{f} \right)  \frac{1}{2}  \partial_\mu \psi \partial_\nu \psi   + \frac{f^2}{8}\sin^2\left( \frac{\psi}{f} \right) \nn\\
&& \times \cos^2\left( \frac{\psi}{f} \right)  [ g^2 (W_\mu^1 W_\nu^1 + W_\mu^2 W_\nu^2)  + Z_\mu Z_\nu ]  \bigg]^2 . 
\eeqa
Thus, when $\psi=0$, it is also consistent with Eq.(30) in Ref~\cite{Buchalla:2014eca}, by substituting back the EW chiral Lagrangian in Eq.(\ref{L_CP-even}) and keep the implicit form.


\subsection{Higgs function for NMCHM}

In this section, we express the Higgs functions in the low energy chiral Lagrangian in terms of the Higgs dependence on high energy Lagrangian.

\subsubsection{CP even case}
The low energy EW chiral Lagrangian that describing the CP-even gauge-Goldstone and the gauge-scalar interactions can be written as
\beqa
\mathscr{L}_{\mathrm{low}}=\mathscr{L}_{\mathrm{low}}^{p^{2}}+\mathscr{L}_{\mathrm{low}}^{p^{4}},
\eeqa
where the Lagrangian are up to CP-even operators with at most four derivatives.

For CP-even operators at ${\mathcal O }(p^2)$ order, i.e.,  with at most two derivatives, 
\beqa
\begin{aligned} 
\mathscr{L}_{\text { low }}^{p^2}=& c_C \mathcal{L}_{C} \mathcal{F}_{C}(h)+c_{T} \mathcal{L}_{T} \mathcal{F}_{T}(h)+ c_H \mathcal{L}_{H} \mathcal{F}_{H}(h), \qquad
\end{aligned} \label{L_CP-even_p2}
\eeqa
where ${\mathcal L}_{{T},{C}}$ are the leading order low energy CP-even chiral effective Lagrangian introduced in Eq.(\ref{L_CP-even}) and ${\mathcal L}_H$ is the canonically normalized Higgs singlet kinetic term defined in Eq.(\ref{L_H}). 

The multiplicative terms $\mathcal{F}_{i}(h)$ in terms of the Higgs functions, are the generic polynomial functions of $h$. For composite Higgs model, ${\mathcal F}_i(h)$ is the trigonometric functions of $h/f$. For the NMCHM, the Higgs functions can be read directly from eq.(\ref{LC_Omega}) as
\beqa
c_C {\mathcal F}_C(h) &=& \frac{4}{\xi}  \sin ^{2}\left(\frac{\phi}{2 f}\right) \cos^2\left(\frac{\psi}{f}\right) , \nn\\
c_T {\mathcal F}_T(h) &=& 0, \quad c_H {\mathcal F}_H(h) = 1. 
\label{F_H-F_C-F_T-Omega}
\eeqa


For the CP-even operators at ${ }(p^4)$ order,
\beqa
\mathscr{L}_{\text { low }}^{p^{4}} &=& c_B \mathcal{L}_{B} \mathcal{F}_{B}(h)+ c_W \mathcal{L}_{W} \mathcal{F}_{W}(h)+\sum_{i=1}^{26} c_{i} \mathcal{L}_{i} \mathcal{F}_{i}(h) \nn \\ 
&& +c_{\square H} \mathcal{L}_{\square H} \mathcal{F}_{\square H}(h)+c_{\Delta H} \mathcal{L}_{\Delta H} \mathcal{F}_{\Delta H}(h) \nn\\
&& +c_{D H} \mathcal{L}_{D H} \mathcal{F}_{D H}(h) . \label{L_CP-even_p4}
\eeqa
The low energy CP-even chiral effective Lagrangian ${\mathcal L}_{{B},{W}, \mathcal{L}_{i},}$ are introduced in Eq.(\ref{L_CP-even}) and $\mathcal{L}_{\square H,\triangle H,D H}$ are the higher order operator denoted in Eq.(\ref{L_H}).

From NMCHM, we find the Higgs functions as
\beqa
\begin{array}{lll} 
 c_B {\mathcal F}_B  &=& \bar{c}_{B} - 4 \overline{c}_{B\Sigma}\left[  1-\cos ^2\left(\frac{\psi }{f}\right) \sin ^2\left(\frac{\phi }{2 f}\right)\right] ,  \\
 c_W {\mathcal F}_W  &=& \bar{c}_{W} - 4 \overline{c}_{W\Sigma}\left[  1-\cos ^2\left(\frac{\psi }{f}\right) \sin ^2\left(\frac{\phi }{2 f}\right)\right] ,   \\
 c_{DH}{\mathcal F}_{DH}^{(\phi)}  &=& 4\xi^2(\overline{c}_4+\overline{c}_5),  \\
 c_{DH}{\mathcal F}_{DH}^{(\psi)}  &=& 64 \xi^2  (\overline{c}_4+\overline{c}_5) \sin^4\left(\frac{\phi}{2f} \right) ,    \\ 
 c_{\Box H}{\mathcal  F}_{\Box H}^{(\phi)}  &=& -2\xi \overline{c}_6,  \\
 c_{\Box H}{\mathcal  F}_{\Box H}^{(\psi)}  &=&  - 8  \overline{c}_6 \xi \sin^2\left(\frac{\phi}{2f}\right) ,  \\
 c_1 {\mathcal F}_1   &=& \overline{c}_1 \cos ^2\left(\frac{\psi }{f}\right) \sin ^2\left(\frac{\phi }{2 f}\right) ,  \\
 c_2 {\mathcal F}_2   &=& \overline{c}_2 \cos ^2\left(\frac{\psi }{f}\right) \sin ^2\left(\frac{\phi }{2 f}\right) ,  \\
 c_3 {\mathcal F}_3   &=& 2\overline{c}_3 \cos ^2\left(\frac{\psi }{f}\right) \sin ^2\left(\frac{\phi }{2 f}\right) ,  \\
 c_4 {\mathcal F}_4^{(\phi)}   &=& \overline{c}_2 \sqrt{\xi } \cos ^2\left(\frac{\psi }{f}\right) \sin\left(\frac{\phi }{f}\right) ,  \\
 c_4 {\mathcal F}_4^{(\psi)}   &=&  - 2 \overline{c}_2 \sqrt{\xi } \sin\left(\frac{2\psi }{f}\right) \sin^2\left(\frac{\phi }{2f}\right),   \\
 c_5 {\mathcal F}_5^{(\phi)}   &=& -2 \overline{c}_3 \sqrt{\xi } \cos ^2\left(\frac{\psi }{f}\right) \sin\left(\frac{\phi }{f}\right),   \\
 c_5 {\mathcal F}_5^{(\psi)}   &=& 4 \overline{c}_3 \sqrt{\xi } \sin \left(\frac{2\psi }{f}\right) \sin ^2\left(\frac{\phi }{2 f}\right)  ,   \\
c_6 {\mathcal F}_6   &=&  16 \overline{c}_4 \cos ^4\left(\frac{\psi }{f}\right) \sin ^4\left(\frac{\phi }{2 f}\right)  \\
&& - 2 \overline{c}_6 \cos ^2\left(\frac{\psi }{f}\right)  \sin ^2\left(\frac{\phi }{2 f}\right)  \\
&& \times \left[ 1 - \cos ^2\left(\frac{\psi }{f}\right) \sin ^2\left(\frac{\phi }{2 f}\right) \right] ,   \\
 c_7 {\mathcal F}_7^{(\phi)}   &=& -2 \overline{c}_6 \sqrt{\xi } \cos ^2\left(\frac{\psi }{f}\right) \sin\left(\frac{\phi }{f}\right),  \\
 c_7 {\mathcal F}_7^{(\psi)}   &=& 4 \overline{c}_6 \sqrt{\xi } \sin \left(\frac{2 \psi }{f}\right) \sin^2\left(\frac{\phi }{2 f}\right) ,  \\
c_8 {\mathcal F}_8^{(\phi)}   &=& -16 \overline{c}_5  \xi  \cos ^2\left(\frac{\psi }{f}\right) \sin ^2\left(\frac{\phi }{2 f}\right)  \\
&& + 4 \overline{c}_6 \xi  \cos ^2\left(\frac{\psi }{f}\right) \cos ^2\left(\frac{\phi }{2 f}\right),   \\
c_8 {\mathcal F}_8^{(\psi)}   &=&  -64  \overline{c}_5 \xi  \cos ^2\left(\frac{\psi }{f}\right) \sin ^4\left(\frac{\phi }{2 f}\right)  \\
&& +    16 \overline{c}_6 \xi  \sin ^2\left(\frac{\psi }{f}\right) \sin ^2\left(\frac{\phi }{2 f}\right),   \\
c_9 {\mathcal F}_9   &=& 4 \overline{c}_6 \cos ^2\left(\frac{\psi }{f}\right) \sin ^2\left(\frac{\phi }{2 f}\right),  \\
c_{10} {\mathcal F}_{10}^{(\phi)}   &=&  4 \overline{c}_6  \sqrt{\xi } \cos ^2\left(\frac{\psi }{f}\right) \sin
   \left(\frac{\phi }{f}\right) ,   \\
c_{10} {\mathcal F}_{10}^{(\psi)}   &=&  - 8 \overline{c}_6 \sqrt{\xi } \sin \left(\frac{2 \psi }{f}\right) \sin^2\left(\frac{\phi }{2 f}\right) ,  \\
c_{11} {\mathcal F}_{11}   &=&   16 \overline{c}_5 \cos ^4\left(\frac{\psi }{f}\right) \sin ^4\left(\frac{\phi }{2 f}\right),   \\
c_{20} {\mathcal F}_{20}^{(\phi)}   &=&  -16 \overline{c}_4 \xi  \cos ^2\left(\frac{\psi }{f}\right) \sin ^2\left(\frac{\phi }{2 f}\right),  \\
c_{20} {\mathcal F}_{20}^{(\psi)}   &=& -64 \overline{c}_4\xi  \cos ^2\left(\frac{\psi }{f}\right) \sin ^4\left(\frac{\phi }{2 f}\right) ,
\end{array}
\eeqa 
where we have made abbreviation for ${\mathcal F}(h)$, by dropping the $h$ dependence.
In the absence of the singlet $s$, i.e., $\psi=0$, the results recovers the expression for $c_i{\mathcal F}_i$ custodial preserving operators for the MCHM in Table.1 in Ref.~\cite{Alonso:2014wta}

\subsubsection{CP odd case}

The low energy EW chiral Lagrangian that describing the CP-odd gauge-Goldstone and the gauge-scalar interactions can be written as
\beqa
\widetilde{\mathscr{L}}_{\mathrm{low}}=\widetilde{\mathscr{L}}_{\mathrm{low}}^{p^{2}}+\widetilde{\mathscr{L}}_{\mathrm{low}}^{p^{4}},
\eeqa
where
\beqa
\begin{aligned} 
\widetilde{\mathscr{L}}_{\text { low }}^{p^2}=& c_{\widetilde{T}} \mathcal{L}_{\widetilde{T}} \mathcal{F}_{\widetilde{T}}(h) ,\\ 
\widetilde{\mathscr{L}}_{\text { low }}^{p^{4}}=& \mathcal{L}_{\widetilde{B}} \mathcal{F}_{\widetilde{B}}(h)+\mathcal{L}_{\widetilde{W}} \mathcal{F}_{\widetilde{W}}(h)+\sum_{i=1}^{16} c_{\widetilde{i}} \mathcal{L}_{\widetilde{i}} \mathcal{F}_{\widetilde{i}}(h),
\end{aligned} \label{L_CP-odd_p2-p4}
\eeqa
where $\mathcal{F}_{\widetilde{i}}(h)$ encoded a generic dependence on $h$ but without derivative of $h$. The low energy CP-odd chiral effective Lagrangian ${\mathcal L}_{\widetilde{T},\widetilde{B},\widetilde{W},\widetilde{i}}$ are introduced in Eq.(\ref{L_CP-odd}) or more explicitly shown in Eq.(\ref{LCEFT_CP-odd}) in Appendix.~\ref{app:LCEFT}.

From the above relations, we read the Higgs functions as
\beqa
\begin{array}{lll} 
 c_{\widetilde{B}} {\mathcal F}_{\widetilde{B}} &=&  -4 \overline{c}_{\widetilde{B}\Sigma}\left[1-\cos ^2\left(\frac{\psi }{f}\right) \sin ^2\left(\frac{\phi }{2 f}\right)\right]  \\
 && + 4 \overline{c}_{\widetilde{2}} \cos ^{4}\left(\frac{\psi}{ f}\right)\sin ^{4}\left(\frac{\phi}{2 f}\right),  \\
 c_{\widetilde{W}} {\mathcal F}_{\widetilde{W}} &=& \overline{c}_{\widetilde{W}} - 4 \overline{c}_{\widetilde{W}\Sigma}\left[1-\cos ^2\left(\frac{\psi }{f}\right) \sin ^2\left(\frac{\phi }{2 f}\right)\right]   \\
 && -  4 \overline{c}_{\widetilde{2}} \cos ^{4}\left(\frac{\psi}{ f}\right)\sin ^{4}\left(\frac{\phi}{2 f}\right) ,  \\
c_{\widetilde{1}}{\mathcal F}_{\widetilde{1}} &=& \frac{1}{2} \overline{c}_{\widetilde{1}} \cos ^2\left(\frac{\psi }{f}\right) \sin ^2\left(\frac{\phi }{2 f}\right),   \\
c_{\widetilde{2}}{\mathcal F}_{\widetilde{2}} &=& 2 \overline{c}_{\widetilde{2}}  \sqrt{\xi} \cos^4 \left(\frac{\psi}{f}\right) \cos \left(\frac{\phi}{2 f}\right)   \sin ^{3}\left(\frac{\phi}{2 f}\right),  \\
c_{\widetilde{3}}{\mathcal F}_{\widetilde{3}} &=& 4 \overline{c}_{\widetilde{2}}  \sqrt{\xi} \cos^4 \left(\frac{\psi}{f}\right) \cos \left(\frac{\phi}{2 f}\right)   \sin ^{3}\left(\frac{\phi}{2 f}\right). 
\end{array} 
\eeqa
When the CP-odd singlet ($s$ or $h_5$) is absent, i.e., $\psi=0$, this just recovers the results in Table.1 in Ref.~\cite{Hierro:2015nna}.

\section{Higgs functions at the EW scale}
\label{sec:HEFT-NMCHM}


\subsection{$p^2$ order: Kinetic terms of Higgs}

\subsubsection{The $(\phi,\psi)$ basis: Polar coordinates }


The CP-even ${ }(p^2)$ order high energy effective Lagrangian with at most two derivatives is custodial preserving one as given in Eq.(\ref{LC_Omega}) for NMCHM, in the $\Sigma$ parametrization, becomes

\beqa
\overline{\mathcal{L}}_{C}(\Sigma) 
&=& \frac{1}{2} (\partial_\mu \phi)(\partial^\mu \phi) 
+ \sin^2\left( \frac{\phi}{f} \right)   (\partial_\mu \psi)(\partial^\mu \psi)  \nn\\ 
&& + \frac{1}{\xi}  \sin ^{2}\left(\frac{\phi}{f}\right) \cos^2\left(\frac{\psi}{f}\right)  \mathcal{L}_{C}, \nn\\
\overline{\mathcal{L}}_{T}(\Sigma) &=& 0 ,
\label{L_C-NMCHM} 
\eeqa
where $\mathcal{L}_{C}$ is the custodial preserving two derivative operators as defined in Eqs.(\ref{L_C-T}) with explicit expression given in Eq.(\ref{LCEFT_CP-even}). 

After EWSB, the Higgs obtains vev $\phi\to \hat\phi + \vev\phi$ and $\psi\to \hat\psi + \vev\psi $. 
The SM gauge bosons $W$ and $Z$, obtains masses from the last term in $\overline{\mathcal{L}}_{C}(\Sigma)$ in Eq.(\ref{L_C-NMCHM}), 
\beqa
{\mathcal L}_{\text{mass}} &=& \frac{1}{8} f^2 \xi  \left[ g^2 \left( (W_\mu^1)^2+(W_\mu^2)^2\right) +  (g^\prime B_\mu -g W_\mu^3 )^2 \right] \nn\\
&\equiv& \frac{1}{2} m_W^2 [  \left( (W_\mu^1)^2+ (W_\mu^2)^2\right)  +   \cdots] ,
\eeqa
where the $W$ gauge boson mass by definition is
\beqa
m_W 
= \frac{1}{2}g f \sqrt\xi = \frac{1}{2}g f \cos\left(\frac{ \vev\psi}{f}\right) \sin\left(\frac{ \vev\phi}{f}\right) . 
\qquad 
\eeqa

To be consistent with the definition of the EW scale $v$, defined by the $W$ mass $m_W^2\equiv g^2v^2/4$, it is entailed to impose that
\beqa
\xi \equiv \frac{\vev{h}^2}{f^2} &=& \sin ^{2}\left(\frac{\vev\phi}{f}\right) \cos^2\left(\frac{\vev\psi}{f}\right) ,
\eeqa
where $\xi$ is the parameter quantifying the d.o.f of the non-linearity of the Higgs dynamics as defined in Eq.(\ref{xi}) or (\ref{xi_psi-phi}).

In the absence of the singlet $s$ (or $h_5$), one has $\mathcal{L}_{\psi}=0$, and $\mathcal{L}_{\phi}={\mathcal L}_H$, the above effective Lagrangian just recovers that in MCHM~\cite{Agashe:2004rs,Csaki:2008zd,Contino:2010rs}
\beqa
\overline{\mathcal{L}}_{C}(\Sigma) 
& = &  
\mathcal{L}_{H} + \frac{1}{\xi} \sin ^{2}\left(\frac{h}{ f}\right) \mathcal{L}_{C} . 
  \label{LC_Sigma}
\eeqa


Equivalently, from linear representation of the fundamental scalar of $SO(6)$ in Eq.(\ref{Omega-Psi_h-s-SO(6)}), one can also write down the kinetic terms of both a scalar $\phi$ and a pseudo scalar $\psi$, which correspond to a sphere with standard metric as~\cite{Redi:2012ha}
\beqa 
{\mathcal L}_{kin} &=&  \frac{1}{2} \partial_\mu \Phi^T \partial^\mu \Phi = \frac{1}{2f^2}[ (\partial_\mu\phi)^2 + s_{\phi}^2  (\partial_\mu\psi)^2 ] , \qquad \label{L_C-phi-psi}
\eeqa
which is consistent with the leading $p^2$ order of the nonlinearly realized effective chiral Lagrangian from the NMCHM, as shown in Eq.(\ref{L_C-NMCHM}) in the next section. 

\subsubsection{The $(h,s)$ basis}

In the projection of the $h,s$ parametrization, the kinetic terms in Eq.(\ref{L_C-phi-psi}) becomes
\beqa
{\mathcal L}_{kin} =  \frac{1}{2f^2} \left( (\partial_\mu h)^2 + (\partial_\mu s)^2 + \frac{\left(h  \partial_\mu h +s  \partial_\mu s \right)^2}{f^2-h ^2-s ^2} \right). \qquad
\eeqa
by making a rescaling of $(h,s)\to (fh,fs)$, so that $(h,s)$ are dimensionless, the Lagrangian above just recovers that in the NMCHM
~\cite{Gripaios:2009pe}.

After EWSB, one obtains the splitting parameter as
\beqa
\xi \equiv \frac{v^2}{f^2} 
&\overset{\vev{\psi}=0}{=}& \sin ^{2}\left(\frac{\vev{h_4}}{f}\right) \overset{\xi\ll 1}{\approx} \frac{{\vev{h_4}}^2}{f^2},  \label{v4_v-MCHM}
\eeqa
which is a model-dependent relation for three quantities: the EW scale $v$, the vev $\vev{h}=v$ for Higgs singlet and the strong dynamics breaking scale $f$. 
Consider the physical Higgs excitation $h$ around $v$, one has
\beqa
\tan\left(\frac{h+v}{f}\right) = \frac{\sqrt{1-\xi} s_h  + \sqrt\xi c_h}{\sqrt{1-\xi} c_h + \sqrt\xi s_h},
\eeqa
where $s_h= \sin \left({{h}}/{f}\right) $ and $c_h = \cos \left({{h}}/{f}\right)$. 


In the NMCHM, one can also change from the polar coordinate $(\phi,\psi)$ to the Cartesian $(h,s)$ coordinate with the explicit expression as below:
\beqa
 && \mathcal{L}_{\phi}   
 \to \frac{1}{1-\xi}  {\mathcal L}_H ,\quad 
  {\mathcal L}_\psi  
 \to  \frac{1}{\xi}{\mathcal L}_S , 
\eeqa
where we have used Eq.(\ref{phi-psi_h-s}) to obtain the following useful identities 
\beqa
&& \sin\frac{\phi}{f} = \frac{h}{f} \to \frac{v}{f}=\sqrt{\xi}, \nn\\
&& \cos\frac{\phi}{f}  \partial_\mu\phi = \frac{h\partial_\mu h +s \partial_\mu s}{\sqrt{h^2+s^2}} \overset{s=0}{=} \partial_\mu h  , \nn\\
&& \partial_\mu \phi \overset{s=0}{=} \frac{f}{\sqrt{f^2-h^2}} \partial_\mu h = \frac{ \partial_\mu h}{\sqrt{1-\xi}} , \nn\\
&& \partial_\mu \psi = \frac{f \left(h \partial_\mu s-s \partial_\mu h \right)}{h^2+s^2} 
\overset{s=0}{=} \frac{f}{h} \partial_\mu s \to   \frac{\partial_\mu s}{\sqrt\xi}. 
\label{phi-psi_h-s-2}
\eeqa
It turns out that, in the $(h,s)$ field basis, the custodial preserving terms in the high energy effective Lagrangian in Eq.(\ref{L_C-NMCHM}) becomes
\beqa
\overline{\mathcal{L}}_{C}(\Sigma)  
%
&=& \frac{1}{2}\left( {(\partial_\mu h)}^2 + (\partial_\mu s)^2 + \frac{(h\partial_\mu h + s \partial_\mu s)^2}{f^2 - h^2 - s^2} \right) \nn\\
&+& \frac{1}{8}h^2 \big[ g^2[ (W_\mu^1)^2+(W_\mu^2)^2] + (gW_\mu^3 - g' B_\mu)^2 \big],   \qquad \quad \label{L_C-h-s}
\eeqa
where the first terms are kinetic terms are not canonically normalized yet as
\beqa
{\mathcal L}_{kin}^{(h,s)} 
&=& (\partial_\mu h,\partial_\mu s) K(h,s) (\partial^\mu h,\partial^\mu s)^T , \label{L_kin-h-s}
\eeqa
where $K(h,s)$ is the matrix of kinetic terms with non-diagonal mixing terms of $\partial_\mu h \partial^\mu s$,
\beqa
K(h,s) = \frac{1}{2}\left(
\begin{array}{cc}
 \frac{f^2-s^2}{f^2- h^2- s^2} & \frac{h s}{f^2-h^2-s^2} \\
 \frac{h s}{f^2-h^2-s^2} & \frac{f^2-h^2}{f^2- h^2- s^2} \\
\end{array}
\right). \quad
\eeqa
After EWSB, $\vev{s}=v_s \ne 0$ and $\vev{h}=v_h$, 
it can cause mixing and the eigenvalue matrix is~\cite{Daza:2019dqx}
\beqa
{\mathcal L}_{kin}^{(h,s)} &=& \partial_\mu (h, s)  \left(
\begin{array}{cc}
 F_{11} & F_{12} \\
 F_{21} & F_{22} \\
\end{array}
\right)  \partial^\mu \left(
\begin{array}{c}
 h  \\
 s\\
\end{array}
\right)  \nn\\
& \to &  \lambda_+ (\partial_\mu h)^2 +  \lambda_- (\partial_\mu s)^2, \nn
\eeqa
where
\beqa
&& F_{11} = \frac{1}{2}\left( 1 + \frac{v_h^2}{f^2 - v_h^2 - v_s^2} \right), \nn\\
&& F_{12} = F_{21} = \frac{1}{2} \frac{v_h v_s}{f^2 - v_h^2 - v_s^2}, \nn\\
&& F_{22} = \frac{1}{2}\left( 1 + \frac{v_s^2}{f^2 - v_h^2 - v_s^2} \right). \qquad
\eeqa
The kinetic matrix can be diagonalized with the eigenvalues are
\beqa
&& \lambda_+ = \frac{1}{2} \frac{1}{1-\frac{v_h^2 + v_s^2}{f^2}}= \frac{1}{2 c_{ \vev\phi}^2} ,
\quad \lambda_- = \frac{1}{2}.\qquad
\eeqa
In the EWSB case, $s=\vev{s}=v_s=0$, 
\beqa
s_{\vev\phi}=\sqrt\xi, \quad c_{\vev\phi}=\sqrt{1-\xi},
\eeqa
one has~\cite{Marzocca:2014msa}
\beqa
\lambda_+ = \frac{1}{2(1-\xi)}, \quad \lambda_- = \frac{1}{2}. \qquad
\eeqa



Then after rescaling, and the physical singlets can be expressed as
\beqa
&& h \to \frac{v_h \frac{h}{\sqrt{2\lambda_+}} - v_s \frac{s}{\sqrt{2\lambda_-}} }{\sqrt{v_h^2+v_s^2}} = \frac{v_h h c_{\vev\phi}- v_s s}{\sqrt{v_h^2+v_s^2}}  \overset{v_s=0}{=} h \sqrt{1-\xi}, \nn\\
&& s \to \frac{v_h \frac{s}{\sqrt{2\lambda_-}} + v_s \frac{h}{\sqrt{2\lambda_+}} }{\sqrt{v_h^2+v_s^2}} = \frac{v_h s + v_s hc_{\vev\phi}}{\sqrt{v_h^2+v_s^2}} \overset{v_s=0}{=} s, \label{h-s_eigen-state}
\eeqa
In the following, for the simplify of our physical results, we will use this eigen basis from the $\Sigma$ parametrization to deduce the physical observables, such as the anomalous couplings of the Higgs Boson to two gauge bosons, etc.
In the EW vacuum, one needs to impose the 
\beqa
h \to v + \sqrt{1-\xi} h, \quad s \to s.
\eeqa 
After redefinition, the kinetic terms are canonically normalized as
\beqa
&& {\mathcal L}_{kin}^{(h,s)} \to {\mathcal L}_{h} + {\mathcal L}_{s} = \frac{1}{2}\left[ (\partial_\mu h)^2 + (\partial_\mu s)^2 \right],
\eeqa
where the kinetic mixing terms are vanishing.

\subsubsection{The $(h_4,h_5)$ basis: Cartesian coordinates}

When going back to the original ($h_4,h_5$) basis  instead of the ($h,s$) and ($\phi, \psi$) basis, according to Eq.(\ref{h-s_h4-h5}), or refer to Table.\ref{tab_phi-psi_h4-h5_h-s}, the kinetic term in Eqs.(\ref{L_C-phi-psi}) or (\ref{L_C-h-s}) can be expressed more explicitly as
\beqa
{\mathcal L}_{kin}^{(h_4,h_5)}
&=& \frac{\left(h_4 \partial_\mu h_4+h_5 \partial_\mu h_5 \right){}^2}{2\left(h_4^2+h_5^2\right)} \nn\\
&& +  \frac{ \left(h_5 \partial_\mu h_4-h_4 \partial_\mu h_5\right){}^2 }{2\left(h_4^2+h_5^2\right){}^2} f^2 \sin ^2\left(\frac{\sqrt{h_4^2+h_5^2}}{f}\right) \nn\\
&=& (\partial_\mu h_4,\partial_\mu h_5) K(h_4,h_5) (\partial^\mu h_4,\partial^\mu h_5)^T , \label{L_kin-h4-s5}
\eeqa
where 
$K(h_4,h_5)$ is the kinetic matrix with non diagonal mixing term of $\partial_\mu h_4 \partial^\mu h_5$ as
\beqa
K(h_4,h_5) =  \frac{1}{2}\left(
\begin{array}{cc}
 K_{11} & K_{12} \\
 K_{21} & K_{22}  \\
\end{array}
\right), \quad
\eeqa
with
\beqa
K_{11} &=& \frac{h_4^2}{h_4^2+h_5^2} +  \frac{ h_5^2 f^2 \sin ^2\left(\frac{\sqrt{h_4^2+h_5^2}}{f}\right)}{(h_4^2+h_5^2)^2}, \nn\\
K_{12} &=&K_{21} = \frac{h_4 h_5}{h_4^2+h_5^2} - \frac{ h_4 h_5 f^2 \sin ^2\left(\frac{\sqrt{h_4^2+h_5^2}}{f}\right)}{(h_4^2+h_5^2)^2} ,\nn\\
K_{22} &=& \frac{h_5^2}{h_4^2+h_5^2} +   \frac{ h_4^2 f^2 \sin ^2\left(\frac{\sqrt{h_4^2+h_5^2}}{f}\right)}{(h_4^2+h_5^2)^2} . \qquad
\eeqa

In the absence of $h_5$, 
the kinetic term just recovers that of MCHM in leading $p^2$ order high energy effective Lagrangian of MCHM in Omega representation in Eq.(\ref{Lkin_phi-psi}).

In the present of SM GBs, one can promote the $h_4$ to be the $SO(4)$ global invariants as
\beqa
\abs{H}^2  = h_1^2 + h_2^2 + h_3^2 + h_4^2 = (h+v)^2, 
\eeqa 
and relabel the singlet $h_5=\eta$.
In this case, the kinetic terms in Eq.(\ref{L_kin-h4-s5}) are promoted to be a more generic one with
\beqa
{\mathcal L}_{kin}^{(H,\eta)}
&=& \frac{f^2\sin^2\left({\sqrt{|H|^2+\eta^2}}/{f} \right) }{2(|H|^2+\eta^2)} [ (\partial_\mu H)^\dag \partial^\mu H + \partial_\mu \eta \partial^\mu \eta ] \nn\\
&+&  \frac{[\partial_\mu (|H|^2+\eta^2) ]^2}{8(|H|^2+\eta^2)}\left( 1 - \frac{f^2\sin^2\left({\sqrt{|H|^2+\eta^2}}/{f} \right) }{|H|^2+\eta^2}\right) . \nn\\
&&
\eeqa
It is also interesting to observe that there the metric in front of the Higgs kinetic terms ${\mathcal L} = g_{ab} \partial_\mu h^a \partial h^b/2 $, can be expressed with a metric 
\beqa
g_{ab} =  \frac{f^2\sin^2\left({\sqrt{|H|^2+\eta^2}}/{f} \right) }{2(|H|^2+\eta^2)}\delta_{ab}.
\eeqa
In the strong coupling limit $f\to \infty$, the second term disappear, while the first term just recovers the kinetic terms of SM Higgs doublet and $\eta$
\beqa
{\mathcal L}_{kin}^{(H,\eta)} &\overset{f\to\infty}{=}&  \frac{1}{2}[(\partial_\mu H)^\dag \partial^\mu H + \partial_\mu \eta \partial^\mu \eta].
\eeqa
By gauging the theory with replacing $\partial_\mu \to D_\mu$, after EWSB, in the $f \ll 1$ limit, one would expect to obtain the Lagrangian of SM effective field theory (SMEFT) as a series of expansion. 
In the unitary gauge, $H^\dag H =|H|^2=(v+h)^2$ and $\partial_\mu (H^\dag H) = 2(v+h)\partial_\mu h$. This leads to kinetic term of Higgs singlet as well as gauge boson mass term as 
\beqa
{\mathcal L}_{h} 
&=& \frac{1}{2}\bigg[ \partial_\mu h \partial^\mu h + 2m_W^2\left(\frac{h}{v}+1\right)^2 W_{\mu}^+W^{\mu -} \nn\\
&& + m_Z^2\left(\frac{h}{v}+1\right)^2 Z_\mu Z^\mu \bigg], \label{Higgs-function-SM}
\eeqa
where the charged weak gauge bosons are defined as $W_\mu^\pm \equiv (W_\mu^1\mp i W_\mu^2)/\sqrt{2}$, the neutral weak gauge boson is $Z_\mu = g W_\mu^3 - g' B_\mu$ and  after EWSB, the bosons $W_\mu^\pm $ and $Z_\mu$ obtains mass 
\beqa
m_W^2 = \frac{1}{4}g^2v^2, \quad m_Z^2 = \frac{1}{4}(g^2+g'^2)v^2.
\eeqa



\subsection{$p^4$ order: EW chiral Lagrangian coefficients}



The EW chiral Lagrangian can be written as 
\beqa
\mathscr{L}_{\mathrm{EWCL}} \equiv \mathscr{L}^{(2)}+  \mathscr{L}^{(4)} ,
\eeqa
where ${\mathscr L}^{(2)}$ denotes the $p^2$ order terms in Eq.(\ref{L_CP-even_p2}) and (\ref{L_CP-odd_p2-p4}), and $\mathscr{L}^{(4)}$ denotes $p^4$ order terms in Eqs.(\ref{L_CP-even_p4}) and (\ref{L_CP-odd_p2-p4}), which
accounts for new interactions and derivations from the leading order operators.


\subsubsection{CP-even case}

From above equations in Eq.(\ref{L_C-NMCHM}), or with the aid of Eq.(\ref{F_H-F_C-F_T-Omega}), we can read the Higgs functions corresponding to the Lagrangians, respectively, as
\beqa
\begin{aligned}
c_C {\mathcal F}_C(h) &= \frac{1}{\xi}  \sin ^{2}\left(\frac{\phi}{f}\right) \cos^2\left(\frac{\psi}{f}\right) = \frac{h^2}{v^2}, \\
c_T {\mathcal F}_T(h)  &= 0, \quad c_H {\mathcal F}_H(h) = 1. \\
\end{aligned} \label{F_H-F_C-F_T-Sigma}
\eeqa
Since the custodial breaking operators at the order $p^2$ is absent, there is no constraint upon $c_T$ for NMCHM.

One can read the expansion coefficients of these Higgs functions in the series up to the order $v^{-2}$ as will be discussed in Eq.(\ref{Fh_i-c-a-b}) in the next section as below:
\beqa
&& c_C = 1, \quad \hat{a}_C = 0, \quad \hat{b}_C = 1 ; \nn\\
&& c_T = 0, \quad \hat{a}_T = 0, \quad \hat{b}_T = 0 ; \nn\\
&& c_H = 1, \quad \hat{a}_H = 0, \quad \hat{b}_H = 0 ,
\eeqa
where $c_T$ is related to the EW oblique $T$-parameter as denoted in Eq.(\ref{S-T}).

%
%
%


For the Higgs functions associated with CP-even operators at the order ${ }(p^4)$, one has
\beqa
\begin{array}{lll} 
 c_B {\mathcal F}_B  &=& \bar{c}_{B} -  4\overline{c}_{B\Sigma} \big[ 1 - \xi^2 \left( 1 + \frac{h}{v} \right)-  \xi(1-\xi) \left( 1 + \frac{h}{v} \right)^2  \big] ,  \\
 c_W {\mathcal F}_W  &=& \bar{c}_{W} - 4\overline{c}_{W\Sigma}\big[ 1 - \xi^2 \left( 1 + \frac{h}{v} \right) -  \xi(1-\xi) \left( 1 + \frac{h}{v} \right)^2  \big]   ,  \\
 c_{DH}{\mathcal F}_{DH}^{(\phi)}  &=& 64\xi^2(\overline{c}_4+\overline{c}_5) ,  \\
 c_{DH}{\mathcal F}_{DH}^{(\psi)}  &=& 1024 (\overline{c}_4+\overline{c}_5) \xi^4 \big[ \left( 1 + \frac{h}{v} \right)^4 + 2 \left( 1 + \frac{h}{v} \right)^2\frac{s^2}{v^2} \big]   \\ 
 c_{\Box H}{\mathcal  F}_{\Box H}^{(\phi)}  &=& -8\overline{c}_6 \xi ,  \\
  c_{\Box H}{\mathcal  F}_{\Box H}^{(\psi)}  &=&  - 8 \overline{c}_6 \xi  \big[  \xi \left( 1 + \frac{h}{v} \right)+ (1-\xi)\left( 1 + \frac{h}{v} \right)^2 +  \frac{s^2}{v^2} \big]  ,  \\
 c_{1,2} {\mathcal F}_{1,2}   &=& \overline{c}_{1,2} \xi \left[ \xi \left( 1 + \frac{h}{v} \right) + (1-\xi) \left( 1 + \frac{h}{v} \right)^2 \right],  \\
 c_3 {\mathcal F}_3   &=& 2 \overline{c}_3  \xi \left[ \xi \left( 1 + \frac{h}{v} \right) + (1-\xi) \left( 1 + \frac{h}{v} \right)^2 \right],  \\
 c_4 {\mathcal F}_4^{(\phi)}   &=& 2\overline{c}_2 \xi \left[ \xi + (2-\xi) \left( 1 + \frac{h}{v} \right) - \xi  \left( 1 + \frac{h}{v} \right)^3 \right],  \\
 c_4 {\mathcal F}_4^{(\psi)}   &=&  - 8\overline{c}_2 \xi^{ 3/2} \left( 1 + \frac{h}{v} \right) \frac{s}{v} ,  \\
 c_5 {\mathcal F}_5^{(\phi)}   &=& -4\overline{c}_3 \xi \left[ \xi + (2-\xi) \left( 1 + \frac{h}{v} \right) - \xi  \left( 1 + \frac{h}{v} \right)^3 \right],  \\
 c_4 {\mathcal F}_4^{(\psi)}   &=&  16\overline{c}_3 \xi^{ 3/2} \left( 1 + \frac{h}{v} \right) \frac{s}{v} ,  \\
 c_6 {\mathcal F}_6   &=&  16 \overline{c}_4 \xi^2  \left( 1 + \frac{h}{v} \right)^4 - 2 \overline{c}_6 \xi \left( 1 + \frac{h}{v} \right)   \\
 && \times \left[ \xi   + (1-\xi)\left( 1 + \frac{h}{v} \right) - \xi \left( 1 + \frac{h}{v} \right)^3 \right] ,  \\
 c_7 {\mathcal F}_7^{(\phi)}   &=& - 4 \overline{c}_6 \xi \left[ \xi +  (2-\xi) \left( 1 + \frac{h}{v} \right) - \xi \left( 1 + \frac{h}{v} \right)^3  \right],  \\
 c_7 {\mathcal F}_7^{(\psi)}   &=& 16 \overline{c}_6 \xi^{3/2} \left( 1 + \frac{h}{v} \right) \frac{s}{v} ,  \\
 c_8 {\mathcal F}_8^{(\phi)}   &=& - 64 \overline{c}_5 \xi^2 \left[ \xi \left( 1 + \frac{h}{v} \right) + (1-\xi)\left( 1 + \frac{h}{v} \right)^2 \right]  \\
 &+& 16 \overline{c}_6 \xi \left[ 1 - \xi^2 \left( 1 + \frac{h}{v} \right) - \xi(1-\xi)\left( 1 + \frac{h}{v} \right)^2 \right],  \\
 c_8 {\mathcal F}_8^{(\psi)}   &=& - 64  \overline{c}_5 \xi^3 \left( 1 + \frac{h}{v} \right)^4 
 +    16 \overline{c}_6 \xi^2 \frac{s^2}{v^2},  \\
 c_9 {\mathcal F}_9   &=& 4 \overline{c}_6 \xi \left[ \xi \left( 1 + \frac{h}{v} \right) + (1-\xi) \left( 1 + \frac{h}{v} \right)^2  \right],  \\
 c_{10} {\mathcal F}_{10}^{(\phi)}   &=&  8 \overline{c}_6 \xi   \left[ \xi + (2-\xi) \left( 1 + \frac{h}{v} \right) - \xi \left( 1 + \frac{h}{v} \right)^3 \right] ,   \\
 c_{10} {\mathcal F}_{10}^{(\psi)}   &=&  - 32 \overline{c}_6 \xi^{3/2} \frac{s}{v} ,  \\
 c_{11} {\mathcal F}_{11}   &=&   16 \overline{c}_5 \xi^2 \left( 1 + \frac{h}{v} \right)^4,  \\
 c_{20} {\mathcal F}_{20}^{(\phi)}   &=&  -16 \overline{c}_4 \xi^2 \left( 1 + \frac{h}{v} \right)^2,  \\
 c_{20} {\mathcal F}_{20}^{(\psi)}   &=& -64 \overline{c}_4 \xi^3 \left( 1 + \frac{h}{v} \right)^2\frac{s^2}{v^2} -64 \overline{c}_4 \xi^3 \left( 1 + \frac{h}{v} \right)^4 . \\
\end{array}
\eeqa


The CP-even anomalous couplings for HEFT from NMCHM can be obtained from expanding the Higgs functions as in the series up to order $v^{2}$ as~\cite{Alonso:2012px}
\beqa
{\mathcal F}_i(h) 
&=& 1 + 2 \frac{\hat{a}_i}{c_i } \frac{h}{v} + \frac{\hat{b}_i}{c_i} \frac{h^2}{v^2} + \cdots, \label{Fh_i-c-a-b} 
\eeqa
where $c_i$ are the global operator coefficients, $\hat{a}_i\equiv a_i c_i$ and $\hat{b}_i \equiv b_i c_i$. $c_i$ are independent of the Higgs functions ${\mathcal F}_i$ 
while  $a_i$ and $b_i$ are related to anomalous couplings beyond the SM, which are related to a three- or four-point function, respectively, e.g, a single or double Higgs scalar, couplings to two gauge bosons. 

For the NMCHM, we find the explicit expression of all non-vanishing coefficients as
\beqa
\begin{array}{lll}
&& c_B = \overline{c}_B - 4 \overline{c}_{B\Sigma}(1-\xi),  \\
&& \hat{a}_B =  2 \overline{c}_{B\Sigma}\xi(2-\xi),   \\
&& \hat{b}_B = 4  \overline{c}_{B\Sigma}\xi(1-\xi) ,  \\
&& c_W = \overline{c}_W - 4 \overline{c}_{W\Sigma}(1-\xi),  \\
&& \hat{a}_W =  2 \overline{c}_{W\Sigma}\xi(2-\xi),   \\
&& \hat{b}_W = 4  \overline{c}_{W\Sigma}\xi(1-\xi) ,  \\
&& c_{DH} = 64\xi^2(\overline{c}_4+\overline{c}_5),  ~ \hat{a}_{DH} =  0,  ~ \hat{b}_{DH} = 0 ;  \\
&& c_{\Box H} = -8 \xi \overline{c}_6,  \\
&&  \hat{a}_{\Box H} = -4 \overline{c}_6  \xi(2-\xi),  \\
&& \hat{b}_{\Box H} = -8\overline{c}_6 \xi(1-\xi) ;  \\
&& c_{1,2} = \overline{c}_{1,2}\xi,  \\
&& \hat{a}_{1,2} = \overline{c}_{1,2}\xi(2-\xi)/2,  \\
&& \hat{b}_{1,2} = \overline{c}_{1,2} \xi(1-\xi) ;  \\
&& c_{3} = 2\overline{c}_{3}\xi,  \\
&& \hat{a}_3 = \overline{c}_{3}\xi(2-\xi),  \\
&& \hat{b}_3 = 2\overline{c}_{3} \xi(1-\xi) ;  \\
&& c_4 = 2\overline{c}_2 \xi(2-\xi), ~ \hat{a}_4 = 2\overline{c}_2\xi(1-2\xi), ~ \hat{b}_4 = - 6\xi^2 \overline{c}_2 ;  \\
&& c_5 = -4\overline{c}_3 \xi(2-\xi), ~ \hat{a}_5 = -4\overline{c}_3\xi(1-2\xi), ~ \hat{b}_5 = 12\xi^2 \overline{c}_3 ;  \\
&& c_6 = 16\xi^2 \overline{c}_4 - 2\xi (1-\xi) \overline{c}_6,  \\
&& \hat{a}_6 = 32 \xi^2 \overline{c}_4 - \xi(2- 5\xi) \overline{c}_6,  \\
&& \hat{b}_6 = 96\xi^2 \overline{c}_4 - 2 \xi(1-7\xi)\overline{c}_6 ;   \\
&& c_7 = -4 \overline{c}_6 \xi(2-\xi), ~ \hat{a}_{7} = - 4 \overline{c}_6 \xi(1-2\xi), ~ \hat{b}_{7} = 12 \overline{c}_6 \xi^2 ;  \\
&& c_8 = - 64 \overline{c}_5 \xi^2 + 16\overline{c}_6 \xi(1-\xi) ,  \\
&&  \hat{a}_8 = - 32 \overline{c}_5 (2-\xi)\xi^2 - 8 \overline{c}_6 (2-\xi)\xi^2,  \\
&&  \hat{b}_8 = - 64\overline{c}_5 (1-\xi)\xi^2 - 16 \overline{c}_6 (1-\xi)\xi^2;   \\
&& c_9 = 4 \overline{c}_6 \xi , ~ \hat{a}_9 = 2 \overline{c}_6 \xi(2-\xi), ~ \hat{b}_9 = 4 \overline{c}_6 \xi(1-\xi);  \\
&& c_{10} = 8 \overline{c}_6 \xi(2-\xi), ~ \hat{a}_{10} = 8 \overline{c}_6 \xi(1-2\xi), ~ \hat{b}_{10} = -24 \overline{c}_6 \xi^2; \\
&& c_{11} = 16 \overline{c}_5 \xi^2, ~ \hat{a}_{11} = 32 \overline{c}_5 \xi^2, ~  \hat{b}_{11} = 96 \overline{c}_5 \xi^2; \\
&& c_{20} = -64 \overline{c}_4 \xi^2, ~ \hat{a}_{20} = - 64 \overline{c}_4  \xi^2, ~  \hat{b}_{20} = -64 \overline{c}_4 \xi^2;\\
\end{array}
\eeqa
where all higher order than ${\mathcal O}(\xi^2)$ are dropped.

\subsubsection{CP-odd case}

By doing transformation from $\Omega$ to $\Sigma$, i.e., by making redefinition $f\to f/2$ (or $\xi\to 4\xi$) and $\psi\to \psi/2$. In the canonically normalized basis of $h,s$, by changing from $d$ after EWSB, i.e., $h\to \sqrt{1-\xi} h+v_h$ and $s\to s$, expand it $\xi$ up to 2nd order and $s$ up to 2nd order too, we have
{
\beqa
\begin{array}{lll} 
c_{\widetilde{B}}  {\mathcal F}_{\widetilde{B}}(h) &=& -4 \overline{c}_{\widetilde{B}} \big[ 1 - \xi^2 \left( 1 + \frac{h}{v} \right) - (1-\xi)\xi \left( 1 + \frac{h}{v} \right)^2 \big]  \\
&& + 4 \overline{c}_{\widetilde{2}}  \xi^2 \left( 1 + \frac{h}{v} \right)^4   ,     \\
c_{\widetilde{W}}  {\mathcal F}_{\widetilde{W}}(h) &=& \overline{c}_{\widetilde{W}}  -4 \overline{c}_{\widetilde{W}} \big[ 1 - \xi^2 \left( 1 + \frac{h}{v} \right)  \\
&& - (1-\xi)\xi \left( 1 + \frac{h}{v} \right)^2 \big] + 4 \overline{c}_{\widetilde{2}}  \xi^2 \left( 1 + \frac{h}{v} \right)^4,      \\
c_1{\mathcal F}_{\widetilde{1}}(h) &=&  \frac{1}{2} c_{\widetilde{1}} \xi \left( 1 + \frac{h}{v} \right) \left[ \xi + (1-\xi) \left( 1 + \frac{h}{v} \right) \right],    \\
c_2{\mathcal F}_{\widetilde{2}}(h) &=&  4 c_{\widetilde{2}} \xi^2 \left( 1 + \frac{h}{v} \right) \big[   \left( 1 + \frac{h}{v} \right)^2 - \frac{1}{2}  \frac{s^2}{v^2}  \big],    \\
c_3{\mathcal F}_{\widetilde{3}}(h) &=& 8 c_{\widetilde{2}} \xi^2 \left( 1 + \frac{h}{v} \right) \big[   \left( 1 + \frac{h}{v} \right)^2 - \frac{1}{2}   \frac{s^2}{v^2} \big].
\end{array} \quad
\eeqa
}


The CP-odd anomalous couplings for NMCHM can be obtained from expanding the CP-odd Higgs functions in Eq.(\ref{Fh_at-bt-ct-NMCHM}) as
\beqa
{\mathcal F}_{\widetilde{i}}(h) 
&& = 1 + 2 \frac{\hat{a}_{\widetilde{i}}}{c_{\widetilde{i}} } \frac{h}{v} + \frac{\hat{b}_{\widetilde{i}}}{c_{\widetilde{i}}} \frac{h^2}{v^2} + \cdots, \label{Fh_at-bt-ct-NMCHM}
\eeqa
where $\hat{a}_{\widetilde{i}} \equiv a_{\widetilde{i}} c_{\widetilde{i}} $, and $\hat{b}_{\widetilde{i}} \equiv b_{\widetilde{i}} c_{\widetilde{i}} $. We find the non-vanishing coefficients $\hat{a}_{\widetilde{i}}$ from NMCHM to be
\beqa
\begin{array}{lll} 
&& c_{\widetilde{B}} = 4[ - (1-\xi) \overline{c}_{\widetilde{B}\Sigma} + \xi^2 \overline{c}_{\widetilde{2}} ] ,   \\
&& \hat{a}_{\widetilde{B}} =  2\xi [ (2-\xi)\overline{c}_{\widetilde{B}\Sigma} +  4 \xi \overline{c}_{\widetilde{2}}  ],  \\
&& \hat{b}_{\widetilde{B}} = 4 \xi[(1-\xi) \overline{c}_{\widetilde{B}\Sigma} + 6\xi \overline{c}_{\widetilde{2}} ] ,  \\
&& c_{\widetilde{W}} = \overline{c}_{\widetilde{W}}  + 4[ - (1-\xi) \overline{c}_{\widetilde{W}\Sigma} + \xi^2 \overline{c}_{\widetilde{2}} ],   \\
&&  \hat{a}_{\widetilde{W}} =  2\xi [ (2-\xi)\overline{c}_{\widetilde{W}\Sigma} +  4 \xi \overline{c}_{\widetilde{2}}  ],  \\
&& \hat{b}_{\widetilde{W}} = 4 \xi[(1-\xi) \overline{c}_{\widetilde{W}\Sigma} + 6\xi \overline{c}_{\widetilde{2}} ] ,  \\
&& c_{\widetilde{1}} = \frac{1}{2} \xi \overline{c}_{\widetilde{1}} , \quad   \hat{a}_{\widetilde{1}} = \frac{1}{4} \xi ( 2 - \xi )  \overline{c}_{\widetilde{1}}  , \quad \hat{b}_{\widetilde{1}} = \frac{1}{2}\xi(1-\xi)\overline{c}_{\widetilde{1}} ;  \\
&& c_{\widetilde{2}} = 4 \xi^2 \overline{c}_{\widetilde{2}} , \quad   \hat{a}_{\widetilde{2}} = 6 \xi^2  \overline{c}_{\widetilde{2}}  , \quad \hat{b}_{\widetilde{2}} = 12 \xi^2  \overline{c}_{\widetilde{2}}  ;  \\
&& c_{\widetilde{3}} = 8 \xi^2 \overline{c}_{\widetilde{2}} , \quad   \hat{a}_{\widetilde{3}} = 12 \xi^2  \overline{c}_{\widetilde{2}}  , \quad \hat{b}_{\widetilde{3}} = 24 \xi^2  \overline{c}_{\widetilde{2}}  .
\end{array} 
\eeqa

\subsection{Geometric meaning of Higgs functions}

\subsubsection{Coset space curvature}

In the $(h_4,h_5)$ basis above, when the SM GBs are restoring, e.g.,
\beqa
\varphi_{\hat{a}} \equiv \frac{h_{\hat{a}}}{h}\sin\frac{h}{f},  \quad h = \sqrt{h_{\hat{a}}h_{\hat{a}}} \in SO(5),
\eeqa
with $\hat{a}=1,2,3,4,5$. Then the scalar field becomes
\beqa
\Phi = \sin\frac{h}{f} \left( \frac{h_1}{h}, \frac{h_2}{h},\frac{h_3}{h},\frac{h_4}{h},\frac{h_5}{h}, \cot\frac{h}{f}  \right)^T,
\eeqa
where $h=\sqrt{h_{\hat{a}}h_{\hat{a}}}$ is $SO(5)$ invariant if $h^2=f^2$ or $\hat{h}=1$. The $\Phi$ is $SO(6)$ invariant since $\Phi^T.\Phi = 1$. In the absence of the SM GBs, i.e., $h_a=0$, it just recovers  .
\beqa
\Phi = \left(0,0,0,\frac{h_4 s_h}{\sqrt{h_4^2+h_5^2}},\frac{h_5 s_h}{\sqrt{h_4^2+h_5^2}},c_h\right)^T,
\eeqa
where $h=\sqrt{h_4^2+h_5^2}$. Then, the metric of the scalar field space is~\cite{Alonso:2015fsp}
\beqa
g_{\hat{a}\hat{b}}(h) &\equiv & f^2 \frac{\partial \Phi }{\partial h^{\hat{a}}} \cdot \frac{\partial \Phi }{\partial h^{\hat{b}}}  \nn\\
&=& \frac{f^2}{h^2} \sin^2\frac{h}{f} \delta_{\hat{a}\hat{b}} + \frac{h_{\hat{a}} h_{\hat{b}}}{h^2}\left( 1 - \frac{f^2}{h^2}\sin^2\frac{h}{f} \right), \quad \label{gab_SO6vsSO5_h4-h5}
\eeqa
where $\hat{a}=1,\ldots,5$. The first term are the diagonal terms, while the second term includes also non-diagonal terms. In the strong couplings limit, the metric just becomes flat, i.e., $\delta_{\hat{a}\hat{b}}$, in the $f\to \infty$ limit. In the $(\phi,\psi)$ basis, when the SM GBs are restoring, i.e., $\varphi_a \sim h_a \ne 0$ with $a=1,2,3$, the $SO(6)$ invariant scalar can be parameterized as
\beqa
\Phi = \sin\frac{h}{f} \left( \frac{h_1}{h}, \frac{h_2}{h},\frac{h_3}{h},\frac{\phi}{h}\cos\frac{\psi}{f},\frac{\phi}{h}\sin\frac{\psi}{f}, \cot\frac{h}{f}  \right)^T , \qquad \label{Phi_phi-psi}
\eeqa
where $h=\sqrt{h_a h_a + \phi^2}$ is $SO(5)$ invariant if $h^2=f^2$ or $\hat{h}=1$.
Considering that $\phi^2=h_4^2+h_5^2$. When the GBs are vanishing, i.e., $h_a=0$, it just recovers that in Eq.(\ref{Omega-Psi_h-s-SO(6)}) in $\phi-\psi$ basis.
Then, the metric of the scalar field space manifold is
\beqa
g_{\hat{a}\hat{b}}(h) &=& \frac{f^2}{h^2}\sin^2\frac{h}{f} \delta_{\alpha\beta}    + \frac{\phi^2}{h^2}\sin^2\frac{h}{f}\delta_{\psi\psi}  \nn\\
&& + \frac{h_{\alpha} h_{\beta}}{h^2}\left( 1 - \frac{f^2}{h^2}\sin^2\frac{h}{f} \right) ,
\eeqa
where $\alpha=(a,\phi), \beta=(b,\phi)$, and $h_\phi\equiv \phi$. 


In the $(h,s)$ basis, when the SM GBs are restoring, i.e., $\varphi_a \equiv h_{a} \ne 0$ with $a=1,2,3$, the scalar becomes
\beqa
\Phi = \frac{1}{f} \left( h_1, h_2, h_3, h_4, h_5, \sqrt{f^2 - h_{\hat{a}} h_{\hat{a}} } \right),
\eeqa
with $\hat{a}=(a,4,5)$ and  $h_4 \equiv h$ and $h_5 \equiv s$. In the absence of the GBs, i.e., $h_a=0$, it just recovers that in Eq.(\ref{Omega-Psi_h-s-SO(6)}) in $h-s$ basis. Then, the metric of the scalar field space is
\beqa
g_{\hat{a}\hat{b}}(h) &=& 
\delta_{\hat{a}\hat{b}} +  \frac{h_{\hat{a}} h_{\hat{b}}}{f^2-h_{\hat{c}} h_{\hat{c}}},  \label{gab_SO6vsSO5}
\eeqa
where $\hat{a}=1,\ldots,5$. The metric is the non-linear transformation of the $SO(5)$ invariant metric on the GB coset space $SO(6)/SO(5)\simeq S^5$.
This metric will be more convenient for our purpose of studying the intrinsic curvature of the scalar space manifold. 

Once the metric is obtained, the kinetic terms of the scalars in $SO(6)/SO(5)$ can be expressed as
\beqa
{\mathcal L} &=& \frac{1}{2} g_{\hat{a}\hat{b}}  \partial_\mu \Phi^{\hat{a}} \partial^\mu \Phi^{\hat{b}}.  \label{L_coset}
\eeqa
Take the last metric above as an example, the non-vanishing Christoffel symbols are
\beqa
&& \Gamma^{\hat{a}}_{\hat{b}\hat{c}} = \frac{h_{\hat{a}} h_{\hat{b}} h_{\hat{c}}}{f^2 (f^2 - h^2)}, \quad \hat{b}\ne \hat{c}, \nn\\
&& \Gamma^{\hat{a}}_{\hat{b}\hat{b}} = \frac{h_{\hat{a}} ( f^2 - h^2 + h_{\hat{b}}^2 ) }{f^2(f^2-h^2)}. 
\eeqa
The Ricci tensor are
\beqa
&& R_{\hat{a}\hat{b}} = \frac{ 4 h_{\hat{a}} h_{\hat{b}} }{f^2(f^2-h^2)} , \quad  \hat{a}\ne \hat{b}, \nn\\
&& R_{\hat{b}\hat{b}} = \frac{ 4(f^2 - h^2 +  h_{\hat{b}}^2 ) }{f^2(f^2-h^2)}.
\eeqa
The extrinsic curvature of the manifold with normal vector $n_{\hat{c}}$ along $\hat{c}$ direction is
\beqa
K^{\hat{a}}_{~\hat{b}} = \frac{ h_{\hat{c}} }{\sqrt{f^2-h_{\hat{c}}^2}}  \delta^{\hat{a}}_{~\hat{b}} .
\eeqa 
Thus, the extrinsic curvature is flat along arbitrary $\hat{c}$ direction.
The intrinsic Ricci curvature of the GBs scalar coset time is ${\mathcal G}/{\mathcal H}=SO(6)/SO(5)\simeq S^5$
\beqa
R = \frac{20}{f^2}>0,
\eeqa
where the number $20=5\times 4$ takes account of the degrees of freedom of the $5$ independent scalars. This positive sign of $R$ indicates that the scalar under the symmetry of compact group $SO(6)$ is in a curved spacetime in an analogy to the de Sitter spacetime with a positive cosmological constant.


\subsubsection{Higgs function in coset $SO(4)/SO(3)$}

Similarly, if the SM GBs are introduced  in the coset $SO(4)/SO(3)$ with the non-linear realization, as parameterized in Eq.(\ref{square root-parametrization}), the same story above would happen, namely the GBs also origins from the non-flat metric induced by the EW scale $v$ as shown in Eq.(\ref{R_SO4vsSO3}). 

The kinetic energy term of the Higgs can be proposed from that in the coset as Eq.(\ref{L_kin-S3}) to the $SO(4)$ manifold, by introducing additional Higgs in fundamental $SO(4)$ $H$ as the quantum fluctuations along the radial field direction, are
\beqa
{\mathcal L}  
& \equiv & \frac{1}{2} g_{\alpha\beta}(h) \partial_\mu \phi^\alpha \partial^\mu \phi^\beta , \label{L_pi-h}
\eeqa
where $\phi^\alpha=(\pi^{a},h)$ and $g_{\alpha\beta}(h)$ is a general metric on $SO(4)$, 
\beqa
g_{\alpha\beta} = \text{diag}(F(h)^2 g_{ab},g_{hh}),
\eeqa
where $F(h)$ is an arbitrary function of radial coordinate filed $h$ canonically normalized so that allows one to set $F(0)=1$ and $g_{hh}=1$. 
$g_{ab}$ is the $SO(4)$ invariant metric on the scalar coset manifold $SO(4)/SO(3)=S^3$ in Eq.(\ref{gab_SO4vsSO3}). In this case, the Riemann tensor $R_{\alpha\beta\gamma\delta}$ are
\beqa
&& R^a_{~bcd} \propto (1-v^2 F'^2)(g^a_{~c}g_{bd}-g^a_{~d}g_{bc}), \nn\\
&& R^a_{~ h b h} \propto FF'' g^a_{~b},  \label{Riemann}
\eeqa
where $F'\equiv \partial F/\partial h$, $F'' \equiv \partial^2 F/\partial h^2$ etc., are derivatives of $F$ with respect to $h$. The Ricci tensor $R_{\alpha\beta}$ are
\beqa
&& R_{ab} \propto \frac{1}{F^2}\left(\frac{2}{v^2} - 2 F'^2 - F F''\right) g_{ab}, \nn\\
&& R_{hh} \propto \frac{F''}{F} , \label{Ricci}
\eeqa
and the intrinsic Ricci scalar curvature $R$ turns out to be
\beqa
R =  \frac{6}{v^2 } \left( \frac{1}{F^2} - \frac{v^2}{F^2} (F'^2 + FF'')  \right).
\eeqa
The exterior curvature with a normal vector along the radial coordinate direction gives
\beqa
K_{~b}^{a}(n^c) =    \frac{\pi^c}{F} \delta^a_{~b} , 
\eeqa
where $ \delta^a_{~b}$ are diagonal components of the induced metric. Thus, the extrinsic curvature is flat as before. While the minimization of the extrinsic curvature along radial direction will entail that $F\ne 0$.
The Einstein equations $G_{ab}=0$ and $G_{hh}=0$ impose the constraints, respectively, as
\beqa
 2FF'' + F'^2 = \frac{1}{v^2}, \quad F' = \frac{1}{v}, 
\eeqa
where the second one origins from the vanishing of Riemann tensor curvature $R^a_{~bcd}$ in Eq.(\ref{Riemann}) above. These equations together gives the solution to $F(h)$ for the Higgs kinetic energy term
\beqa
F_{\text{SM}}(v) = C + \frac{h}{v} , \label{F_SM}
\eeqa
where integral constant $C=1$ is determined by imposing $F(0)=1$. This just reproduce the Higgs function in front of the kinetic terms of SM would-be goldstone bosons in Eq.(\ref{Higgs-function-SM}), where the d.o.f of GBs are transferred to the gauge boson masses after EWSB due to Higgs mechanism. 



When expanding the non-flat metric of SM GBs in Eq.(\ref{L_pi-h}) in power series of $1/v$, we just obtains the
\beqa
{\mathcal L} &=& \frac{1}{2}(\partial_\mu h)(\partial^\mu h) + \frac{1}{2}\left( 1 + \frac{h}{v} \right)^2 \partial_\mu \vec\pi \cdot \partial^\mu \vec\pi  \nn\\
&&+  \frac{1}{v^2} \left( 1 + \frac{h}{v} \right)^2 (\vec\pi \cdot \partial_\mu \vec\pi)^2 + \ldots,
\eeqa 
where $\ldots$ denotes the higher order multi-interactions terms. It worthy of noticing that the GBs are derivatively coupled.


\subsubsection{Higgs function beyond the SM}

For NMCHM, the kinetic terms of the Higgs sector in Eq.(\ref{L_pi-h}) will be
\beqa
{\mathcal L}  &=& \frac{1}{2} F(h)^2 g_{ab}\partial_\mu \pi^{a}\partial^\mu \pi^{b} + \frac{1}{2}\partial_\mu h \partial^\mu h , \label{L_pi-h}
\eeqa
where the Higgs function $F$ will not be that for the SM as in Eq.(\ref{F_SM}). For NMCHM, the coset space $SO(6)/SO(5)\simeq S^5$, the scalar in Eq.(\ref{Phi_phi-psi}) becomes
\beqa
\Phi = f \sin\frac{h}{f} \left(\hat{h}_1, \hat{h}_2, \hat{h}_3, \hat{h}_4, \hat{h}_5, \cot\frac{h}{f} \right)^T , \qquad \label{Phi_phi-psi=0}
\eeqa
where $h\equiv\sqrt{h_{\hat{a}}h_{\hat{a}}}$ with $\hat{a}=1,\ldots, 5$. The vacuum of $SO(5)$ is $\vev\Phi=(0,0,0,0,0,f)^T$. The kinetic action is
\beqa
{\mathcal L} = \frac{f^2}{2}\sin^2\left(\frac{h}{f}\right) \partial_\mu h^{\hat{a}} \partial^\mu h^{\hat{a}} + \frac{1}{2}\partial_\mu h \partial^\mu h.
\eeqa
By setting set $h_5=0$, the scalar above recovers those in coset space $SO(5)/SO(4)\simeq S^4$ as~\cite{Alonso:2016btr,Alonso:2016oah}, 
\beqa
\Phi = f \sin\frac{h}{f} \left(\hat{h}_1, \hat{h}_2, \hat{h}_3, \hat{h}_4, 0, \cot\frac{h}{f} \right)^T , \qquad \label{Phi_phi-psi=0}
\eeqa
where $h\equiv \sqrt{h_\alpha h_\alpha}$ with $\alpha=1,2,3,4$. 
The kinetic terms becomes
\beqa
{\mathcal L} = \frac{f^2}{2}\sin^2\left(\frac{h}{f}\right) \partial_\mu \hat{h}^\alpha \partial^\mu \hat{h}^\alpha + \frac{1}{2}\partial_\mu h \partial^\mu h.
\eeqa
By comparing with the kinetic terms of Higgs and GBs with those in Eq.(\ref{L_pi-h}), one obtains the identity that
\beqa
F(h)^2 = \frac{f^2}{v^2}\sin^2\left(\frac{h_4}{f}\right) , 
\eeqa
where $ h_4 = h + v_4$ in the $f\gg 1$ limit and $h$ is the quantum fluctuations of Higgs. By using the vevs identities denoted in Eq.(\ref{v4_v-MCHM}) as
\beqa
\frac{v_4}{f} = \arcsin\left(\frac{v}{f}\right) , 
\eeqa
where $v_4=\vev{h_4}$. One can re-express the Higgs function and expand it as
\beqa
F(h) 
&=& \cos\left( \frac{h}{f} \right) + \frac{f}{v}\sqrt{1-\frac{v^2}{f^2}} \sin\left( \frac{h}{f} \right)  \nn\\
& \approx & 1 + \frac{h}{v}\sqrt{1-\frac{v^2}{f^2}} - \frac{h^2}{2f^2} - \frac{h^3}{6f^2v}\sqrt{1-\frac{v^2}{f^2}} + \ldots, \qquad \quad 
\eeqa
which in the strong coupling limit $f\to \infty$, the scalar manifold tends to be flat, the result above just reduces to the SM Higgs function in Eq.(\ref{F_SM}).





Thus, the matching of the High energy effective Lagrangian of the model to the low energy effective EW chiral Lagrangian, leads to the Higgs functions, encoding the information of Higgs non-linearity.


\section{Connection to physical observables}
\label{sec:HEFT-obs}


In this section, in order to compare our results to the physical observables, we adopt to the following procedure:

Firstly, we need to change the GB matrix in Eq.(\ref{LC_Omega}) in the $\Omega$ parametrization to those in the $\Sigma$ parametrization.
This can be realized by making the shift $f\to f/2$ (or $\xi\to 4\xi$)  and $\psi\to \psi/2$ as Eq.(\ref{Omega-to-Sigma}). 

Secondly, for the NMCHM, since the kinetic terms of $h, s$ are not canonically normalized at the stage, we need to make eigenstate basis transformation as in Eq.(\ref{h-s_eigen-state}), in order to obtain the canonically normalized kinetic terms of Higgs $h$ and $s$. 


Thirdly,  in order to compare our results to the physical observables below $\tev$ scale, we need to consider the EW vacuum expectation value $v$ after EWSB. Therefore, we make $h\to v+ \sqrt{1-\xi}h$ and $s\to s$, and expand up to 2nd order of $\xi$. 


%



\subsection{Phenomenology of scalar sector}

The scalar is a PNGB singlet with no EW charge, which could be light. At the low energy, the phenomenology of scalar sector of the model is very similar to that of the singlet extended standard model. The singlet scalar $s$ could be a dark matter candidate, if it satisfies a $Z_2$ symmetry in the Lagrangian below. 

\subsubsection{Higgs kinetics and self interactions}

The kinetic terms of $h$ and $s$, as well as mass term of the gauge boson, can be obtained from $p^2$ order Lagrangian in Eq.(\ref{L_kin-h-s}) after EWSB,
by transforming $(h,s)$ into the canonically normalized eigen-state with the aid of Eq.(\ref{h-s_eigen-state}), one obtains
\beqa
{\mathcal L}_{eign} &=& \frac{1}{2}(\partial_\mu h)(\partial^\mu h) \left( 1 + 2 a_{hh} \frac{h}{v} + b_{hh} \frac{h^2}{v^2} + b_{hs} \frac{s^2}{v^2} + \cdots \right) \nn\\
&&+ \frac{1}{2}(\partial_\mu s)(\partial^\mu s) \left( 1 + 2 a_{sh} \frac{h}{v}  + b_{sh} \frac{h^2}{v^2} + b_{ss} \frac{s^2}{v^2} + \cdots \right) \nn\\
&& + \partial_\mu h \partial^\mu s \left( c_s \frac{s}{v} + d_{hs} \frac{hs}{v^2} + \cdots \right) - V_{\text{eff}}(h,s) \nn\\
&& + \left( 1+ 2a_{Vh}\frac{h}{v} + b_{Vh} \frac{h^2}{v^2} + b_{Vs} \frac{s^2}{v^2} + b_{Vhs} \frac{hs}{v^2} + \cdots \right) \nn\\
&& \times \left( m_W^2 M_\mu^+ W^{-\mu} + \frac{1}{2}m_Z^2 Z_\mu Z^\mu \right) ,  \label{L_HEFT-h-s}
\eeqa
where $V_{\text{eff}}(h,s)$ is the effective Higgs potential given in Eq.(\ref{Veff_h-s}), and $\cdots$ denotes higher dimensional terms. From above expansion, one can read the 
\beqa
&& a_{hh} = \frac{\xi}{\sqrt{1-\xi}} , \quad b_{hh} = \frac{\xi(1+3\xi)}{1-\xi} , \quad b_{hs}= \frac{\xi^2}{1-\xi} ; \nn\\
&& a_{sh} = 0, \quad b_{sh} = 0, \quad b_{ss} = \frac{\xi}{1-\xi}; \nn\\
&& c_{s} = \frac{\xi}{\sqrt{1-\xi}}, \quad d_{hs} = \frac{\xi(1+\xi)}{1-\xi} ; \nn\\
&& a_{Vh} = \sqrt{1-\xi}, ~ b_{Vh} = 1-\xi,~ b_{Vs} = 0, ~ b_{Vhs} = 0, \qquad \quad
\eeqa
which are consist with those list in Table.1 in Ref.~\cite{Marzocca:2014msa}. Since $b_{Vhs}=0$, the above action has a $Z_2$ symmetry for the singlet scalar $s$.

\subsubsection{Higgs potential}

Given an effective Higgs portential $V_{\text{eff}}(h,s)$ is dynamically generated, e.g., through the Coleman-Weinberg potential approach, the Higgs potential can be parameterized as
\beqa
V_{\text{eff}}(h,s) = \frac{\mu_h^2}{2} h^2 + \frac{\lambda_h}{4}h^4 + \frac{\mu_s^2}{2} s^2 + \frac{\lambda_s}{4}s^4 + \frac{\lambda}{2}h^2 s^2. \qquad
\eeqa
After making transformation to eigen-state, one has
\beqa
V_{\text{eff}}  &=& \frac{m_h^2}{2} h^2 + \frac{\lambda_{h^3}}{2} h^3 v + \frac{\lambda_{h^4}}{4}h^4 + \frac{m_s^2}{2} s^2 \nn\\
&& + \frac{\lambda_{hs^2}}{2} hs^2+ \frac{\lambda_{s^4}}{4}s^4 + \frac{\lambda_{h^2s^2}}{2}h^2 s^2,  \label{Veff_h-s}
\eeqa
where
\beqa
&& m_h^2 = (\mu_h^2+3v^2\lambda_h) (1-\xi), \quad m_s^2 = \mu_s^2 + \lambda v^2, \nn\\
&& 
\lambda_{h^3} = 2\lambda_h (1-\xi)^{3/2}, \quad \lambda_{h^4} = \lambda_h (1-\xi)^2, \quad  \lambda_{s^4} = \lambda_s,\nn\\
&& \lambda_{hs^2} = 2\lambda\sqrt{1-\xi}, \quad \lambda_{h^2s^2} = 2\lambda(1-\xi). 
\eeqa
By using the model dependent relations as those in Eq.(\ref{c_psi-s_psi}),
the effective Higgs portal in the NMCHM up to quartic order can be parameterized as~\cite{Marzocca:2014msa}
\beqa
V_{\text{eff}} &=& -\gamma \cos^2\left( \frac{\psi}{f}\right)\sin^2\left( \frac{\phi}{f}\right) + \beta \cos^4\left( \frac{\psi}{f}\right)\sin^4\left( \frac{\phi}{f}\right) \nn\\
&& + \delta \sin^2\left( \frac{\phi}{f}\right) + \sigma \cos^2\left( \frac{\psi}{f}\right)\sin^4\left( \frac{\phi}{f}\right) + \chi \sin^4\left( \frac{\phi}{f}\right) \nn\\
&\overset{\psi=0}{=}& 
- \bar\gamma  \sin^2\left( \frac{\phi}{f}\right) + \bar\beta \sin^4\left( \frac{\phi}{f}\right), 
\eeqa
where $\bar\gamma \equiv \gamma-\delta$, and $\bar\beta \equiv \beta+\sigma+\chi$ with
\beqa
&& \gamma = - \frac{f^2}{2}(\mu_h^2-\mu_s^2), \quad \beta = \frac{f^4}{4}(\lambda_h+\lambda_s-2\lambda), \nn\\
&& \delta = \frac{f^2}{2}\mu_s^2, \quad \sigma = \frac{f^4}{2}(\lambda-\lambda_s) , \quad \chi =\frac{f^4}{4}\lambda_s.
\eeqa
From which, the parameters can also re-express the free parameters of the Higgs potential as
\beqa
&& \mu_h^2 = -\frac{2}{f^2}\bar\gamma, \quad \mu_s^2 = \frac{2\delta}{f^2}, \nn\\
&& \lambda_h = \frac{4\bar\beta}{f^4}, \quad \lambda = \frac{2(\sigma+2\chi)}{f^4}, \quad \lambda_s = \frac{4\chi}{f^4}. \qquad
\eeqa
By minimizing the Higgs potential after EWSB, one obtains EW vev,
\beqa
v^2 = - \frac{\mu_h^2}{\lambda_h} , \quad \xi = \frac{\bar\gamma}{2\bar\beta}.
\eeqa
Thus, by making the transformation to the mass eigen state, one can read the physical Higgs $h$ and the singlet $s$ mass as
\beqa
\left\{\begin{array}{l}
 m_h^2 = 2 \lambda_h v^2 (1-\xi) = \frac{8\bar\beta}{v^2}(1-\xi)\xi^2 , \\
 m_s^2 = \mu_s^2 + \lambda v^2 = \frac{2}{v^2}[\delta + (\sigma+2\chi)\xi]\xi,
\end{array}\right. \label{mh-ms}
\eeqa
and there is no mixing term, i.e., $m_{hs} = 0$.

\subsection{Low energy EWPT oblique parameters}

The Wilson coefficients $c_i$ (or $c_{\widetilde{i}}$ for CP-odd case) in the low energy effective chiral Lagrangian, are related to the most significant EW oblique parameters from EWPT due to two-point functions, i.e. the $S$ and $T$ parameter~\cite{Peskin:1991sw}, which parameterize new physics contributions to electroweak radiative corrections.
For examples, the EW oblique parameters $S$ and $T$ are related to $c_1$ and $c_T$, respectively, as~\cite{Grojean:2006nn} 
\beqa
\alpha_{em} \Delta S = - 8 e^2 c_1, \quad \alpha_{em} \Delta T = 2 c_T, \label{S-T}
\eeqa
where $\alpha_{em}=e^2/(4\pi)$. In the NMCHM, we have
\beqa
&& c_{1} = \overline{c}_{1}\xi, \quad c_T = 0.
\eeqa

\subsection{Triple anomalous gauge-boson couplings}

\subsubsection{CP-even case}

The triple gauge-boson couplings in the pure weak boson sector~\cite{Hagiwara:1996kf,Brivio:2013pma,Brivio:2014pfa} can  be parameterized as
\beqa
\begin{array}{lll} 
\mathscr{L}^{W W V}_{\text{eff},CP}&=&-i g_{W W V} \big[ \kappa_{V} W_{\mu}^{+} W_{\nu}^{-} V^{\mu \nu}  \\
&& + g_{1}^{V}\left(W_{\mu \nu}^{+} W^{-\mu} V^{\nu}-W_{\mu}^{+} V_{\nu} W^{-\mu \nu}\right)  \\
&& -i g_{5}^{V} \epsilon^{\mu \nu \rho \sigma}\left(W_{\mu}^{+} \partial_{\rho} W_{\nu}^{-}-W_{\nu}^{-} \partial_{\rho} W_{\mu}^{+}\right) V_{\sigma}  \\
&& +g_{6}^{V}\left(\partial_{\mu} W^{+\mu} W^{-\nu}-\partial_{\mu} W^{-\mu} W^{+\nu}\right) V_{\nu} \big], 
\end{array}\qquad \quad
\eeqa
where $V\equiv \{\gamma, Z\}$ and 
are
\beqa
g_{WW\gamma}=g s_\theta \equiv e, \quad g_{WWZ} = g c_\theta, 
\eeqa 
where $e$ is the electric charge.
The chiral Lagrangian contributes to the anomalous cubic couplings as~\cite{Brivio:2014pfa}
\beqa
\begin{array}{lll} 
\Delta\kappa_\gamma 
&=& -2 \frac{e^2}{s_\theta^2}  \big(\bar{c}_1-\bar{c}_2-\bar{c}_3\big) \xi ,  \\
\Delta \kappa_Z 
&=& 2  e^2  \big(   \frac{2 }{c_{2 \theta }}\bar{c}_1 - \frac{1}{c_{\theta }^2} \bar{c}_2 +\frac{1}{s_{\theta }^2} \bar{c}_3  \big) \xi  ,  \\ 
\Delta g_1^Z 
&=& \frac{2e^2}{s_\theta^2} \frac{1}{c_\theta^2} \big(  \frac{ s_{\theta }^2}{c_{2 \theta }} \bar{c}_1+\bar{c}_3  \big)  \xi  ,  \\
\Delta g_5^Z 
&=& 0 , \\
\Delta g_6^\gamma 
&=& -4 \frac{e^2}{s_\theta^2} \bar{c}_6 \xi ,  \\
\Delta g_6^Z 
&=& 4\frac{e^2}{c_\theta^2}   \bar{c}_6  \xi   . 
\end{array}
\eeqa
It is obvious that, from the high-energy view point, the anomalous triple gauge couplings of SM comes from $\overline{{\mathcal L}}_{1,2,3,6}$ for NMCHM.

\subsubsection{CP-odd case}

The CP-odd triple gauge boson couplings can be parameterized as~\cite{Hagiwara:1986vm,Gavela:2014vra}
\beqa
\begin{array}{lll} 
{\mathcal{L}_{\mathrm{eff}, \cancel{CP} }^{W W V} }
&=& g_{W W V} \big[ -i \widetilde{\kappa}_{V} W_{\mu}^{\dag} W_{\nu} \tilde{V}^{\mu \nu} \\ 
&& + g_{4}^{V} W_{\mu}^{\dag} W_{\nu} \left(\partial^{\mu} V^{\nu}+\partial^{\nu} V^{\mu}\right)  \\
&& +\widetilde{g}_{6}^{V} \left(W_{\nu}^{\dag} \partial_{\mu} W^{\mu}+W_{\nu} \partial_{\mu} W^{\dag \mu}\right) V^{\nu} \\
&& +\widetilde{g}_{7}^{V} W_{\mu}^{\dag} W^{\mu} \partial^{\nu} V_{\nu} \big] , \\
\mathcal{L}_{\mathrm{eff}, \cancel{CP}}^{ZZZ} &=& \widetilde{g}_{3 Z} Z_{\mu} Z^{\mu} \partial_{\nu} Z^{\nu} , \\ 
\end{array} \qquad
\eeqa
where~\cite{Gavela:2014vra}
\beqa
\begin{array}{lll}
\Delta \widetilde{\kappa}_{\gamma} 
&=& -\frac{2 e^2 }{s_{\theta }^2}  \bar{c}_{\tilde{1}} \xi ,  \quad
\Delta \widetilde{\kappa}_{Z} 
= \frac{2 e^2  }{c_{\theta }^2}  \bar{c}_{\tilde{1}} \xi , \\ 
\Delta g_{4}^{\gamma}&=& 0, \quad
\Delta g_{4}^{Z} 
= 0, \\
\Delta\widetilde{g}_{6}^{\gamma}&=& 0 , \quad
\Delta\widetilde{g}_{6}^{Z} 
= 0 , \\ 
\Delta\widetilde{g}_{7}^{\gamma} &=& 0 , \quad
\Delta\widetilde{g}_{7}^{Z} 
= 0, \\
\Delta\widetilde{g}_{3Z} 
&=& 0.\\
\end{array} \qquad
\eeqa
It is obvious that, from the high-energy view point, the CP violating anomalous triple gauge couplings comes only from $\overline{{\mathcal L}}_{\widetilde{1}}$ for NMCHM.

\subsection{Quartic anomalous gauge-boson couplings}

The effective Lagrangian of quartic gauge bosons can be parameterized as~\cite{Brivio:2013pma,Brivio:2014pfa}
\beqa
\begin{array}{lll} 
\mathscr{L}^{V^4}_{\text{eff}}
&=& g^{2} \big[ g_{Z Z}^{(1)}\left(Z_{\mu} Z^{\mu}\right)^{2}  +g_{W W}^{(1)} W_{\mu}^{+} W^{+\mu} W_{\nu}^{-} W^{-\nu} \\
&& -g_{W W}^{(2)}\left(W_{\mu}^{+} W^{-\mu}\right)^{2}  \\
&& +g_{V V^{\prime}}^{(3)} W^{+\mu} W^{-\nu}\left(V_{\mu} V_{\nu}^{\prime}+V_{\mu}^{\prime} V_{\nu}\right) \\
&& -g_{V V^{\prime}}^{(4)} W_{\mu}^{+} W^{-\mu} V^{\nu} V_{\nu}^{\prime}  +i g_{V V^{\prime}}^{(5)} e^{\mu \nu \rho \sigma} W_{\mu}^{+} W_{\nu}^{-} V_{\rho} V_{\sigma}^{\prime}\big] ,
\end{array}
\eeqa
where non-vanishing at the tree level in the SM are
\beqa
\begin{array}{lll}
&& g_{W W}^{(1) S M}=\frac{1}{2}, \quad g_{W W}^{(2) S M}=\frac{1}{2}, \\
&& g_{Z Z}^{(3) S M}=\frac{c_{\theta}^{2}}{2}, \quad g_{\gamma \gamma}^{(3) S M}=\frac{s_{\theta}^{2}}{2} \\ 
&& g_{Z \gamma}^{(3) S M}=\frac{s_{2 \theta}}{2}, \quad g_{Z Z}^{(4) S M}=c_{\theta}^{2}, \\
&& g_{\gamma \gamma}^{(4) S M}=s_{\theta}^{2}, \quad g_{Z \gamma}^{(4) S M}=s_{2 \theta}.\\
\end{array}
\eeqa
The chiral Lagrangian contributes to the anomalous quartic couplings as~\cite{Brivio:2013pma,Brivio:2014pfa}
\beqa
\begin{array}{lll}
 \Delta g^{(1)}_{WW} 
 &=& \frac{e^2}{4s_\theta^2} \big[ 8 \big(\frac{ s_{\theta }^2}{c_{2 \theta }}\bar{c}_1+\bar{c}_3\big)\xi + 32 \bar{c}_5 \xi^2 \big],  \\
 \Delta g^{(2)}_{WW} &=& \frac{e^2}{4s_\theta^2}\big(  \frac{s_{2 \theta}^{2}}{e^{2} c_{2 \theta}} c_{T}+\frac{8 s_{\theta}^{2}}{c_{2 \theta}} c_{1}+4 c_{3}-4 c_{6}-\frac{1}{2} c_{\square H} \\
  && -2 c_{11}-16 c_{12}+8 c_{13} \big) ,\\ 
 &=& \frac{e^2}{4s_\theta^2} \big[ \big( \frac{8 s_{\theta }^2}{c_{2 \theta }}\bar{c}_1 +8 \bar{c}_3 + 12 \bar{c}_6 \big) \xi  \\
 && -8 \big(8 \bar{c}_4+4 \bar{c}_5+\bar{c}_6\big) \xi^2 \big],\\
 \Delta g^{(1)}_{ZZ}  &=& \frac{e^2}{4s_\theta^2} \frac{1}{c_\theta^4} \big( c_{6}+\frac{1}{8} c_{\square H}+c_{11}+2 c_{23}+2 c_{24}+4 c_{26}  \big), \\
  &=& \frac{e^2}{4s_\theta^2} \frac{1}{c_\theta^4} \big[ - 3 \bar{c}_6 \xi + 2 \left(8
   \left(\bar{c}_4+\bar{c}_5\right)+\bar{c}_6\right) \xi^2 \big] , \\ 
 \Delta g^{(3)}_{ZZ} &=& \frac{e^2}{4s_\theta^2} \frac{1}{c_\theta^2}  \big( \frac{s_{2\theta}^{2} c_{\theta}^{2}}{e^{2} c_{2 \theta}} c_{T}+\frac{2 s_{2 \theta}^{2}}{c_{2 \theta}} c_{1}+4 c_{\theta}^{2} c_{3}-2 s_{\theta}^{4} c_{9}  +2 c_{11}   \\
 && +4 s_{\theta}^{2} c_{16}+2 c_{24} \big) ,\\ 
 &=& \frac{e^2}{4s_\theta^2} \frac{1}{c_\theta^2} \big[ \big( \frac{2  s_{2 \theta }^2}{c_{2 \theta}}\bar{c}_1 + 8  c_{\theta }^2 \bar{c}_3 -8  s_{\theta }^4 \bar{c}_6 \big) \xi + 32 \bar{c}_5 \xi^2 \big] ,\\
 \Delta g^{(4)}_{ZZ} &=& \frac{e^2}{4s_\theta^2} \frac{1}{c_\theta^2}  \big( \frac{2 s_{2 \theta}^{2} }{e^{2} } \frac{c_{\theta}^{2}}{c_{2 \theta} } c_{T}+\frac{4 s_{2 \theta}^{2}}{c_{2 \theta}} c_{1}+8 c_{\theta}^{2} c_{3}  \\
 && -4 c_{6} -\frac{1}{2} c_{\square H}-4 c_{23} \big) ,\\
 &=& \frac{e^2}{4s_\theta^2} \frac{1}{c_\theta^2} \big[ 4 \big( \frac{ s_{2 \theta }^2}{c_{2 \theta }} \bar{c}_1 + 4 c_{\theta }^2 \bar{c}_3  + 3 \bar{c}_6\big)\xi \\
 &&  -8 \left(8\bar{c}_4+\bar{c}_6\right)\xi^2  \big] , \\
 \Delta g^{(3)}_{\gamma\gamma} &=&\frac{e^2}{4s_\theta^2}  {s_\theta^2}({-2 c_{9}}) , \\ 
 &=&\frac{e^2}{4s_\theta^2}  {s_\theta^2}( -8 \bar{c}_6 )\xi , \\
 \Delta g^{(3)}_{\gamma Z}  &=& \frac{e^2}{4s_\theta^2}\frac{s_\theta}{c_\theta} \big( {\frac{s_{2 \theta}^{2}}{e^{2} c_{2 \theta}} c_{T}+\frac{8 s_{\theta}^{2}}{c_{2 \theta}} c_{1}+4 c_{3}+4 s_{\theta}^{2} c_{9}-4 c_{16}} \big) , \\ 
 &=&   \frac{2e^2}{s_\theta^2}\frac{s_\theta}{c_\theta}   \big(  \frac{s_{\theta }^2}{c_{2 \theta }}  \bar{c}_1 +2 s_{\theta }^2
   \bar{c}_6 + \bar{c}_3  \big) \xi ,\\
 \Delta g^{(4)}_{\gamma Z} 
 &=& \frac{4e^2}{s_\theta^2}\frac{s_\theta}{c_\theta} \big( \frac{ s_{\theta }^2}{c_{2 \theta }} \bar{c}_1 +\bar{c}_3\big) \xi ,  \\
 \Delta g^{(5)}_{\gamma Z} 
 &=&0. \\
\end{array}
\eeqa
From the high-energy view point, the anomalous triple gauge couplings of SM comes from $\overline{{\mathcal L}}_{1,2,3,4,5,6}$ for NMCHM.

\subsection{$HVV$ anomalous couplings}
\label{sec:EWPT_HVV}

\subsubsection{CP-even case}

The anomalous couplings of the Higgs bosons can be parameterized as~\cite{Hagiwara:1993ck,Brivio:2013pma,Brivio:2014pfa}
\beqa
\begin{array}{lll}
\mathcal{L}_{\text {eff }, C P}^{\mathrm{HVV}} 
& =& 
g_{H \gamma \gamma} A_{\mu \nu} A^{\mu \nu} h \\
&& +g_{H Z \gamma}^{(1)} A_{\mu \nu} Z^{\mu} \partial^{\nu} h+g_{H Z \gamma}^{(2)} A_{\mu \nu} Z^{\mu \nu} h \\ 
&&+g_{H Z Z}^{(1)} Z_{\mu \nu} Z^{\mu} \partial^{\nu} h+g_{H Z Z}^{(2)} Z_{\mu \nu} Z^{\mu \nu} h \\
&& +g_{H Z Z}^{(3)} Z_{\mu} Z^{\mu} h+g_{H Z Z}^{(4)} Z_{\mu} Z^{\mu} \Box h \\ 
&&+g_{H Z Z}^{(5)} \partial_{\mu} Z^{\mu} Z_{\nu} \partial^{\nu} h+g_{H Z Z}^{(6)} (\partial_{\mu} Z^{\mu})^2 h \\ 
&&+g_{H W W}^{(1)}\left(W_{\mu \nu}^{+} W^{-\mu} \partial^{\nu} h+\mathrm{h.c.}\right)\\
&&+g_{H W W}^{(2)} W_{\mu \nu}^{+} W^{-\mu \nu} h \\
&&+g_{H W W}^{(3)} W_{\mu}^{+} W^{-\mu} h +g_{H W W}^{(4)} W_{\mu}^{+} W^{-\mu} \square h \\
&&+g_{H W W}^{(5)}\left(\partial_{\mu} W^{+\mu} W_{\nu}^{-} \partial^{\nu} h+\mathrm{h.c.}\right)\\
&&+g_{H W W}^{(6)} \partial_{\mu} W^{+\mu} \partial_{\nu} W^{-\nu} h,
\end{array} \qquad
\eeqa
where the contribution of the anomalous couplings are separated from that of the SM one as
\beqa
g_{HVV}^{(i)} \approx g_{HVV}^{(i)SM} + \Delta g_{HVV}^{(i)},
\eeqa
where $V=\{\gamma, Z, W_\pm \}$. All the tree-level SM coupling above vanish, except the followings:
\beqa
&& g_{HZZ}^{(3)SM} = \frac{e^2}{4v}\frac{m_Z^2}{e^2}( -2 c_H)= - \frac{m_Z^2}{2v}, \nn\\
&& g_{HWW}^{(3)SM} = \frac{e^2}{4v}\frac{m_Z^2 c_\theta^2}{e^2}[-4(c_H+c_C)]  = - 2\frac{m_W^2}{v}. 
\eeqa
The anomalous coefficients are related to the coefficients of low energy effective chiral Lagrangian as~\cite{Hagiwara:1993ck,Qi:2008ex,Brivio:2013pma,Brivio:2014pfa}
\beqa
\begin{array}{lll}
 \Delta g_{H\gamma\gamma} 
 &=& \frac{e^2}{4v} \big[  -8   \left(\bar{c}_{B \Sigma }+\bar{c}_{W \Sigma}-\bar{c}_1\right) \xi \\
 && + 4  \left(\bar{c}_{B \Sigma }+\bar{c}_{W \Sigma}-\bar{c}_1\right) \xi ^2\big], \\
 \Delta g_{HZ\gamma}^{(1)} 
 &=& \frac{e^2}{4v}\frac{1}{s_{2\theta}}\left[  -32   \left(\bar{c}_2-\bar{c}_3\right) \xi + 64  \left(\bar{c}_2-\bar{c}_3\right) \xi ^2 \right], \\
 \Delta g_{HZ\gamma}^{(2)} 
 &=& \frac{e^2}{4v}\frac{c_\theta}{s_{\theta}}\big[ \frac{8   }{c_{\theta }^2} \left(2 s_{\theta }^2 \bar{c}_{B \Sigma }-2 c_{\theta }^2 \bar{c}_{W \Sigma} + c_{2 \theta } \bar{c}_1  \right) \xi \\
 &&  - \frac{4}{c_{\theta }^2} \left( 2 s_{\theta }^2 \bar{c}_{B \Sigma } - 2 c_{\theta }^2 \bar{c}_{W \Sigma} +  c_{2 \theta } \bar{c}_1 \right) \xi ^2 \big],  \\
\Delta g_{HZZ}^{(1)} 
&=& \frac{e^2}{4v} \frac{1}{c_{\theta}^2}\big[ 16 \big( \bar{c}_2 + \frac{ c_{\theta }^2}{s_{\theta}^2}\bar{c}_3 \big)  \xi   - 32  \big( \bar{c}_2 + \frac{ c_{\theta }^2}{s_{\theta }^2} \bar{c}_3 \big) \xi ^2 \big], \\
\Delta g_{HZZ}^{(2)} 
&=& -\frac{e^2}{4v}\frac{c_\theta^2}{s_{\theta}^2}\big[ 8 \big(\frac{s_{\theta }^4 }{c_{\theta }^4}\bar{c}_{B \Sigma }  + \bar{c}_{W \Sigma} +\frac{ s_{\theta }^2}{c_{\theta }^2}\bar{c}_1 \big) \xi  \\
&& -4  \big(\frac{ s_{\theta }^4  }{c_{\theta }^4}\bar{c}_{B \Sigma } + \bar{c}_{W \Sigma} + \frac{  s_{\theta }^2}{c_{\theta }^2}\bar{c}_1 \big) \xi ^2 \big],  \\
\Delta g_{HZZ}^{(3)} &=& 0  , \\
\Delta g_{HZZ}^{(4)} 
&=& -\frac{e^2}{4v}\frac{1}{s_{2\theta}^2 }\left( -64 \bar{c}_6 \xi  + 128 \bar{c}_6 \xi ^2  \right),  \\
\Delta g_{HZZ}^{(5)} 
&=& -\frac{e^2}{4v}\frac{1}{s_{2\theta}^2 }\left( 128 \bar{c}_6 \xi -256 \bar{c}_6 \xi ^2 \right),  \\
\Delta g_{HZZ}^{(6)} 
&=& -\frac{e^2}{4v}\frac{1}{s_{2\theta}^2 }\left( 64 \bar{c}_6 \xi -32 \bar{c}_6 \xi ^2  \right),  \\
\Delta g_{HWW}^{(1)} 
&=& \frac{e^2}{4v}\frac{1}{s_\theta^2}( 64 \bar{c}_3 \xi -32 \bar{c}_3  \xi ^2) , \\
\Delta g_{HWW}^{(2)} 
&=& \frac{e^2}{4v}\frac{1}{s_\theta^2}( -16   \bar{c}_{W \Sigma} \xi + 8  \bar{c}_{W \Sigma} \xi ^2 ) , \\
\Delta g_{HWW}^{(3)} 
&=& \frac{e^2}{4v}\frac{m_Z^2 c_\theta^2}{e^2}   \frac{32 e^2   }{c_{2 \theta }} \bar{c}_1 \xi  , \\
\Delta g_{HWW}^{(4)} 
&=& -\frac{e^2}{4v}\frac{1}{s_\theta^2}( -32 \bar{c}_6 \xi + 64 \bar{c}_6  \xi ^2 ) , \\
\Delta g_{HWW}^{(5)} 
&=& -\frac{e^2}{4v}\frac{1}{s_\theta^2}( 32 \bar{c}_6 \xi   -64  \bar{c}_6 \xi ^2 ) , \\
\Delta g_{HWW}^{(6)} 
&=& -\frac{e^2}{4v}\frac{1}{s_\theta^2}( 32 \bar{c}_6  \xi  -16  \bar{c}_6 \xi ^2  ) . \\
 \end{array}
\eeqa
It is obvious that, the CP-even anomalous couplings $\Delta g_{HVV}^{(1,2,3)}$ of SM comes from operators $\overline{{\mathcal L}}_{B\Sigma,W\Sigma, 1,2,3}$, while the additional $\Delta g_{HVV}^{(4,5,6)}$ all attribute to one operator $\overline{{\mathcal L}}_{6}$ in the high energy.

\subsubsection{CP-odd case}

The anomalous couplings of the CP-odd interactions involving the Higgs to two gauge bosons can be parameterized as~\cite{Hagiwara:1986vm,Gavela:2014vra}
\beqa
\begin{array}{lll}
\mathcal{L}_{\text {eff }, \cancel{C P}}^{\mathrm{HVV}} &=& 
\widetilde{g}_{H \gamma \gamma} h A_{\mu \nu} \widetilde{A}^{\mu \nu}+\widetilde{g}_{H Z \gamma} h A_{\mu \nu} \widetilde{Z}^{\mu \nu} \\ 
&& +\widetilde{g}_{H Z Z}^{(2)} h Z_{\mu \nu} \widetilde{Z}^{\mu \nu}+\widetilde{g}_{H W W}^{(2)} h W_{\mu \nu}^{+} \widetilde{W}^{-\mu \nu} \\ 
&&+\left[\widetilde{g}_{H W W}^{(1)}\left(W_{\mu \nu}^{+} W^{-\mu} \partial^{\nu} h\right)+\mathrm{h.c.}\right]  \\ 
&& +\left[\widetilde{g}_{H W W}^{(5)}\left(\partial_{\mu} W^{+\mu} W_{\nu}^{-} \partial^{\nu} h\right)+\mathrm{h.c.}\right], \\
 \end{array}\qquad \qquad
 \eeqa
where the anomalous CP-odd $HVV$ couplings at the tree level are~\cite{Gavela:2014vra}
\beqa
\begin{array}{lll} 
 \widetilde{g}_{H \gamma \gamma} 
 &=&\frac{4 e^{2}}{v}  \big[ \frac{1}{2} \left(-2 \bar{c}_{\widetilde{B} \Sigma  }-\bar{c}_{\widetilde{W} \Sigma  }+\bar{c}_{\tilde{1}}\right)  \xi  \\
   && +  \frac{1}{4}  \left(2 \bar{c}_{\widetilde{B} \Sigma  }+\bar{c}_{\widetilde{W} \Sigma  }-\bar{c}_{\tilde{1}}-24 \bar{c}_{\tilde{2}}\right) \xi ^2 \big] \\
\widetilde{g}_{H Z \gamma} 
&=&-\frac{8 e^{2} s_{\theta}}{v c_{\theta}} \big[ -\frac{1}{4 s_{\theta }^2} \left(4 s_{\theta }^2 \bar{c}_{\widetilde{B} \Sigma  } -2 c_{\theta }^2 \bar{c}_{\widetilde{W} \Sigma  } +c_{2 \theta } \bar{c}_{\tilde{1}}\right) \xi   \\
&& + \frac{1}{8 s_{\theta }^2}  \big(4 s_{\theta }^2 \bar{c}_{\widetilde{B} \Sigma  } -2 c_{\theta }^2 \bar{c}_{\widetilde{W} \Sigma  } +c_{2 \theta } \bar{c}_{\tilde{1}}  \\
&& +8 \bar{c}_{\tilde{2}} \left( 6 c_{\theta }^2-1 \right) \big) \xi ^2  \big] \\
\widetilde{g}_{H Z Z}^{(2)} 
&=& \frac{4 e^{2} s_{\theta}^{2}}{v c_{\theta}^{2}} \big[ -\frac{1}{2 s_{\theta }^4}   \left( 2 s_{\theta }^4 \bar{c}_{\widetilde{B} \Sigma  }+c_{\theta }^4 \bar{c}_{\widetilde{W} \Sigma  } +c_{\theta }^2 s_{\theta }^2 \bar{c}_{\tilde{1}} \right)  \xi  \\
&& + \frac{1}{4 s_{\theta}^4}(2 s_{\theta }^4 \bar{c}_{\widetilde{B} \Sigma  } +c_{\theta }^4 \bar{c}_{\widetilde{W} \Sigma  } +c_{\theta }^2 s_{\theta }^2 \bar{c}_{\tilde{1}}   \\
&& +\left[-8 c_{2 \theta }-3 \left(c_{4 \theta }+3\right) \right] \bar{c}_{\tilde{2}}) \xi ^2 \big]   \\ 
\widetilde{g}_{H W W}^{(1)} 
&=& 0 , \\
\widetilde{g}_{H W W}^{(2)} 
&=&-\frac{2 e^{2}}{v s_{\theta}^{2}}\big( 2 \bar{c}_{\widetilde{W} \Sigma  }  \xi  +  \left(20 \bar{c}_{\tilde{2}}-\bar{c}_{\widetilde{W} \Sigma  }\right) \xi ^2 \big) , \\
\widetilde{g}_{H W W}^{(5)} 
&=& 0 . 
 \end{array} \qquad \quad
\eeqa
From the high energy viewpoint, the CP-odd anomalous couplings $\Delta g_{HVV}^{(1,2,3)}$ of SM comes from operators $\overline{{\mathcal L}}_{\widetilde{B}\Sigma,\widetilde{W}\Sigma, \widetilde{1},\widetilde{2},\widetilde{3}}$ for NMCHM.

\subsubsection{$HVVV$ anomalous couplings}

The  anomalous couplings of the Higgs to the three gauge bosons, i.e., ${HVVV}$ can be parameterized as~\cite{Brivio:2014pfa}
\beqa
\begin{array}{lll}
\mathcal{L}_{\text {eff }, C P}^{\mathrm{HVVV}} 
& = &  g_{HWWZ}^{(1)} [ i (\partial_\mu W^{-\mu}) (Z_\nu W^{+\nu}) h + \text{h.c.} ]  \\
&& + g_{HWWA}^{(1)} [ i (\partial_\mu W^{-\mu}) (A_\nu W^{+\nu}) h + \text{h.c.}  ] \\
&& + g_{HWWZ}^{(2)} [ i (Z_\mu W^{+\mu}) (W_\nu^- \partial^\nu h) + \text{h.c.} ] \\
&& + g_{HWWA}^{(2)} [ i (A_\mu W^{+\mu}) (W_\nu^- \partial^\nu h) + \text{h.c.} ] ,\\
\end{array} \qquad \quad
\eeqa
where the anomalous couplings $g_{HVVV}$ turns out to be~\cite{Brivio:2014pfa}
\beqa
 \begin{array}{lll}
\Delta g_{HWWZ}^{(1)} &=& 
\frac{e^2g^2}{c_\theta^2}( \frac{8 }{v} \bar{c}_6 \xi -\frac{4 }{v} \bar{c}_6 \xi^2 )  , \\
\Delta g_{HWWA}^{(1)} &=& 
-eg^2 (\frac{8 }{v} \bar{c}_6 \xi -\frac{4 }{v} \bar{c}_6 \xi^2 ) , \\
\Delta g_{HWWZ}^{(2)} &=& 
-\frac{e^2 g}{2c_\theta} (  \frac{16}{v}  \bar{c}_6 \xi -\frac{32 }{v} \bar{c}_6 \xi^2)  , \\
\Delta g_{HWWA}^{(2)} &=& 
\frac{eg^2}{2} ( \frac{16 }{v} \bar{c}_6 \xi -\frac{32 }{v} \bar{c}_6 \xi^2 ) .
 \end{array}
\eeqa
All of the $HVVV$ anomalous couplings origins from one singlet operator $\overline{\mathcal L}_6$ from a high-energy view point, instead of depending upon different EWCL operators, such as ${\mathcal L}_{9,10}$, respectively.

\subsubsection{$HHVV$ anomalous couplings}

The anomalous couplings of the two Higgs to two gauge bosons, i.e., ${HHVV}$ can be parameterized as~\cite{Brivio:2014pfa}
\beqa
\begin{array}{lll}
\mathcal{L}_{\text {eff }, C P}^{\mathrm{HHVV}} 
& = &  g_{HHWW}^{(1)} W_\mu^+ W^{-\mu} h^2 
+ g_{HHWW}^{(2)} W_\mu^+ W^{-\mu} \Box (h^2) \ii\ii \\
&& + g_{HHWW}^{(3)} W_\mu^+ W_\nu^- \partial^\mu h \partial^\nu h \\
&& + g_{HHWW}^{(4)} (\partial_\mu W^{+\mu} )(\partial_\nu W^{-\nu} )h^2 \\
&& + g_{HHWW}^{(5)} [ (\partial_\mu W^{+\mu} )(W_\nu^- \partial^\nu h)h + \text{h.c.} ] \ii\ii \\
&& + g_{HHZZ}^{(1)} Z_\mu Z^\mu h^2 + g_{HHZZ}^{(2)} Z_\mu Z^\mu \Box h^2 \\
&& +  g_{HHZZ}^{(3)} (Z_\mu \partial^\mu h)^2 + g_{HHZZ}^{(4)} (\partial_\mu Z^\mu)^2 h^2  \\
&& + g_{HHZZ}^{(5)} (\partial_\mu Z^\mu)(Z_\nu \partial^\nu h)  h^2  \\
\end{array} \qquad
\eeqa
where the only non-vanishing couplings at the tree level in the SM are
\beqa
g_{HHWW}^{(1)SM}  = \frac{g^2}{4}, \quad g_{HHZZ}^{(1)SM} = \frac{g^2}{8c_\theta^2}. 
\eeqa
The anomalous couplings of $g_{HHVV}$ turns out to be~\cite{Brivio:2014pfa}
\beqa
 \begin{array}{lll}
\Delta g_{HHWW}^{(1)} &=& 
 10 g^2  m_h^2 \bar{c}_6  \xi , \\
\Delta g_{HHWW}^{(2)} &=& 
-\frac{12 g^2 }{v^2} \bar{c}_6 \xi^2  , \\
\Delta g_{HHWW}^{(3)} &=& 
0, \\
\Delta g_{HHWW}^{(4)} &=& 
-\frac{4 g^2 }{v^2} \bar{c}_6 \xi + \frac{4 g^2 }{v^2} \bar{c}_6\xi^2 , \\
\Delta g_{HHWW}^{(5)} &=& 
\frac{24 g^2}{v^2}   \bar{c}_6 \xi^2 ; \\
\Delta g_{HHZZ}^{(1)} &=& 
\frac{5 g^2  m_h^2}{c_{\theta }^2}  \bar{c}_6 \xi, \\
\Delta g_{HHZZ}^{(2)} &=& 
-\frac{6 g^2 }{v^2 c_{\theta }^2} \bar{c}_6 \xi^2, \\
\Delta g_{HHZZ}^{(3)} &=& 
0, \\
\Delta g_{HHZZ}^{(4)} &=& 
-\frac{2 g^2 }{v^2 c_{\theta }^2} \bar{c}_6 \xi + \frac{2 g^2 }{v^2 c_{\theta }^2} \bar{c}_6 \xi^2, \\
\Delta g_{HHZZ}^{(5)} &=& 
\frac{24 g^2 }{v^2 c_{\theta }^2}  \bar{c}_6 \xi^2.
 \end{array}
\eeqa
It is intriguing to observe that all of the $HHVV$ anomalous couplings also origins from one singlet operator $\overline{\mathcal L}_6$ from a high-energy view point. Even though they may obtains contributions from different EWCL operators, such as ${\mathcal L}_{\Box H,7,8,9,10}$.

\section{
Conclusions}
\label{sec:dis-con}

In summary, we have studied the effective field theory in the minimal composite Higgs model based upon the $SO(6)/SO(5)$ symmetry breaking pattern up to ${ }p^4$ order in the CCWZ formalism. The vacuum misalignment is parametrized by a rotation angle in the coset space in the effective Lagrangian framework. 
In order to match with the electroweak chiral Lagrangian, we re-parametrize the pseudo-Goldstone boson fields as $(h, s)$, and obtain the connection
between the high energy effective Lagrangian and the low energy effective electroweak chiral Lagrangian with the Higgs function dependences. By expanding in the decoupling limit, i.e., $f\to\infty$, one would expect to recover the linearly realized SM Higgs theory. 

Regarding to the connection between the high energy effective Lagrangian for the NMCHM and the low energy effective chiral Lagrangian, several main results are in order:
\bit
\item Both CP-even and CP-odd operators, and both the Omega and Sigma parametrizations are considered.
\item Exact relations between definitions of polar, Cartesian and mixing coordinate of fields are provided.
\item The complete explicit expression of CCWZ formalism are provided.
\item All the exact Higgs functions in the electroweak chiral Lagrangian are presented, which exactly recover the Higgs function in the $SO(5)/SO(4)$ composite Higgs model, in the absent of singlet $s$.
\item Higgs functions are analytic functions of Lorentz invariant scalar, observables in ultra-relativistic limit.
\item The Higgs functions in the effective Lagrangian origin from the Riemann curvature on the non-flat pseudo-Nambu-Goldstone boson scalar manifold, and incorporate the Higgs non-linearity/vacuum misalignment effects in the next minimal composite Higgs model.
\item Linear and nonlinear representation are related by a field redefinition, which implies diffeomorphism covariance, or general covariance, thus the invariant measure as curvature in curved group manifold.
\item After canonical normalization of the kinetic term, we obtain all the Higgs couplings deviated from the standard model ones extracted from the Higgs functions. 
\item Higgs self couplings, anomalous triple and quartic gauge couplings, anomalous couplings of Higgs to gauge bosons are given, as criteria for detecting the physics beyond the SM.
\item The singlet can be light and plays a role a dark matter candidate, in the singlet extended standard model with an additional $Z_2$ symmetry in scalar sector.
\eit


In the Higgs effective theory, we could obtain the Wilson coefficients derived from the corresponding Higgs functions in the $v \to \infty $ limit. At the $p^2$ order, the corresponding expansion coefficients are related to the EW oblique parameter $S,T$ constrained by the EWPT. While at the $p^4$ order, they are related to the anomalous couplings of triple or quartic gauge bosons, the anomalous couplings coefficients of Higgs to gauge bosons, etc. 
Some of these couplings are constrained by the precision measurements of the Higgs self couplings at the LHC. 
The future colliders will also be able to explore some of these couplings, such as VVH and VVHH couplings. We will leave this study to the future work. Finally the phenomenology of the scalar singlet could be quite interesting, such as the collider searches for such scalar, and its cosmological implications, etc.

\vspace{0.2in}  \centerline{\bf{Acknowledgements}} \vspace{0.2in}

We appreciate Hao-Lin Li for early collaboration and discussion of this work. We would like thank Brando Bellazzini, Csaba Csaki and Jing Shu for valuable discussions. In particularly, Y.~H.~Qi would like to thank Yu-Ping Kuang for instructing him theoretical elementary particle physics. He would like to thank Xing-Chang Song and Zhong-Qi Ma for teaching him Lie groups and Lie algebras. 
Y.~H.~Qi is supported by the Korean Ministry of Education, Science and Technology, Gyeongsangbuk-do Provincial Government and Pohang City Government for Independent Junior Research Groups at the Asia Pacific Center for Theoretical Physics (APCTP). J.~H.~Y. is supported in part by the National Science Foundation of China under Grants No.~11875003 and the Chinese Academy of Sciences (CAS) Hundred-Talent Program. S.H.Z. is supported in part by the National Science Foundation of China under Grants No.~11635001,~11875072. 

\appendix

\section{SO(N) generators and Higgs representation}
\label{app:rep}

\subsection{Generators of $SO(6)$ group}

In SO(6), there are $15$ generators denoted with 10 generators of the unbroken $SO(5)$ as $T^a = \{T^a_{L/R}, T^\alpha \}$ with $a=1,2,3$, $\alpha=1,2,3,4$ and $5$ generators of coset $SO(6)/SO(5)$ as $T^{\hat{a}} = \{ T^{\hat\alpha}, T^{\hat{5}} \}$ with $\hat{\alpha}=1,2,3,4$. Among these generators, $\{T^a_{L/R}\}$ belongs to the $SU(2)_L\times SU(2)_R \simeq SO(4)\subset SO(5)$, $\{T^\alpha\}$ belong to the coset $SO(5)/SO(4)$. $T^{\hat{5}}$ belongs to the $SO(2)\subset SO(6)$. These generators satisfy the normalization conditions as
\beqa
\text{Tr}[T^{\hat{a}}T^{\hat{b}}]=\delta^{\hat{a}\hat{b}}, ~ \text{Tr}[T^{\hat{a}}T^{b}]=\delta^{\hat{a}b},~\text{Tr}[T^{a}T^{b}]=\delta^{ab}. \label{T_norm}
\eeqa

The $5$ generators of the coset $SO(6)/SO(5)$ can be written in a compact form as~\cite{Georgi:1982jb,Marzocca:2014msa}
\beqa
\left(T^{\hat{a}}\right)_{i j} \equiv \frac{1}{\sqrt{2}} \left(T^{\hat{a}6}\right)_{i j} = \frac{i}{\sqrt{2}}\left(\delta_{i }^6  \delta_{j }^{\hat{a}}-\delta_{j }^6  \delta_{i }^{\hat{a}}\right), \label{T_ahat}
\eeqa
where $\hat{a}=1, \ldots, 5$, $i,j=1,\ldots 6$. It can also be expressed as $\textbf{5}=\textbf{4}+\textbf{1}_S \in SO(4) \oplus SO(2)$ as
\beqa
&& (T^{\hat{\alpha}})_{i j} \equiv \frac{1}{\sqrt{2}} (T^{\hat{\alpha}6})_{i j} =\frac{i}{\sqrt{2}}\left(\delta_{i }^6  \delta_{j }^{\hat{\alpha}}-\delta_{j }^6  \delta_{i }^{\hat{\alpha}}\right), \quad \hat{\alpha}=1, 2,3 , 4, \nn\\
&& (T^{\hat{5}})_{ij} \equiv \frac{1}{\sqrt{2}}(T^{\hat{5}\hat{6}})_{i j}=\frac{i}{\sqrt{2}}\left(\delta_{i }^6  \delta_{j }^{\hat{5}}-\delta_{j }^6  \delta_{i }^{\hat{5}}\right),  \label{T_alpha-T_5}
\eeqa
where $T^{\hat{5}}\equiv T_S$ is the generators of $SO(2)$, with $S$ denotes singlet. If the unbroken symmetry is $SO(4)\times SU(2)$ with $9$ generators or $SO(4)\times SO(2)$ with $7$ generators instead of $SO(5)$ with $10$ generators, then there are $6$, $8$ instead of $5$ NGBs will be present in the end.  
If one breaks $SO(6)$ down to instead $SO(5)$ but $SO(4)\times SO(2)$, 
the symmetry breaking pattern $SO(6)/SO(4)\times SO(2)$ leads to $\textbf{4}+\textbf{4}$ consists the elements of a composite two Higgs doublet model (2HDM)~\cite{Mrazek:2011iu}. 
In this case, $T^{\hat{\alpha}}$ become unbroken generators.

The $5$ generators of the coset $SO(6)/SO(5)$ in $6$ by $6$ matrix includes $\textbf{3} \in SU(2)_\alpha\subset SO(4)$ as 
\beqa
&& T^{\hat{1}}=\frac{i}{\sqrt{2}}\left(\begin{array}{cccc}{0_2} & {0_2} &  & {-e_{1}} \\ 
{0_2} & {0_2} & {} &\\ 
 &  & 0 & \\
{e_{1}^{T}} &&&0
\end{array}\right), \nn\\
&& T^{\hat{2}}=\frac{i}{\sqrt{2}}\left(\begin{array}{cccc}{0_2} & {0_2} &  & {-e_{2}} \\ 
{0_2} & {0_2} & {} &\\ 
 &  & 0 & \\
 {e_{2}^{T}} & &&0
\end{array}\right), \nn\\
&& T^{\hat{3}}=\frac{i}{\sqrt{2}}\left(\begin{array}{cccc}{0_2} & {0_2} &  &  \\ 
{0_2} & {0_2} & {} & {-e_{1}}\\ 
 &  & 0 & \\
 & {e_{1}^{T}} &&0
\end{array}\right),  \label{SO6-Tc_1-3}
\eeqa
where ${0}_2 \equiv \text{diag}(0,0)$, and $e_{1,2}$ are eigen-vectors of Pauli matrix as
\beqa
e_1= (1,0)^T, \quad e_2=(0,1)^T. \label{e_1-e_2}
\eeqa
The broken generator along the EW symmetry breaking direction is denoted as $T^{\hat{4}}$ as a $6$ by $6$ matrix as
\beqa
&& T^{\hat{4}}=\frac{i}{\sqrt{2}}\left(\begin{array}{cccc}{0_2} & {0_2} &  &  \\ 
{0_2} & {0_2} & {} & {-e_{2}}\\ 
 &  & 0 & \\
 & {e_{2}^{T}} &&0
\end{array}\right), \nn\\
&& T^{\hat{5}}=\frac{i}{\sqrt{2}}\left(\begin{array}{cccc}{0_2} & {0_2} &  &  \\ 
{0_2} & {0_2} & {} & {}\\ 
 &  & 0 & -1\\
 & {} &1 &0
\end{array}\right) , \qquad \label{SO6-Tc_4-5}
\eeqa
where $T^{\hat{5}}$ is the generator of $SO(2)$ and ${0}_2 \equiv \text{diag}(0,0)$. 


It can be check that all generators are normalized as
\beqa
\text{Tr}[T_L^aT_L^b] = \delta^{ab}, ~\text{Tr}[T_R^aT_R^b] = \delta^{ab}, ~ \text{Tr}[T^{\hat{a}}T^{\hat{b}}] = \delta^{\hat{a}\hat{b}}, \qquad
\eeqa
and they satisfy the commutation relations as below:
\beqa
&& \left[T^{a }_L, T^{b }_R \right]=0, \quad\left[T^{a }_L, T^{b }_L \right]=i \epsilon^{a b c} T^{c }_L, ~\left[T^{a }_R, T^{b }_R \right]=i \epsilon^{a b c} T^{c }_R, \nn\\
&& \left[T^{\hat{a}}, T^{\hat{4}}\right]=\frac{i}{2} \delta^{\hat{a} }_b\left(T^{b}_{L}-T^{b}_{R}\right), ~\left[T^{\hat{a}}, T^{\hat{b}}\right]=\frac{i}{2} \epsilon^{\hat{a} \hat{b} c}\left(T^{c}_{L}+T^{c}_{R} \right), \nn\\
&& \left[T^{\hat{a}}, T^{b}_{L,R} \right]= \frac{i}{2} \left( \epsilon^{\hat{a} b \hat{c}} T^{\hat{c}} \mp  \delta^{\hat{a} b} T^{\hat{4}} \right),  ~ 
 \left[T^{\hat{4}}, T^{a}_{L,R} \right]= \pm \frac{i}{2} \delta^{a\hat{b}} T^{\hat{b}}. \nn\\
&& 
\eeqa
It is obvious that in the $SO(6)/SO(5)$ case, the unbroken generators transform as a the reducible representations $(\textbf{3},\textbf{1})+(\textbf{1},\textbf{3})\in SO(4)\sim SU(2)_L\times SU(2)_R$, respectively.

Under the unitary transformation $P$ as will be discussed in Eq.(\ref{P}), the $5$ generator of coset $SO(6)/SO(5)$ in Eqs.(\ref{SO6-Tc_1-3}) and (\ref{SO6-Tc_4-5}) can be expressed explicitly as
\beqa
&& t^{\hat{1}}= - \frac{1}{2}\left(\begin{array}{cccc}{0_2} & {0_2} &  & e_2 \\ 
{0_2} & {0_2} & {} &  e_1 \\ 
 &  & 0 & \\
e_2^T & e_1^T &&0
\end{array}\right), \nn\\
&& t^{\hat{2}}=\frac{i}{2}\left(\begin{array}{cccc}{0_2} & {0_2} &  & {-e_2} \\ 
{0_2} & {0_2} & {} & e_1\\ 
 &  & 0 & \\
 e_2^T & -e_1^T &&0
\end{array}\right), \nn\\
&& t^{\hat{3}}=\frac{1}{2}\left(\begin{array}{cccc}{0_2} & {0_2} &  & -e_1 \\ 
{0_2} & {0_2} & {} & e_2\\ 
 &  & 0 & \\
-e_1^T & e_2^T &&0
\end{array}\right),  \nn \\ 
&& t^{\hat{4}}=\frac{i}{2}\left(\begin{array}{cccc}{0_2} & {0_2} &  & e_1 \\ 
{0_2} & {0_2} & {} & e_2 \\ 
 &  & 0 & \\
-e_1^T & - e_2^T &&0
\end{array}\right), \nn\\
&& t^{\hat{5}}=\frac{i}{2}\left(\begin{array}{cccc}{0_2} & {0_2} &  &  \\ 
{0_2} & {0_2} & {} & {}\\ 
 &  & 0 & \sqrt{2} \\
 & {} &-\sqrt{2} &0
\end{array}\right) , \qquad \label{SO6-tc_1-5}
\eeqa
where $T^{\hat{5}}$ is the generator of $SO(2)$ and ${0}_2 \equiv \text{diag}(0,0)$. In the unitary of NMCHM, $h_{1,2,3}=0$, the Higgs PNGBs can be expressed explicitly as
{ 
\small
\beqa
U &=& \exp{\left( i \frac{\sqrt{2} }{f} (h_4 t^{\hat{4}} + h_5 t^{\hat{5}}) \right)} \nn\\
&=& \left(
\begin{array}{cccccc}
  \frac{c_{\phi }+1}{2}  c_{\psi }^2+s_{\psi }^2 & 0 & 0 &   \frac{c_{\phi }-1}{2}  c_{\psi }^2 & -\frac{\left(1-c_{\phi }\right) c_{\psi }
   s_{\psi }}{\sqrt{2}} & -\frac{c_{\psi } s_{\phi }}{\sqrt{2}} \\
 0 & 1 & 0 & 0 & 0 & 0 \\
 0 & 0 & 1 & 0 & 0 & 0 \\
   \frac{c_{\phi }-1}{2}  c_{\psi }^2 & 0 & 0 &  \frac{c_{\phi }+1}{2}  c_{\psi }^2+s_{\psi }^2 & -\frac{\left(1-c_{\phi }\right) c_{\psi }
   s_{\psi }}{\sqrt{2}} & -\frac{c_{\psi } s_{\phi }}{\sqrt{2}} \\
 -\frac{\left(1-c_{\phi }\right) c_{\psi } s_{\psi }}{\sqrt{2}} & 0 & 0 & -\frac{\left(1-c_{\phi }\right) c_{\psi } s_{\psi }}{\sqrt{2}} & c_{\psi }^2+c_{\phi } s_{\psi
   }^2 & -s_{\phi } s_{\psi } \\
 \frac{c_{\psi } s_{\phi }}{\sqrt{2}} & 0 & 0 & \frac{c_{\psi } s_{\phi }}{\sqrt{2}} & s_{\phi } s_{\psi } & c_{\phi } \\
\end{array}
\right), \nn\\
&&
\eeqa
}
where $c_\phi=\cos(\phi/f)$, $s_\phi=\sin(\phi/f)$, with $\phi=h_4^2+h_5^2$ and $\hat{h}_4=c_\psi$, $\hat{h}_5=s_\psi$. In the absence of the singlet $h_5=0$, $\psi=0$, $\phi=h_4=h$, or $h_4=0$, $\phi=h_5=s$, then
\beqa
&& U (h) = \left(
\begin{array}{cccccc}
 \frac{1}{2} \left(c_h+1\right) & 0 & 0 & \frac{1}{2} \left(c_h-1\right) & 0 & -\frac{s_h}{\sqrt{2}} \\
 0 & 1 & 0 & 0 & 0 & 0 \\
 0 & 0 & 1 & 0 & 0 & 0 \\
 \frac{1}{2} \left(c_h-1\right) & 0 & 0 & \frac{1}{2} \left(c_h+1\right) & 0 & -\frac{s_h}{\sqrt{2}} \\
 0 & 0 & 0 & 0 & 1 & 0 \\
 \frac{s_h}{\sqrt{2}} & 0 & 0 & \frac{s_h}{\sqrt{2}} & 0 & c_h \\
\end{array}
\right), \nn\\
&& U(s) = \left(
\begin{array}{ccc}
 \textbf{1}_4 & 0 & 0  \\
 0 & c_s & -s_s \\
 0 & s_s & c_s \\
\end{array}
\right), 
\eeqa
where $\textbf{1}_4 = \text{diag}(1,1,1,1)$. They just recovers the GBs matrix for the coset $SO(6)/SO(4)$ or $SO(6)/SO(2)$, respectively.

\subsection{Generators of $SO(5)$}

There are $10$ unbroken generator of $SO(5)\subset SO(6)$ are
\beqa
 (T^{a}_{L / R})_{i j} 
 &=& -\frac{i}{2}\left(\frac{1}{2} \epsilon^{a b c}\left(\delta^{b }_i \delta^{c }_j-\delta^{b }_j \delta^{c }_i \right) \pm\left(\delta^{a }_i \delta^{4 }_j-\delta^{a }_j \delta^{4 }_i \right)\right), \nn\\
 \left(T^{\alpha}\right)_{i j} & \equiv & \frac{1}{\sqrt{2}} (T^{\alpha 5})_{i j} = - \frac{i}{\sqrt{2}}\left(\delta_{j }^5  \delta_{i }^{\alpha} - \delta_{i }^5  \delta_{j }^{\alpha} \right),  \label{T_a}
\eeqa
where $a=1,2,3$, $\alpha=1,2,3,4$ and $i,j=1,\ldots,6$. These gives the coset space $SO(5)/SO(4)$ for MCHM~\cite{Agashe:2004rs,Csaki:2008zd,Contino:2010rs,vonGersdorff:2015fta} .
  
To be more explicitly, the $6$ generators span the representation
$\textbf{6} = (\textbf{3},1)+ (1,\textbf{3})\in SU(2)_L\times SU(2)_R\simeq SO(4)\subset SO(5)\subset SO(6)$ can be chosen as
\beqa
&& T^{1}_{L}=\frac{i}{2}\left(\begin{array}{ccc}{0} & {-\sigma_{1}} &\\ 
{\sigma_{1}} & {0} & \\
      &  & {0}_2 
\end{array}\right) , \quad T^{2}_{L}=\frac{i}{2}\left(\begin{array}{ccc}{0} & {\sigma_{3}} & \\ {-\sigma_{3}} & {0}& \\
      &  & {0}_2 
\end{array}\right) , \nn\\
&& T^{3}_{L}=\frac{i}{2}\left(\begin{array}{ccc}{-i \sigma_{2}} & {0} &\\ 
{0} & {-i \sigma_{2}}& \\
&  & {0}_2 \end{array}\right) ; \nn\\
&& T^{1}_{R}=\frac{i}{2}\left(\begin{array}{ccc} 0 & i \sigma_{2} & \\ 
i \sigma_{2} & 0 & \\
      &  & {0}_2 \end{array}\right) , \quad T^{2}_{R}=\frac{i}{2}\left(\begin{array}{ccc}{0} & {1}_2 & \\ 
      -{1}_2 & 0 & \\
      &  & {0}_2 \end{array}\right) , \nn\\
&& T^{3}_{R}=\frac{i}{2}\left(\begin{array}{ccc}{-i \sigma_{2}} & {0} & \\ 
      {0} & {i \sigma_{2}} & \\
      &  & {0}_2 \end{array}\right),  \label{T_L-R}
\eeqa
where ${0}_2 \equiv \text{daig}(0,0)$, ${1}_2 \equiv \text{daig}(1,1)$, and it can be checked that they satisfy the commutation relations as
\beqa
\left[T_{a}^{L}, T_{b}^{R}\right]=0, \quad T_{a}^{L, R} T_{b}^{L, R}=\frac{1}{4} \delta_{a b}+\frac{i}{2} \varepsilon_{a b c} T_{c}^{L, R} . \qquad
\eeqa

There are $4$ residue unbroken generators span the representation $(\textbf{2},\textbf{2})\in SU(2)_L\times SU(2)_R \simeq SO(4)$ are
\beqa
&& T_\alpha^{1}=\frac{i}{\sqrt{2}}\left(\begin{array}{cccc}{0_2} & {0_2} & {-e_{1}}  & \\ 
{0_2} & {0_2} & {} &\\ 
 {e_{1}^{T}} &  & 0 & \\
 &&&0
\end{array}\right), \nn\\
&& T_\alpha^{2}=\frac{i}{\sqrt{2}}\left(\begin{array}{cccc}{0_2} & {0_2} & {-e_{2}}  & \\ 
{0_2} & {0_2} & {} &\\ 
 {e_{2}^{T}} &  & 0 & \\
 &&&0
\end{array}\right), \nn\\
&& T_\alpha^{3}=\frac{i}{\sqrt{2}}\left(\begin{array}{cccc}{0_2} & {0_2} &   & \\ 
{0_2} & {0_2} & {-e_{1}} &\\ 
 &  {e_{1}^{T}} & 0 & \\
 &&&0
\end{array}\right), \nn\\
&& T_\alpha^{4}=\frac{i}{\sqrt{2}}\left(\begin{array}{cccc}{0_2} & {0_2} &   & \\ 
{0_2} & {0_2} & {-e_{2}} &\\ 
 &  {e_{2}^{T}} & 0 & \\
 &&&0
\end{array}\right),  \label{SO6-Talpha_1-4}
\eeqa
where ${0}_2 \equiv \text{diag}(0,0)$.

\subsection{Generators of $SO(4)$}

The generator of $SU(2)_L\times SU(2)_R\simeq SO(4)$ in Eq.(\ref{T_a}) can also expressed as
\beqa
 (T^{a}_{L / R})_{i j} &=& \frac{1}{2} [(J^a)_{ij}\pm (K^a)_{ij}], 
\eeqa
where the generators $J^a$ and $K^a$ are those corresponding to the angular momentum and boosts as
\beqa
(J^a)_{ij} & \equiv  & \frac{1}{2}\epsilon_{abc}(T^{bc})_{ij} = -i \frac{1}{2}\epsilon_{abc} \left(\delta^{b }_i \delta^{c }_j-\delta^{b }_j \delta^{c }_i \right) , \nn\\
(K^a)_{ij} & \equiv & (T^{a4})_{ij} = - i \left(\delta^{a }_i \delta^{4 }_j-\delta^{a }_j \delta^{4 }_i \right).
\eeqa

To be more explicitly,
\beqa
&& J^1 = i\left(
\begin{array}{cccc}
 0 & 0 & 0 &  \\
 0 & 0 & -1 &  \\
 0 & 1 & 0 &  \\
  &  &  & 0_3 \\
\end{array}
\right), \quad J^2 = i\left(
\begin{array}{cccc}
 0 & 0 & 1 &  \\
 0 & 0 & 0 &  \\
 -1 & 0 & 0 &  \\
  &  &  & 0_3 \\
\end{array}
\right), \nn\\
&& J^3 = i\left(
\begin{array}{cccc}
 0 & -1 & 0 &  \\
 1 & 0 & 0 &  \\
 0 & 0 & 0 &  \\
  &  &  & 0_3 \\
\end{array}
\right); \nn\\
&& K^1 = i\left(
\begin{array}{ccccc}
 0 & 0 & 0 & -1 &  \\
 0 & 0 & 0 & 0 &  \\
 0 & 0 & 0 & 0 &  \\
 1 & 0 & 0 & 0 &  \\
  &  &  &  & 0_2 \\
\end{array}
\right), \quad K^2 = i\left(
\begin{array}{ccccc}
 0 & 0 & 0 & 0 &  \\
 0 & 0 & 0 & -1 &  \\
 0 & 0 & 0 & 0 &  \\
 0 & 1 & 0 & 0 &  \\
  &  &  &  & 0_2 \\
\end{array}
\right), \nn\\
&& K^3 = i\left(
\begin{array}{ccccc}
 0 & 0 & 0 & 0 &  \\
 0 & 0 & 0 & 0 &  \\
 0 & 0 & 0 & -1 &  \\
 0 & 0 & 1 & 0 &  \\
  &  &  &  & 0_2 \\
\end{array}
\right),  \label{SO6-J-K_1-3}
\eeqa
where $0_3 \equiv \text{diag}(0,0,0)$ and $0_2 \equiv \text{diag}(0,0)$.

The representation in Eq.(\ref{T_L-R}) can be transformed to those one are more familiar with through a similar transformation with a unitary matrix $P$,
\beqa
t^{a}_{L,R} &=&  P T^{a}_{L,R}  P^{-1}, \nn\\ 
P &=& \frac{1}{\sqrt{2}}\left(
\begin{array}{cccccc}
 0 & 0 & i & 1 & 0 & 0 \\
 i & -1 & 0 & 0 & 0 & 0 \\
 i & 1 & 0 & 0 & 0 & 0 \\
 0 & 0 & -i & 1 & 0 & 0 \\
 0 & 0 & 0 & 0 & \sqrt{2} & 0 \\
 0 & 0 & 0 & 0 & 0 & -\sqrt{2} \\
\end{array}
\right). \label{P}
\eeqa
To be more explicitly, we have
\beqa
&& t^1_L =\frac{1}{2}\left(\begin{array}{ccc}{\sigma_{1}} & {} &\\ 
{} & {\sigma_{1}} & \\
      &  &  0_2 
\end{array}\right), \quad t^2_L =\frac{1}{2}\left(\begin{array}{ccc}{\sigma_{2}} & {} &\\ 
{} & {\sigma_{2}} & \\
      &  &  0_2 
\end{array}\right)
, \nn\\
&& t^3_L =\frac{1}{2}\left(\begin{array}{ccc}{\sigma_{3}} & {} &\\ 
{} & {\sigma_{3}} & \\
      &  &  0_2 
\end{array}\right),  \nn\\
&& t^1_R = - \frac{1}{2}\left(\begin{array}{ccc}{} & {  1_2 } &\\ 
{ 1_2} & {} & \\
      &  &  0_2 
\end{array}\right), \quad t^2_R =\frac{i}{2}\left(\begin{array}{ccc}{} & -{  1_2 } &\\ 
{ 1_2} & {} & \\
      &  &  0_2 
\end{array}\right), \nn\\
&& t^3_R =\frac{1}{2}\left(\begin{array}{ccc}{- 1_2} & {} &\\ 
{} & { 1_2} & \\
      &  &  0_2 
\end{array}\right), \label{t_L-R}
\eeqa
where $ 1_2 \equiv\text{diag}(1,1)$ and $0_2 \equiv \text{diag}(0,0)$. Under above unitary transformation $P$ in Eq.(\ref{P}), the fundamental representation of scalar in $SO(5)\subset SO(6)$ becomes
\beqa
\phi = \left(\begin{array}{c}
h_1 \\ 
h_2 \\ 
h_3 \\
h_4 \\
s \\
0 \\
\end{array}\right)  \overset{P}{\to} 
\frac{1}{\sqrt{2}} \left(\begin{array}{c}
{h_4+i h_3}  \\
{i \left(h_1+i h_2\right)}  \\
{h_2+i h_1}  \\
{h_4-ih_3} \\
\sqrt{2} s\\
0\\
\end{array}\right) = \left(\begin{array}{c}
h^{0\star}  \\
-h^- \\
h^+  \\
h^0 \\
 s\\
0\\
\end{array}\right) , \qquad \quad
\eeqa
which consists of the Higgs bi-doublet as denoted in Eq.(\ref{H-h-U}) as
\beqa
\frac{1}{\sqrt{2}}\mathbf{H}=\frac{1}{\sqrt{2}}(H^c,H) = 
\left(\begin{array}{cc}
h^{0\star} & h^+ \\ 
h^- & h^0 \\ 
\end{array}\right).
\eeqa
where $H$ and $H^c$ are SM Higgs doublet and its complex-conjugate with notation defined in Eqs.(\ref{Phi-H}) and (\ref{Phi-Hc}), respectively.

Under above unitary transformation $P$ in Eq.(\ref{P}), the $4$ generator in the coset $SO(5)/SO(4)$ in Eq.(\ref{SO6-talpha_1-4}) can be expressed more explicitly as
\beqa
&& t_\alpha^{1}= \frac{1}{2}\left(\begin{array}{cccc}{0_2} & {0_2} & {e_2}  & \\ 
{0_2} & {0_2} & {e_1} &\\ 
 e_2^T & e_1^T & 0 & \\
 &&&0
\end{array}\right), \nn\\
&& t_\alpha^{2}= \frac{i}{2}\left(\begin{array}{cccc}{0_2} & {0_2} & {e_2}  & \\ 
{0_2} & {0_2} & {-e_1} &\\ 
 -e_2^T & e_1^T & 0 & \\
 &&&0
\end{array}\right), \nn\\
&& t_\alpha^{3}= \frac{1}{2} \left(\begin{array}{cccc}{0_2} & {0_2} & e_1  & \\ 
{0_2} & {0_2} & {-e_2} &\\ 
e_1^T &  -e_2^T & 0 & \\
 &&&0
\end{array}\right), \nn\\
&& t_\alpha^{4}=  \frac{i}{2} \left(\begin{array}{cccc}{0_2} & {0_2} & -e_1  & \\ 
{0_2} & {0_2} & -e_2 &\\ 
e_1^T &  e_2^T & 0 & \\
 &&&0
\end{array}\right),  \label{SO6-talpha_1-4}
\eeqa
where ${0}_2 \equiv \text{diag}(0,0)$. In the unitary of MCHM, $h_{1,2,3}=0, h_4=h$, the Higgs PNGBs can be expressed explicitly as
\beqa
U &=& \exp{\left( i \frac{\sqrt{2} }{f} h t_\alpha^4\right)} \nn\\
&=& \left(
\begin{array}{cccccc}
 \frac{1}{2} \left(c_h+1\right) & 0 & 0 & \frac{1}{2} \left(c_h-1\right) & \frac{s_h}{\sqrt{2}} & 0 \\
 0 & 1 & 0 & 0 & 0 & 0 \\
 0 & 0 & 1 & 0 & 0 & 0 \\
 \frac{1}{2} \left(c_h-1\right) & 0 & 0 & \frac{1}{2} \left(c_h+1\right) & \frac{s_h}{\sqrt{2}} & 0 \\
 -\frac{s_h}{\sqrt{2}} & 0 & 0 & -\frac{s_h}{\sqrt{2}} & c_h & 0 \\
 0 & 0 & 0 & 0 & 0 & 1 \\
\end{array}
\right),
\eeqa
where $c_h=\cos(h/f)$ and $s_h=\sin(h/f)$. 

Under the unitary transformation $P$ in Eq.(\ref{P}), the $3$ generators of angular momentum and boost in Eq.(\ref{SO6-J-K_1-3}) can be respectively, re-expressed as
\beqa
&& {\mathrm J}^1 = \frac{1}{2}\left(
\begin{array}{ccccc}
 0 & 1 & -1 & 0 &  \\
 1 & 0 & 0 & -1 &  \\
 -1 & 0 & 0 & 1 &  \\
 0 & -1 & 1 & 0 &  \\
  &  &  &  & 0_2 \\
\end{array}
\right), \nn\\
&&  {\mathrm J}^2 =\frac{i}{2} \left(
\begin{array}{ccccc}
 0 & -1 & -1 & 0 &  \\
 1 & 0 & 0 & -1 &  \\
 1 & 0 & 0 & -1 &  \\
 0 & 1 & 1 & 0 &  \\
  &  &  &  & 0_2 \\
\end{array}
\right) , \quad {\mathrm J}^3 = \left(
\begin{array}{ccccc}
 0 & 0 & 0 & 0 &  \\
 0 & -1 & 0 & 0 &  \\
 0 & 0 & 1 & 0 &  \\
 0 & 0 & 0 & 0 &  \\
  &  &  &  & 0_2 \\
\end{array}
\right); \nn\\
&& {\mathrm K}^1 = \frac{1}{2}\left(
\begin{array}{ccccc}
 0 & 1 & 1 & 0 &  \\
 1 & 0 & 0 & 1 &  \\
 1 & 0 & 0 & 1 &  \\
 0 & 1 & 1 & 0 &  \\
  &  &  &  & 0_2 \\
\end{array}
\right), \quad {\mathrm K}^2 = \frac{i}{2} \left(
\begin{array}{ccccc}
 0 & -1 & 1 & 0 &  \\
 1 & 0 & 0 & 1 &  \\
 -1 & 0 & 0 & -1 &  \\
 0 & -1 & 1 & 0 &  \\
  &  &  &  & 0_2 \\
\end{array}
\right), \nn\\
&& {\mathrm K}^3 = \left(
\begin{array}{ccccc}
 1 & 0 & 0 & 0 &  \\
 0 & 0 & 0 & 0 &  \\
 0 & 0 & 0 & 0 &  \\
 0 & 0 & 0 & -1 &  \\
  &  &  &  & 0_2 \\
\end{array}
\right), \label{SO6-j-k_1-3}
\eeqa
where $0_2 \equiv \text{diag}(0,0)$. 
One can make the recombination to give

\subsection{Higgs Representations}

\subsubsection{Custodial symmetry $ SO(3)\subset SO(4)$}

As we will show in the following, the unbroken ${\mathcal H}=SO(3)\simeq SU(2)_V$ symmetry after EW symmetry breaking origins from an enlarged ${\mathcal G}=SO(4)\simeq SU(2)_L\times SU(2)_R$ global symmetry.

At first, it is interesting to observing that the unitary product of the Higgs doublet is a $SO(4)$ invariant.
\beqa
H^\dag H = h_1^2 + h_2^2 + h_3^2 + h_4^2 = h^T h.
\eeqa
with $h=(h_1,h_2,h_3,h_4)^T$ is $\textbf{4} \in SO(4)$.
Thus the SM Higgs action can also be re-expressed as $SO(4)\subset SO(5)$ invariant form as
\beqa
{\mathcal L}_h &=&  \frac{1}{2}(D_\mu h^T) (D_\mu h)  +  \frac{1}{2}\mu^2 (h^T h) - \frac{1}{4}\lambda (h^T h)^2 
 , \label{L_h-SO(4)}
\eeqa
with $h=(h_1,h_2,h_3,h_4)^T$ transforms linearly as the four-dimensional vector representation $\textbf{4}$ of global symmetry group $SO(4)$, from which, it is obvious that the Higgs potential is also invariant under $SO(4)$ symmetry. The Higgs potential can be re-expressed as
\beqa
V_h = - \frac{\lambda}{4} (h^T h - v^2 )^2,
\eeqa
which has a minimum as the three sphere $S^3$ with radius $v$ 
\beqa
\vev{h^T h} = v^2
\eeqa
where $v\approx 246\gev$. It is convenient to choose the vacuum expectation value of $h$ in vector representation $\textbf{4}$ of $SO(4)$ as
\beqa
\vev{h} = v \left(\begin{array}{c}
0 \\ 
0\\
0\\
1\\
\end{array}\right),  \label{h_vev-SO(4)}
\eeqa
Considering the shift field $v \to v+h$ and non-shifted fields as $h_a$ with $a=1,2,3$, then
\beqa
h = \left(\begin{array}{c}
h_1 \\ 
h_2\\
h_3\\
v+h\\
\end{array}\right),
\eeqa
where $v$ is the vacuum expectation value (Vev) of the Higgs singlet $h_4$ and $h\equiv\tilde{h}_4$ is the quantum fluctuation around the v.
Thus, the $SO(4)$ symmetry acts linearly on the Cartesian coordinates in terms of field $h_a$.

The vacuum $\vev{h}$ simultaneously breaks the global $SO(4)$ symmetry down to the unbroken $O(3)$ global symmetry, under which, the would GBs $\varphi_a$ with $a=1,2,3$ transforms linearly as a triplet $\textbf{3}$ under $O(3)$, while the Higgs $h$ transforms as a singlet $\textbf{1}\in O(3)$.
The four-real component of the Higgs vector representation can be re-organized as the usual SM Higgs complex doublet with two complex scalar fields $h^+$ and $h_0$ as
\beqa
\Phi &=& \frac{1}{\sqrt{2}} \left(\begin{array}{c}{ h_{2}+i h_{1}} \\ { h_4-i h_{3}}\end{array}\right)   \equiv \left(\begin{array}{c}{ h^+ } \\ { h_0 }\end{array}\right) = \frac{1}{\sqrt{2}}H \nn\\
& \overset{\varphi=0}{=} & \frac{1}{\sqrt{2}} \left(\begin{array}{c}{ 0 } \\ { h + v }\end{array}\right)  , \label{Phi-H}
\eeqa
where $H$ is up to a normalization constant to $\Phi=H/\sqrt{2}$, due to the kinetic term, and in the last equality, we have adopt to the unitary gauge. 

After EWSB, $SO(4)$ breaks down to a $SO(3)\simeq SU(2)_V$ symmetry, in terms of custodial symmetry, which is invariant for the three goldstone bosons, or as equivalent physical degree of freedom, the three gauge bosons $W_\mu^a$ in $\textbf{3}$, the fundamental representation of $SO(3)$ or adjoint representation of $SU(2)_V$. The unbroken global symmetry $SO(3)$ leads to the well-know gauge boson mass relation that $m_W=m_Z\cos\theta_W$ as the prediction of the SM, which can also be expressed through the ratio 
\beqa
\rho \equiv \frac{m_W^2}{m_Z^2c_\theta^2}, \quad \label{rho_SM_2}
\eeqa
when $g'\ne0$, where $c_\theta=\cos\theta_W$ is the Weinberg mixing angle. The gauge boson mass relation implies that custodial is a good approximate symmetry of the SM, which is exact at least at tree level. Thus, the custodial symmetry ensures $\rho=1$ at tree level, and also ensures small corrections to $\rho$ at quantum level as stated before in Eq.(\ref{rho_SM_0}). It also guarantees that $m_W^2/m_Z^2=1$ in the absent of $U(1)_Y$, i.e., $g'\to0$. 

To be brief, the unbroken $SO(3)$ symmetry of the SM vacuum can be origins from an $SO(4)$ invariant fixed point.

\subsubsection{From Cartesian to Polar}

It is worthy of observing a fact that after $h_4$ obtains vev, the SM Higgs doublet with Cartesian coordinate field $h_a$ in Eq.(\ref{Phi-H}) can be reexpressed as below~\cite{Hagiwara:1993ck,Buchalla:2013rka,Alonso:2016oah}
\beqa
\Phi 
&=& \frac{1}{\sqrt{2}} \left(\begin{array}{c}{ h_{2}+i h_{1}} \\ { h+v-i h_{3}}\end{array}\right) \nn\\
&=& \left(\textbf{1}_2 + \frac{h \textbf{1}_2 + i\sigma^a h_a}{v} \right)   \left(\begin{array}{c}{ 0 } \\ { \frac{ v}{\sqrt{2}}  }\end{array}\right)  \nn\\
&\overset{v\gg 1}{\approx }& \frac{v}{\sqrt{2}}  \exp{\left( \frac{h \textbf{1}_2 + i\sigma^a h_a}{v} \right)}   \left(\begin{array}{c}{ 0 } \\ { 1  }\end{array}\right) \nn\\
&\overset{h \ll v}{\approx }&  \exp{\left( \frac{ i\sigma^a h_a}{v} \right)}  \frac{1}{\sqrt{2}}  \left(\begin{array}{c}{ 0 } \\ { v + h  }\end{array}\right), 
\label{h-varphi}
\eeqa
where $e_2=(0,1)^T$ is one of two eigenvector of Pauli matrix $\sigma^3$ for spin-down state as denoted in Eq.(\ref{e_1-e_2}) and in the last equality, we have made the identification that $\varphi_{a} \approx h_{\hat{a}}$ with $\hat{a}=1,2,3$, which are exact in the limit $v\to \infty $, i.e., the unitary gauge.
Thus, one can map the Higgs doublet in Eq.(\ref{Phi-H}) to that in the polar decomposition, or angular coordinate $\varphi_a$ as
\beqa
\Phi \equiv \frac{1}{\sqrt{2}} h_4 {\mathbf U} e_2 ,  \label{H_polar} 
\eeqa
where $h_4=h+v$ is the magnitude of $\Psi$ with $h$ denotes the radial coordinate, while ${\mathbf U}\in S^3$ is a $4$-dimensional unit vector, denoting the SM Goldstone bosons matrix in $SU(2)$ as defined in Eq.(\ref{U_v}) 
\beqa
{\mathbf U} & \equiv & \exp{\left( i \frac{\sigma^a \varphi_a}{v} \right)} = c_\varphi + i \sigma^a \hat\varphi_a s_\varphi \nn\\
&=& \left(
\begin{array}{cc}
 c_\varphi+\frac{i s_\varphi}{|\varphi| }\varphi_3 & \frac{s_\varphi}{|\varphi| } (i \varphi^1+\varphi^2) \\
 \frac{i s_\varphi}{|\varphi| }  (\varphi^1+i \varphi^2) & c_\varphi-\frac{i s_\varphi }{|\varphi| } \varphi_3 \\
\end{array}
\right) \nn\\
& \overset{ v \gg 1 }{=} & \left(
\begin{array}{cc}
 1+ \frac{i \varphi_3}{v} & \frac{i \varphi_1+\varphi_2}{v} \\
 \frac{i \varphi_1-\varphi_2}{v} & 1-\frac{i \varphi_3}{v} \\
\end{array}
\right) \overset{\varphi = 0 }{=} \textbf{1}_{2} , \label{U_varphi}
\eeqa
where $\hat\varphi_a\equiv \varphi_a/|\varphi|$ are the three dimensionless angular coordinates and we have made the notation $s_\varphi$ and $c_\varphi$ defined in Eq.(\ref{s_varphi-c_varphi}). In the last equality, it is shown that the unitary matrix just reduces to be unity in the unitary gauge, as expected.


With the mapping from the $h$ in doublet $H$ in Eq.(\ref{Phi-H}) to that in Eqs.(\ref{H_polar}) and (\ref{U_varphi}), one can read the one-to-one correspondence as
\beqa
h_a= \frac{h_4}{ v}\varphi_a, \quad a=1,2,3, 
\eeqa
where $h_4=h+v$ with $h=\tilde{h}_4$ as the quantum fluctuation of Higgs field.
In this case, there is a relations between the GBs in Cartesian and polar coordinates as
\beqa
|h|^2 = h_a h_a= \frac{h_4^2 \left(v^2+ |\varphi| ^2\right)}{v^2} \overset{\varphi\to 0 }{=}(h+v)^2 , \quad \label{const_polar}
\eeqa
which origins from the constraint of polar coordinate. Therefore, 
\beqa
h 
= h_4 - v + h_4 \frac{\varphi^2}{2v^2} - h_4 \frac{\varphi^4}{8v^4} + \cdots .  \quad 
\eeqa
Thus, the Higgs filed $h$ is $SO(4)$ invariant, while only three of the components of $\Phi(h_a,h_4)$ are independent due to the constraint above.

In the polar coordinate, the SM Lagrangian can be re-expressed as
\beqa
{\mathcal L}_h &=& \frac{1}{2}(v+h)^2 (\partial_\mu {\mathbf U}^\dag \partial^\mu {\mathbf U}) \nn\\
&& + \frac{1}{2} (\partial_\mu h)^2 - \frac{\lambda}{4}((h+v)^2-v^2)^2.
\eeqa
where the Higgs scalar potential depends only upon the radial coordinate $h$, while the three GB fields $\varphi$ only appears in the kinetic terms, either gauged or not. In this case, it is obvious that the interactions of $h$ with $\varphi^a$ becomes nonlinearly, comparing to that between $h_a$ and $h_4$ in Cartesian coordinate. Nevertheless, by the Lehmann-Symanzik-Zimmermann (LSZ) reduction formula, $h$ and $h_4$ give the same $S$-matrix, so that the the change of coordinates does not affect $S$-matrix elements~\cite{Alonso:2016oah}. 


\subsubsection{Higgs bi-doublet in $SO(4)$}

Since $SO(4) $ is isomorphic to $SU(2)_L\times SU(2)_R$, one can relate the $SO(4)$ vector to $\mathbf{H}\equiv (H^c,H)$, a complex $SU(2)_R \times SU(2)_L$ bi-doublet $(\bar{2},2)\in SU(2)_L\times SU(2)_R$, with the complex conjugate of $H$ in Eq.(\ref{Phi-H}) as
\beqa
H^c \equiv i\sigma^2 H^\star =  \left(\begin{array}{c}{ h_{4}+i h_{3}} \\ { -(h_2-i h_{1}}) \end{array}\right)  = \left(\begin{array}{c}{ h^{0\ast} } \\ { h^- }\end{array}\right), \label{Phi-Hc}
\eeqa
where $h^-=(h^+)^\ast$. More explicitly, the polar decomposition of $\mathbf{H}$ into $h$ and the $SU(2)$ matrix $U$ as
\beqa
&& \mathbf{H} = 
\left(\begin{array}{c}{h_{4}+i h_{3} \quad h_{2}+i h_{1}} \\ {  -h_{2}+i h_{1}  \quad h_{4}-i h_{3}}\end{array}\right)=h_{\hat{\alpha}} \overline{\sigma}_{\hat{\alpha}} \equiv|h| \mathbf{U},  \qquad\quad \label{H-h-U}
\eeqa
with $\overline{\sigma}_{\hat{\alpha}} = \left(i \vec{\sigma}, \textbf{1}_{2}\right)$ with $\hat{\alpha} = (a,4), ~ a=1,2,3$.


From eq.(\ref{H-h-U}), it can be found that
\beqa
\frac{h_{\hat{\alpha}}}{|h|} \equiv \hat{h}_{\hat{\alpha}} = \frac{1}{2}\text{Tr}[\mathbf{U} {\sigma}_{\hat{\alpha}}] .
\qquad \label{h_a-varphi_a}
\eeqa
where $\sigma_{\hat{\alpha}}\equiv (-i\vec{\sigma}, \textbf{1}_2)$ is the complex conjugate of $\overline{\sigma}_{\hat{\alpha}}$.

To be explicitly, $\hat{h}_{\hat\alpha}$ can also be expressed as, respectively,
\beqa
&& \hat{h}_{a} = -\frac{i}{2}\text{Tr}[\mathbf{U}\sigma_{a}] ,
 \quad \hat{h}_4 = c_\varphi . \label{ha-h4_U} 
\eeqa
Combing Eqs.(\ref{U_varphi}), the $h_a$ in the vector representation can be expressed as
\beqa
\hat{h}_{\hat{\alpha}} = \left( \hat\varphi_a s_\varphi, c_\varphi \right) 
\overset{ v \to \infty }{=} \bigg( \frac{\varphi_a}{v}, 1 \bigg) , \label{h_varphi}
\eeqa
where $\hat\varphi_a \equiv \varphi_a/|\varphi|$ and $c_\varphi$, $s_\varphi $ as defined in Eq.(\ref{s_varphi-c_varphi}).
Therefore, the fundamental scalar $h^T$ can be re-expressed as
\beqa
\phi = v \left(
\begin{array}{c}
{ s_\varphi } \vec{\varphi} \\ 
{ c_\varphi }
\end{array}
\right)  , \label{phi_linear}
\eeqa
where $\vec\varphi\equiv (\hat\varphi_1,\hat\varphi_2,\hat\varphi_3)$ is a $3$-dimensional unit vector, which transforms linearly as the $3$-dimensional representation under the unbroken $O(3)$ symmetry. The angle $\varphi$ together with the $\vec\varphi$ have $4$ degrees of freedom.

\subsubsection{Non-linear transformation of $SO(4)$}

Since there are three independent component in polar coordinate, of the constraint in Eq.(\ref{const_polar}), one can make a redefinition of the GBs as below,
\beqa
\pi_a \equiv  v s_\varphi \hat\varphi_a,
\eeqa
so that the fourth non-independent component can be expressed as a non-linear function of these unconstrained fields $\pi_a$. In this case, the linear parametrization in Eq.(\ref{phi_linear}) can be re-expressed in a square-root parametrization as
\beqa
\phi =  \left(
\begin{array}{c}
\vec\pi \\ 
\sqrt{v^2 -\vec\pi\cdot\vec\pi}
\end{array}
\right)  , \label{pi_nonlinear}
\eeqa
where $\vec\pi$ has $3$-components
Thus, the unitary GB matrix ${\mathbf U}$ can also be rewritten in non-linear expression as
\beqa
{\mathbf U}  = \sqrt{1 - \frac{\pi^2}{v^2} } + i  \sigma^a \pi_a, \label{square root-parametrization}
\eeqa
which implies that the fourth component $c_\varphi$ is a non-linear function of the independent three components $\pi^a$ with $a=1,2,3$. The constraint of polar coordinate in Eq.(\ref{const_polar}) turns the $SO(4)$ linear transformation upon $\Phi(h_a,h_4)$ into a non-linear $SO(4)$ transformation when written only in terms of unconstrained fields $\pi^{1,2,3}\in SO(3)$.

Thus, the $SO(4)$ symmetry acts non-linearly on the angular coordinates in terms of field $\pi_a$ as
\beqa
{\mathbf U}  \approx 1 + i  \pi - \frac{\pi^2}{2v^2} + {\mathcal O}(\frac{\pi^4}{v^4}) ,
\eeqa
where $\pi \equiv \sigma^a \pi_a$.

In this definition, the kinetic terms of the non-linear sigma model can be rewritten as
\beqa
{\mathcal L}_{kin} &=& \frac{v^2}{2} \text{Tr}[(\partial_\mu {\mathbf U}^\dag)(\partial^\mu {\mathbf U})] \nn\\
&=&  \frac{1}{2} (\partial_\mu \pi^a)^2 + \frac{v^2}{2} s_\varphi (\partial_\mu \varphi)^2 ,
\eeqa
where the second term denote the the non-linear kinetic terms of $\pi^a$ fields, since 
\beqa
{\mathcal L}^{(\varphi )}_{kin} = \frac{v^2}{2} s_\varphi (\partial_\mu \varphi)^2 = \frac{1}{2} \frac{(\pi^a \partial_\mu \pi^a)^2 }{v^2-\pi^2}.
\eeqa
Together, the kinetic term of $\pi^a$ becomes in curved space $S^3\simeq SO(4)/SO(3)$ as
\beqa
{\mathcal L}_{kin}  = \frac{1}{2} g_{ab} \partial_\mu \pi^a \partial^\mu \pi^b, \label{L_kin-S3}
\eeqa
with a non-flat metric in PNGB field space
\beqa
g_{ab} = \delta_{ab} + \frac{\pi_a \pi_b}{v^2 - \pi^2},  \label{gab_SO4vsSO3}
\eeqa
and the Ricci scalar curvature of the coset space is
\beqa
R = \frac{6}{v^2} >0, \label{R_SO4vsSO3}
\eeqa
where the number $6$ takes accounts of the degrees of freedom of the number of unconstrained $\pi^a$ fields. Therefore, the PNGB, or Higgs nonlinearity origins from the scalar $S^3$ manifold curvature. At the scale $v\gg 1$, the physics at scale $f$ gives non-linear interactions due to the non-flat metric
\beqa
g_{ab} = \delta_{ab} + \frac{\pi_a \pi_b}{v^2 } + \frac{\pi^2}{v^4}\pi_a\pi_b + {\mathcal O}(\frac{\pi^6\delta_{ab}}{v^6}) .
\eeqa 
The kinetic term of the non-linear sigma model can be expressed in a series of expansion as
\beqa
{\mathcal L}^{(\varphi )}_{kin} &=& \frac{1}{2}\partial_\mu \vec\pi \cdot \partial^\mu \vec\pi + \frac{1}{2v^2}(\vec\pi \cdot \partial_\mu\vec\pi)^2 \nn\\
&&+ \frac{1}{2v^4} \vec\pi \cdot \vec\pi (\vec\pi \cdot \partial_\mu \vec\pi)^2 + \ldots . 
\eeqa
where $\ldots $ denotes higher order derivatives coupled interaction terms no less than $1/v^6$ order. The Lagrangian has only three scalar d.o.f.




\section{CCWZ and Chiral Lagrangian}
\label{sec:CCWZ}


\subsection{Nambu Goldstone boson matrix}

\subsubsection{$\Omega$ parametrization}

Assuming the Lagrangian is invariant under global ${\mathcal G}$, with a unbroken symmetry $H \subset {\mathcal G}$, which is a linearly realized subgroup of ${\mathcal G}$. The Nambu Goldstone bosons (NGBs) due to the global spontaneous symmetry breaking pattern ${\mathcal G}\to {\mathcal H}$ can be described by the $\Omega$ field defined as
\beqa
\Omega  \equiv e^{i \frac{\varphi}{2f}\Xi } \equiv e^{i\Pi}, \label{Omega}
\eeqa
where in the last equality, we have absorbed the decay constant $f$ into $\Pi$ so that it is a dimensionless field. The $\Omega$ is under the action of an element of global symmetry $\mathfrak{g}\in {\mathcal G}$ and an element of local symmetry $\mathfrak{h} \in {\mathcal H}$ as~\cite{Coleman:1969sm,Callan:1969sn},
\beqa
\mathfrak{g} \Omega(\Xi) = \Omega(g(\Xi))\mathfrak{h}(\Xi,g),
\eeqa
which transforms in a linear representation for $H$. While it is a non-linear one for the other elements under the global symmetry ${\mathcal G}$ as
\beqa
\Omega  \rightarrow \mathfrak{g} \Omega  \mathfrak{h}^{-1}(\Xi, \mathfrak{g}), \label{Omega_g-h}
\eeqa
which defines the non-linear transformation of the NGBs fields as $\Xi $ denoted as
\beqa
\Xi =\Xi^{\hat{a}}  T_{\hat{a}},
\eeqa
where $T_{\hat{a}}$ are the broken generators in the coset $\mathcal{G} / \mathcal{H}$, with $\hat{a}=1,\cdots,\text{dim}(\mathcal{G} / \mathcal{H})$. To be brief, the Nambu-Goldstone boson fields are associated with the coset or quotient space. The quotient space ${\mathcal G}/{\mathcal H}$ is said to be symmetric if there exists an automorphism of the grading, ${\mathcal R}$, under which the broken generators change sign.


\subsubsection{Symmetric coset and automorphism}

For symmetric coset, there exists an automorphism or ``grading'' symmetry ${\mathcal R}$ that acts upon the generators of ${\mathcal G}$, which changes the sign of the broken generators as
\beqa
\mathcal{R} :
\left\{\begin{array}{l}{T_{a} \rightarrow+T_{a}} \\ 
{T_{\hat{a}} \rightarrow-T_{\hat{a}}}
\end{array}\right.  \label{R_automorphism}
\eeqa
where $T_a$  with $a=1,2,\cdots, \text{dim}(\mathcal{H})$ are the generators of ${\mathcal H}$ and $T_{\hat{a}}$  with $\hat{a}=1,2,\cdots, \text{dim}({\mathcal G}/{\mathcal H})$ are the generators of the coset ${\mathcal G}/{\mathcal H}$. The $(T_a,T_{\hat{a}})$ form an orthonormal basis of ${\mathcal G}$, and they satisfy
\beqa
[T_a,T_b] \propto T_c, \quad [T_a, T_{\hat{b}}] \propto T_{\hat{c}}, \quad [T_{\hat{a}},T_{\hat{b}}] \propto T_{c},
\eeqa
where the fist condition follows from that ${\mathcal H}$ is closed, the second condition is due to the fact that the structures constants are completely antisymmetric for compact groups, while the last condition is one for a symmetric coset. 



Under the discrete automorphism symmetry ${\mathcal R}$, the GB fields transforms as
\beqa
{\mathcal R}: \quad \Omega  \rightarrow \Omega ^{-1} . \label{R_Omega} 
\eeqa 

In the symmetric coset, the non-linear field transformation of $\Omega$ can be read as below by acting on Eq.(\ref{Omega_g-h}) with the grading ${\mathcal R}$ and taking the hermitian conjugate
\beqa
\Omega \rightarrow \mathfrak{h} \Omega \mathfrak{g}_{\mathcal{R}}^{-1} .  \label{Omega_h-gR}
\eeqa


\subsubsection{$\boldsymbol{\Sigma}$ Parametrization}

From Eqs.(\ref{Omega_h-gR}) and (\ref{Omega_g-h}), one can define a new non-linear field $\Sigma$ as square of $\Omega$
\beqa
\boldsymbol{\Sigma}  \equiv \Omega ^{2} = e^{i \frac{\varphi}{f}\Xi } = e^{2i\Pi},
\eeqa
which transforming linearly under the global symmetry $\mathfrak{g} \in {\mathcal G}$ as
\beqa
\boldsymbol{\Sigma}  \rightarrow \mathfrak{g} \boldsymbol{\Sigma}  \mathfrak{g}_{\mathcal{R}}^{-1}.
\eeqa
where $\mathfrak{g}_{\mathcal{R}}$ is a global element of ${\mathcal G}$, obtained by acting on $g$ with ${\mathcal R}$, which is independent of the GBs field $\Pi$. Hence $\Sigma$ transforms linearly under ${\mathcal G}$. This shows explicitly that the transformation on $\Pi$ is a realization of ${\mathcal G}$ and that it is linear when restricted to ${\mathcal H}$. Under the grading symmetry, the GB fields transforms as
\beqa
{\mathcal R}: \boldsymbol{\Sigma} \rightarrow \boldsymbol{\Sigma}^{-1} . 
\eeqa
The transformation also implies that the covariant derivative of the non-linear $\boldsymbol{\Sigma} $ is defined as
\beqa
\mathbf{D}_{\mu} \boldsymbol{\Sigma}=\partial_{\mu} \boldsymbol{\Sigma}+i g_{A} \left(\overline{\mathrm{A}}_{\mu} \boldsymbol{\Sigma}-\boldsymbol{\Sigma} \overline{\mathrm{A}}_{\mu}^{(\mathcal{R})}\right). \label{Dm_Sigma-Ab-AbR}
\eeqa
Under the gauge transformation, it is straightforward to have
\beqa
&& \overline{A}_{\mu} \rightarrow \mathfrak{g} \overline{A}_{\mu} \mathfrak{g}^{-1}-\frac{i}{g_{A}} \mathfrak{g}\left(\partial_{\mu} \mathfrak{g}^{-1}\right), \nn\\
&& \overline{A}_{\mu}^{\mathcal{R}} \rightarrow \mathfrak{g}_{R} \overline{A}_{\mu}^{\mathcal{R}} \mathfrak{g}_{\mathcal{R}}^{-1}-\frac{i}{g_{A}} \mathfrak{g}_{\mathcal{R}}\left(\partial_{\mu} \mathfrak{g}_{\mathcal{R}}^{-1}\right) ,
\eeqa
with $g_A$ denoting the associated gauge couple constant. 

To obtain gauge interactions by formally gauging the symmetry ${\mathcal G}$, one has introduced gauge fields $\overline{A}_{\mu}=(\overline{W}_\mu, \overline{B}_\mu)\in {\mathcal G}$ with
\beqa
\overline{\mathrm{W}}_{\mu} \equiv W_{\mu}^{a} Q_{L}^{a} \quad \text { and } \quad \overline{\mathrm{B}}_{\mu} \equiv B_{\mu} Q_{Y}, \label{W_L-B}
\eeqa
and its automorphism field $\overline{A}_{\mu}^{\mathcal{R}} \equiv \mathcal{R}\left(\overline{A}_{\mu}\right)$. 
$Q_L^a$ and $Q_Y$ denote the embedding in ${\mathcal G}$ of the $SU(2)_L\times U(1)_Y$ generators. 

In the symmetric coset case, it is equivalent to use either $\Omega $ or $\boldsymbol{\Sigma} $ in describing the GBs degrees of freedoms and its interactions. 

For external gauge field $g_A \overline{\mathrm{A}}_\mu=A_\mu$, where the gauge coupling $g_A$ are absorbed into the gauge fields. Consequently, the covariant derivative of $\boldsymbol\Sigma$ in Eq.(\ref{Dm_Sigma-Ab-AbR}) transform under local ${\mathcal G}$ transformation are defined as
\beqa
D_\mu  \boldsymbol{\Sigma} = \partial_\mu  \boldsymbol{\Sigma} + i A_\mu  \boldsymbol{\Sigma} - i  \boldsymbol{\Sigma} A_\mu^{(\mathcal{R})},
\eeqa
where $A_\mu = A_\mu^a T^a + A_\mu^{\hat{a}}T^{\hat{a}}$ is the external gauge field and $A_\mu^{(\mathcal{R})}$ is obtained by acting on $A_\mu$ with the ``grading'' ${\mathcal R}$ as 
\beqa
A_\mu^{(\mathcal{R})} = {\mathcal R}(A_\mu) = A_\mu^a T^a -  A_\mu^{\hat{a}}T^{\hat{a}}.
\eeqa
where $T^{\hat{a}}$ and $T^a$ are the broken and unbroken generators, respectively and normalized as $\text{Tr}(T^A T^B)=\delta^{AB}$. In gauging the standard model group, i.e, $A_\mu^{\hat{a}}=0$. The external gauge field $A_\mu$ transform under the local ${\mathcal H}$ symmetry as
\beqa
&& A_\mu \rightarrow  \mathfrak{g} A_\mu  \mathfrak{g}^{-1} - i  \mathfrak{g} \partial_\mu  \mathfrak{g}^{-1}, \nn\\
&& A_\mu^{{\mathcal R}} \rightarrow  \mathfrak{g}_{{{\mathcal R}}} A_\mu  \mathfrak{g}_{{{\mathcal R}}}^{-1} - i  \mathfrak{g}_{{{\mathcal R}}} \partial_\mu  \mathfrak{g}_{{{\mathcal R}}}^{-1}. \label{A_g-AR_gR}
\eeqa


Therefore, the covariant term of $\boldsymbol\Sigma$ above can be expressed as
\beqa
D_\mu \boldsymbol{\Sigma}= \partial_\mu \boldsymbol{\Sigma}+ i [A_\mu^{a}T^a , \boldsymbol{\Sigma}] + i  \{ A_\mu^{\hat{a}}T^{\hat{a}} , \boldsymbol{\Sigma} \}. \label{Dm_Sigma-Aa-Aat}
\eeqa


\subsection{Omega decomposition}

In the CCWZ's general approach, one can redefine the GB fields as $d_\mu$ and $e_\mu$ through the Maurer-Cartan one form,  which is decomposed along the broken and unbroken directions, respectively,
\beqa
\omega_\mu \equiv - i  \Omega^\dag \partial_\mu \Omega 
=d_{\mu}^{\hat{a}} T^{\hat{a}}+e_{\mu}^{a} T^{a} \equiv d_{\mu}+e_{\mu}, \label{omega_d-E}
\eeqa
where $\Omega$ is defined in Eq.(\ref{Omega_g-h}) 
is defined in Eq.(\ref{Omega-Sigma}) and the kinetic term can be generalized to be covariant derivative term. $T^{\hat{a}}\in {\mathcal G}/{\mathcal H}$ while $T^a\in {\mathcal H}$ are respectively the broken and unbroken generators normalized as $\text{Tr}(T^A T^B)=\delta^{AB}$.

Since $\Omega(\Pi,{\mathfrak g})$ is transformed under global symmetry  ${\mathcal G}$ as Eq.(\ref{Omega_g-h}), 
\beqa
-i \Omega^{-1}\partial_\mu \Omega & \to &  -i \mathfrak{h} \Omega^{-1} \mathfrak{g}^{-1} \partial_\mu( \mathfrak{g} \Omega \mathfrak{h}^{-1} ) = -i \mathfrak{h} \Omega^{-1}  \partial_\mu( \Omega \mathfrak{h}^{-1} ) \nn\\
&& = -i \mathfrak{h} \Omega^{-1} ( \partial_\mu \Omega  ) \mathfrak{h}^{-1} -i \mathfrak{h}  (\partial_\mu \mathfrak{h}^{-1} ) \nn\\
&& =  \mathfrak{h} \omega_\mu \mathfrak{h}^{-1} -i \mathfrak{h}  (\partial_\mu \mathfrak{h}^{-1} ) ,  \qquad 
\eeqa
where $\mathfrak{h} \omega_\mu \mathfrak{h}^\dag \in {\mathcal G}$, while the shift term $i \mathfrak{h} \partial_\mu \mathfrak{h}^\dag \in {\mathcal H}$.

From above transformation, together with Eq.(\ref{omega_d-E}), one deduces that under any element $\mathfrak{g}\in {\mathcal G}$, $d_\mu$ and $e_\mu$ transform as
\beqa
&& d_\mu \rightarrow \mathfrak{h}  (\Pi, \mathfrak{g}) d_\mu \mathfrak{h}^{-1}(\Pi, \mathfrak{g}), \nn\\
&& e_\mu \rightarrow \mathfrak{h}  (\Pi, \mathfrak{g}) e_\mu \mathfrak{h}^{-1}(\Pi, \mathfrak{g})  - i h(\Pi, \mathfrak{g}) \partial_\mu \mathfrak{h}^{-1}(\Pi, \mathfrak{g}).  \qquad \label{d-E_g-h}
\eeqa
Thus, $d_\mu$ transforms linearly under local symmetry ${\mathcal H}$ as the GBs, while $e_\mu$ transforms non-linearly as a gauge field with gauge symmetry ${\mathcal H}$. 

It is obvious that the covariant field $e_\mu$ transforms like a gauge field under ${\mathcal H}$, i.e., transforms as a connection, and it will be more obvious by defining a field strength
\beqa
e_{\mu\nu} \equiv \partial_{\mu} e_{\nu}-\partial_{\nu} e_{\mu}+i\left[e_{\mu}, e_{\nu}\right]  , \qquad \label{emn}
\eeqa
where $e_\mu$ can be viewed as non-linear relation of gauge field with gauge symmetry ${\mathcal H}$. $e_{\mu\nu}$ transforms linearly under the local symmetry ${\mathcal H}$ again, like that of $d_\mu$ as
\beqa
e_{\mu\nu} \rightarrow \mathfrak{h} (\Pi, \mathfrak{g})   e_{\mu\nu} \mathfrak{h}^{-1}(\Pi, \mathfrak{g}).
\eeqa
Therefore, $d_\mu(\Pi)$ and $e_{\mu\nu}(\Pi)$ are covariant variable, and together with those by acting upon the covariant derivative $\nabla_\mu$, consist of the building blocks of the low energy effective Lagrangian. 



Except the covariant derivative $\nabla_\nu$ for local symmetry ${\mathcal H}$, one might be interested in gauging an external local symmetry ${\mathcal H}^\prime \subset {\mathcal G}$, e.g., a subgroup $SU(2)_L\times U(1)_Y$, which plays the role of the SM electroweak group. In this case, one can replace the kinetic derivative with the a covariant derivative of the external gauge field $A_\mu$ as
\beqa
&& D_{\mu}=\partial_{\mu}+i A_{\mu} , 
\eeqa
where the gauge coupling $g_A$ are absorbed into the gauge fields.

\subsubsection{Explicit expression of covariant derivative}
\label{app:cov_deri}


The most general high-energy effective Lagrangian for describing the electroweak interactions of the gauge group ${\mathcal G}$ and of the GBs of a non-linear realization of the symmetric ${\mathcal G}/{\mathcal H}$ basis, up to four-derivative bosonic interactions, can be obtained by gauging only the EW symmetry of SM gauge group, while keep the group ${\mathcal G}=SO(6)$ global.

The gauge covariant derivative as defined in Eq.(\ref{Dm_Sigma-Aa-Aat}) becomes
\beqa
D_\mu \boldsymbol{\Sigma} &=& \partial_\mu \boldsymbol{\Sigma}+ i [W_\mu^{a} Q_L^a + B_\mu^{3} Q_Y , \boldsymbol{\Sigma}],
\eeqa
where the external gauge field as $g_A\overline{\mathrm{A}}_\mu = g\overline{\mathrm{W}}_\mu+g' \overline{\mathrm{B}}_\mu $ 
by using the embedding representations in the SM as Eq.(\ref{Q_L-Y}), we find the explicit expression of the gauge invariant fields strength as 
\beqa
&& g_A\overline{\mathrm{A}}_\mu =
 \frac{1}{2}\left(\begin{array}{ccc}{ \ii\ii \sigma_2 \gamma_\mu } & \ii\ii {-i \sigma_{1} gW_\mu^1 + i \sigma_3 gW_\mu^2  } & {} \\ 
{i \sigma_{1} gW_\mu^1 - i \sigma_3 gW_\mu^2 } & \ii\ii {\sigma_2 Z_\mu} & {} \\ 
{} & {  } & \ii{ {0}_{2}}\end{array}\right),   \qquad \quad \label{Wb-Bb}
\eeqa
where $A_\mu \equiv gW_\mu^3 +  g' B_\mu $, $Z_\mu \equiv gW_\mu^3 - g' B_\mu$ and $0_2=\text{daig}$. Then, we can obtain the field stress tensor of SM $g_A\overline{\mathrm{F}}_{\mu\nu}\equiv(g\overline{\mathrm{W}}_{\mu\nu},g' \overline{\mathrm{B}}_{\mu\nu})$ in a similar expression. 

The explicit expression of the building block in $SO(6)/SO(5)$ with SM gauge group is
\beqa
\overline{\mathrm{V}}_\mu(\Omega) = \left(
\begin{array}{cc}
 0_3 & \overline{\mathrm{V}}^{(W,B)}_\mu   \\
 -\overline{\mathrm{V}}^{(W,B)T}_\mu & \overline{\mathrm{V}}^{(\phi,psi)}_\mu  \\
\end{array}
\right),
\eeqa
where $0_3=\text{diag}(0,0,0)$, and $\overline{\mathrm{V}}_\mu^{(W,B)}$ and $\overline{\mathrm{V}}_\mu^{(\phi,\psi)}$ are both $3$ by $3$ matrix those can be expressed, respectively, as
\begin{widetext}
\beqa
 \overline{\mathrm{V}}^{(W,B)}_\mu 
 &=& \left(
\begin{array}{ccc}
 g c_\psi^2 \sin ^2\left(\frac{\phi }{2 f}\right) W^1_{\mu } &  g \sin ^2\left(\frac{\phi }{2 f}\right)  s_\psi c_\psi W^1_{\mu } & \frac{1}{2} g c_\psi s_\phi W^1_{\mu } \\
 g c_\psi^2 \sin ^2\left(\frac{\phi }{2 f}\right) W^2_{\mu } & g \sin ^2\left(\frac{\phi }{2 f}\right)  s_\psi c_\psi W^2_{\mu } & \frac{1}{2} g c_\psi s_\phi W^2_{\mu } \\
 c_\psi^2 \sin ^2\left(\frac{\phi }{2 f}\right) Z_{\mu } &  \sin ^2\left(\frac{\phi }{2 f}\right)  s_\psi c_\psi Z_{\mu } & \frac{1}{2} c_\psi s_\phi Z_{\mu } \\
\end{array}
\right), \nn\\
\overline{\mathrm{V}}^{(\phi,\psi)}_\mu &=& \frac{1}{f} \left(
\begin{array}{ccc}
 0 & 0 & c_\psi \\
 0 & 0 & s_\psi \\
 -c_\psi & -s_\psi & 0 \\
\end{array}
\right) \phi_\mu + \left(
\begin{array}{ccc}
 0 & c_\phi-1 & -s_\phi s_\psi \\
 1-c_\phi & 0 & c_\psi s_\phi \\
 s_\phi c_\psi & -c_\psi s_\phi & 0 \\
\end{array}
\right) \psi_\mu , \qquad \quad
\eeqa
where $\phi_\mu =\partial_\mu \phi$ and $\psi_\mu =\partial_\mu \psi$.
The scalar chiral field $\overline{\mathbf{T}} $ in Eq.(\ref{Tbar}), becomes
\beqa
\overline{\mathbf{T}} = -\frac{i}{2}\left(
\begin{array}{cccccc}
 0 & 1 & 0 & 0 & 0 & 0 \\
 -1 & 0 & 0 & 0 & 0 & 0 \\
 0 & 0 & 0 & -1+2 c_\psi^2 \sin ^2\left(\frac{\phi }{2 f}\right) & \sin ^2\left(\frac{\phi }{2f}\right) 2 s_\psi c_\psi & c_\psi s_\phi \\
 0 & 0 & 1-2 c_\psi^2 \sin ^2\left(\frac{\phi }{2 f}\right) & 0 & 0 & 0 \\
 0 & 0 & - \sin ^2\left(\frac{\phi }{2f}\right) 2 s_\psi c_\psi & 0 & 0 & 0 \\
 0 & 0 & -c_\psi s_\phi & 0 & 0 & 0\\
\end{array}
\right).\qquad
\eeqa

\end{widetext}

\subsection{Building blocks in CCWZ}

\subsubsection{Building blocks for Sigma}

The global vector chiral field (If the interactions are not gauged) is defined as
\beqa
\overline{\mathrm{V}}_{\mu}=\left(\mathbf{\partial}_{\mu} \boldsymbol{\Sigma}\right) \boldsymbol{\Sigma}^{-1},
\eeqa
and it transforms in the adjoint representation of ${\mathcal G}$ as
\beqa
\overline{\mathrm{V}}_{\mu} \to \mathfrak{g} \overline{\mathrm{V}}_{\mu} \mathfrak{g}^{-1}. 
\eeqa
The effective Lagrangian describing the GB interactions in the context of the non-linearly realized ${\mathcal G}$ breaking mechanism, with symmetric coset ${\mathcal G}/{\mathcal H}$ can be constructed from $\overline{\mathrm{V}}_{\mu}$. Then the gauged version of the chiral vector field of the non-linear sigma model can be defined as through the formally gauging of the the full group as
\beqa
\overline{\mathrm{V}}_{\mu}=\left(\mathbf{D}_{\mu} \boldsymbol{\Sigma}\right) \boldsymbol{\Sigma}^{-1}. \label{Vbar}
\eeqa
The Cartan form for ${\mathcal G}$ of above
\beqa
 (\boldsymbol{\Sigma}^{-1} d \boldsymbol{\Sigma})^{\hat{a}},
\eeqa
after projecting onto the subspace of broken generators, gives a vielbein corresponding to the metric on ${\mathcal G}/{\mathcal H}$. As will be seen, the leading order terms with at most two derivatives are built out of two vielbein. The Wess-Zumino-Witten term is a $5$-form, which can be build out of five vielbein~\cite{Gripaios:2009pe}.

The building blocks for the effective chiral gauge theory includes ${\mathcal G}$ invariants vector chiral field $\overline{\mathrm{V}}_{\mu}$and the gauge field strength $\overline{A}_{\mu \nu} = (\overline{\mathrm{W}}_{\mu \nu},\overline{\mathrm{B}}_{\mu \nu})$ which transform in the adjoint of ${\mathcal G}$.

The building block are
\beqa
\overline{\mathrm{V}}_{\mu}, \quad \overline{\mathrm{F}}_{\mu \nu} , \quad \boldsymbol{\Sigma} \overline{\mathrm{F}}_{\mu \nu}^{R} \boldsymbol{\Sigma}^{-1}.
\eeqa 
with the automorphism symmetry in the symmetric coset, one can introduce the graded vector chiral fields 
\beqa
\overline{\mathrm{V}}_{\mu}^{\mathcal{R}} \equiv \mathcal{R}\left(\overline{\mathrm{V}}_{\mu}\right)=\left(\mathbf{D}_{\mu} \boldsymbol{\Sigma}\right)^{-1} \boldsymbol{\Sigma} = - \boldsymbol{\Sigma}^{-1} \overline{\mathrm{V}}_{\mu}\boldsymbol{\Sigma}.
\eeqa
One additional custodial symmetry breaking source besides the SM ones, in analogy to the scalar chiral field $\boldsymbol{T}$ in the high energy can be embedded in ${\mathcal G}$ of the hypercharge generators  $Q_Y$ as
\beqa
\overline{\mathbf{T}} \equiv \boldsymbol{\Sigma} Q_{Y} \boldsymbol{\Sigma}^{-1}. \label{Tbar}
\eeqa

To be brief, the building block of high-energy effective Lagrangian for a UV complete model, i.e, a symmetric CHM are
\beqa
\overline{\mathrm{V}}_{\mu}, \quad \overline{\mathrm{F}}_{\mu \nu} , \quad \boldsymbol{\Sigma} \overline{\mathrm{F}}_{\mu \nu}^{R} \boldsymbol{\Sigma}^{-1}, \quad \overline{\mathbf{T}} .
\eeqa
Based upon these, we can construct the gauge covariant operators in the following.

\subsubsection{Building blocks for Omega}

With the field strength of the external gauge field $F_{\mu\nu}$, one can also redefine the gauge field as a new covariant structure variables $f_{\mu\nu}^{-,L,R}$ as~\cite{Contino:2011np}
\beqa
 f_{\mu\nu} & \equiv &\Omega^{-1} F_{\mu\nu} \Omega 
 \equiv   f_{\mu\nu}^{-}+ f_{\mu\nu}^{+} 
,  \nn\\
 f_{\mu\nu}^{({\mathcal R})} & = & \Omega F_{\mu\nu}^{({\mathcal R})} \Omega^{-1} 
 \equiv  - f_{\mu\nu}^{-}+ f_{\mu\nu}^{+} . \qquad \quad \label{fmn}
\eeqa
with $\Omega=\Omega^{-1}$, $f_{\mu\nu}^{({\mathcal R})} \equiv {\mathcal R}(f_{\mu\nu}) $ and
\beqa
f_{\mu\nu}^{+}  = f_{\mu\nu}^{L}+f_{\mu\nu}^{R} + f_{\mu\nu}^{(\alpha)}. 
\eeqa
where $f_{\mu\nu}^{-} = f_{\mu \nu}^{-, \hat{a}} T_{\hat{a}} $, $f_{\mu\nu}^{+} = f_{\mu \nu}^{+, a} T_{a}$, $f_{\mu\nu}^{L/R} = f_{\mu \nu}^{L/R, a} T_{a}^{L/R} $, $ f_{\mu\nu}^{(\alpha)} = f_{\mu\nu}^{(\alpha) \alpha} T_\alpha $ with $T_{a}^{L,R}$ are the six generators of $SU(2)_{L,R}$ those embedded into the unbroken generators $T^a$. Thus, the even and odd components of $f_{\mu\nu}$ under the grading ${\mathcal R}$ is
\beqa
f_{\mu\nu}^\pm = \frac{1}{2}(f_{\mu\nu}\pm f_{\mu\nu}^{({\mathcal R})}).   \label{fpm} 
\eeqa
where $f_{\mu\nu}^{\pm}$ coincide respectively with the even and odd components of $f_{\mu\nu}$ under the grading ${\mathcal R}$, since
\beqa
{\mathcal R}[f_{\mu\nu}^\pm] = \pm f_{\mu\nu}^{\pm}.
\eeqa
which implies that the gauge field $f_{\mu\nu}^\pm$ carries even and odd ${\mathcal R}$ parity.

The new covariant structure variable $f_{\mu\nu}^{-,L,R}$ trans forms as a gauge covariant one under the local symmetry ${\mathcal H}$
\beqa
&& f_{\mu\nu}^{\pm} \rightarrow \mathfrak{h} (\Pi, \mathfrak{g})   f_{\mu\nu}^{\pm} \mathfrak{h}^{-1}(\Pi, \mathfrak{g}), \nn\\
&& f_{\mu\nu}^{(\mathcal R)\pm} \rightarrow \mathfrak{h} (\Pi, \mathfrak{g})   f_{\mu\nu}^{(\mathcal R)\pm} \mathfrak{h}^{-1}(\Pi, \mathfrak{g}).
\eeqa


According to Eq.(\ref{fmn}), with the aid of the trace property, one find that the gauge invariant field Lagrangian can be decomposed as
\beqa
\text{Tr}(F_{\mu\nu}F^{\mu\nu}) = \text{Tr}(f^+_{\mu\nu}f^{+\mu\nu}) + \text{Tr}(f^-_{\mu\nu}f^{-\mu\nu}).
\eeqa
Thus, one obtains some building block as below
\beqa
\text{Tr}(f_{\mu\nu}^- f^{-\mu\nu}), \quad \text{Tr}(f_{\mu\nu}^+ f^{+\mu\nu}).
\eeqa


From the gauged version of Eq.(\ref{omega_d-E}), i.e., (\ref{Eq:omegaB_mu}), one expand the ${\mathcal G}$ invariant strength tensor as below:
\beqa
\Omega^\dag F_{\mu\nu} \Omega 
&=& -i \Omega^\dag [D_\mu, D_\nu ] \Omega \nn\\
&=& \Omega^\dag D_\mu (\Omega \omega_\nu) - \Omega^\dag D_\nu (\Omega\omega_\mu) \nn\\
&=& i\omega_\mu \omega_\nu - i \omega_\nu \omega_\mu + \partial_\mu \omega_\nu - \partial_\nu \omega_\mu \nn\\
&=& d_{\mu\nu} + e_{\mu\nu} - i [e_\nu,d_\mu] + i [e_\mu,d_\nu]  \nn\\
&=& \nabla_\mu d_\nu - \nabla_\nu d_\mu  + i[d_\mu, d_\nu] + e_{\mu\nu} ,\label{Fmn_d_emn}
\eeqa
where we have used the factor that $[D_\mu, D_\nu] = i F_{\mu\nu} $, and made the notations as below 
\beqa
&& d_{\mu\nu} \equiv \partial_\mu d_\nu - \partial_\nu d_\mu  + i[d_\mu, d_\nu],\nn\\
&& e_{\mu\nu}\equiv \partial_\mu e_\nu -\partial_\nu e_\mu + i[e_\mu, e_\nu] , \nn\\
&& \nabla_\mu d_\nu \equiv \partial_\mu d_\nu + i [e_\mu,d_\nu], \nn\\
&& \nabla_\nu d_\mu \equiv  \partial_\nu d_\mu + i [e_\nu,d_\mu].
\eeqa
By comping the identities in Eq.(\ref{Fmn_d_emn}) with the decomposition of the gauge field in Eq.(\ref{fmn}), we find the relations between the gauge fields with the $e_{\mu\nu}$ and $d_\mu$ as below
\beqa
&& f_{\mu\nu}^+ \equiv i[d_\mu, d_\nu] + e_{\mu\nu}, \nn\\
&& f_{\mu\nu}^- \equiv  2\nabla_{[\mu} d_{\nu]} = \nabla_\mu d_\nu - \nabla_\nu d_\mu. \label{fmn_d_emn}
\eeqa

To be brief, the CCWZ covariant building block
\beqa
d_\mu, e_{\mu\nu},
\eeqa
can be re-expressed in terms of $\mathbf\Sigma$, which is manifestly invariant under global ${\mathcal G}$
\beqa
d_\mu &=& - \frac{i}{2}\Omega^{-1}(D_\mu \mathbf\Sigma) \Omega^{-1}= \frac{i}{2}\Omega (D_\mu \mathbf\Sigma)^{-1}\Omega , \nn\\
e_{\mu\nu} &=& - \frac{i}{4} \Omega^{-1}(D_\mu \mathbf\Sigma D_\nu \mathbf\Sigma^{-1}- D_\nu \mathbf\Sigma D_\mu \mathbf\Sigma^\dag) \Omega \nn\\
&& + \frac{1}{2}(\Omega^{-1}F_{\mu\nu} \Omega + \Omega F_{\mu\nu}^{{\mathcal R}} \Omega^{-1} ) , 
\eeqa
where in a symmetric coset ${\mathcal G}/{\mathcal H}$. Thus $e_{\mu\nu}$ and it is related to the unbroken sector field strength $f_{\mu\nu}^+$ and the commutator of $d_\mu$ field.

Therefore, in the end we have a set of covariant building blocks as
\beqa
\{d_\mu, f_{\mu\nu}^\pm\}, \quad \text{or} \quad \{ d_\mu, f_{\mu\nu}^-,  e_{\mu\nu} \}.
\eeqa
In summary, we have three independent building blocks of the effective Lagrangian in the CCWZ formalism as
$d_\mu$, $e_{\mu\nu}$ and $\nabla_\mu$, since they are covariant variables, which satisfies the local transformation under ${\mathcal H}$,
\beqa
d_\mu \to \mathfrak{h} d_\mu \mathfrak{h}^{-1} , \quad e_{\mu\nu}\to \mathfrak{h} e_{\mu\nu}\mathfrak{h}^\dag, \quad   f_{\mu\nu}^\pm \to \mathfrak{h} \nabla_\mu \mathfrak{h}^\dag. \qquad 
\eeqa

For $SO(6)/SO(5)$ case, one has
\beqa
&& e_{\mu\nu} = e_{\mu\nu}^{a L}T^{a L} + e_{\mu\nu}^{a R}T^{a R} + e_{\mu\nu}^\alpha T^\alpha, \nn\\
&& f_{\mu\nu} = f_{\mu\nu}^{a L}T^{a L} + f_{\mu\nu}^{a R}T^{a R} + f_{\mu\nu}^\alpha T^\alpha,
\eeqa
where two independent gauge invariant field strength tensors can be defined as
\beqa
e_{\mu\nu}^{L,R~a}  &=& \partial_\mu e_{\nu}^{L,R~a} -  \partial_\nu e_{\mu}^{L,R~a} + i [e_{\mu},e_{\nu}]^{L,R a}, \nn\\
e_{\mu\nu}^{\alpha}  &=& \partial_\mu e_{\nu}^\alpha -  \partial_\nu e_{\mu}^\alpha + i [e_{\mu},e_{\nu}]^\alpha.
\eeqa

Under the automorphism symmetry ${\mathcal R}$ in Eq.(\ref{R_automorphism}), its representation in the coset $SO(6)/SO(5)$ as shown in Eq.(\ref{RR_theta}). As  the broken generators $T^{\hat{a}}$ change sign, while the unbroken ones $T^a$ does not, thus one has
\beqa
&& h^{\hat{a}} \to - h^{\hat{a}}, \quad d_\mu^{\hat{a}} \to - d_\mu^{\hat{a}}, \quad f_{\mu\nu}^{ - \hat{a}} \to - f_{\mu\nu}^{ - \hat{a}} , \nn\\
&& e_{\mu\nu}^a \to  e_{\mu\nu}^a, \quad f_{\mu\nu}^{ + a} \to f_{\mu\nu}^{ + a}.
\eeqa
If $e_{\mu\nu}^\alpha$ and $f_{\mu\nu}^\alpha$ are absent, i.e., $\textbf{6} = (\textbf{3},1) + (1,\textbf{3}) \in SU(2)_L\times SU(2)_R \subset SO(4)$, then under left-right parity symmetry $P_{LR}$ in Eq.(\ref{P_LR}), one has
\beqa
T^{\hat{a}} &:& h^{\hat{a}} \to - \eta^{\hat{a}} h^{\hat{a}}, \quad d_\mu^{\hat{a}} \to - \eta^{\hat{a}} d_\mu^{\hat{a}}, \quad f_{\mu\nu}^{ - \hat{a}} \to - \eta^{\hat{a}}f_{\mu\nu}^{ - \hat{a}} , \nn\\
T^a&:& e_{\mu\nu}^{L/R a} \to  e_{\mu\nu}^{R/L a}, \quad f_{\mu\nu}^{ +L/R a} \to f_{\mu\nu}^{+R/L a},
\eeqa
where $\eta^{\hat{a}}=(1,1,1,-1)^T$.

\subsection{External Gauge Symmetry: SM or beyond}

Once an external gauge symmetry ${\mathcal H}_0\subset {\mathcal G}$ is turned on, e.g., the SM electroweak group ${\mathcal H}_0={\mathcal G}_{SM}=SU(2)_W\times U(1)_Y$, the ordinary derivative $\partial_\mu$ is promoted to be the covariant derivative $D_\mu \equiv \partial_\mu + i A_\mu$, where the external gauge fields $A_\mu =  A_\mu^{\hat{a}} T^{\hat{a}} + A_\mu^a T^a \in  {\mathcal G}$, i.e. still a general element of ${\mathcal G}$, transform under the local ${\mathcal H}_0$ transformation, namely
\beqa
A_\mu \to \mathfrak{g} (A_\mu - i \partial_\mu) \mathfrak{g}^\dag, \quad F_{\mu\nu} \to \mathfrak{g} F_{\mu\nu} \mathfrak{g}^\dag.
\eeqa
where $F_{\mu\nu}\equiv \partial_\mu A_\nu - \partial_\nu A_\mu +i[A_\mu, A_\nu]$ is the field strength of external gauge field $A_\mu$. Some of the above external gauge field source will become dynamical while the others will be turned off, i.e., by setting the others to zero. Explicitly, the dynamical part of $A_\mu$ will be those transforming under the local ${\mathcal H}_0={\mathcal G}_{EW}=SU(2)_{W}\times U(1)_Y\subset {\mathcal H}$,
\beqa
A_\mu &=& \frac{g}{\sqrt{2}}W_\mu^+ T_L^+ + \frac{g}{\sqrt{2}}W_\mu^- T_L^- + g W_\mu^3 T_L^3 + g^\prime B_\mu Y,  \qquad \label{Eq:Amu_H0}
\eeqa
where $T_L^\pm = T_L^1 \pm i T_L^2$. The charged weak boson fields are defined as
\beqa
W_\mu^\pm = \frac{1}{\sqrt{2}} (W_\mu^1 \mp i W_\mu^2).
\eeqa
For SM, $A_\mu = g (W_\mu^1 \sigma^1 + W_\mu^2 \sigma^2 + W_\mu^3 \sigma^2)/2 + g^\prime B_\mu Y$ with $Y=1/2$, $g,g'$ are the $SU(2)_L$ and $U(1)$ gauge coupling, respectively.   
The neutral gauge boson fields are
\beqa
g W_\mu^3 T_L^3 + g^\prime B_\mu Y = g_Z (T^3 - s_W^2 Q) Z_\mu + e Q \gamma_\mu.
\eeqa
\beqa
\left(\begin{array}{c}
W_\mu^3 \\ 
B_\mu \\
\end{array}\right)  = \left(\begin{array}{cc}
 c_W & s_W\\ 
 -s_W & c_W \\
\end{array}\right) \left(\begin{array}{c}
Z_\mu \\ 
\gamma_\mu \\
\end{array}\right) ,
\eeqa
with $c_W = \cos\theta_W \equiv {g}/{\sqrt{g^2+g^{\prime 2}}}$, $s_W = \sin\theta_W  \equiv {g^\prime}/{\sqrt{g^2+g^{\prime 2}}}$ denote the cosine and the since of the weak mixing angle, $\tan\theta_W=g^\prime/g$ and $g,g^\prime$ are the gauge couplings of $SU(2)_L$ and $U(1)_Y$ in the SM. The commutator of the gauge field strength is
\beqa
[{\mathbf D}_\mu, {\mathbf D}_\nu ] = i g \mathbf{W}_{\mu\nu} + i g' \frac{1}{2} B_{\mu\nu} .
\eeqa

Similarly, if one want to turn on the residual ${\mathcal H}$-invariant gauge source, e.g., $SU(2)_R\subset {\mathcal H}$, then one can include an additional term at a high energy scale $f_2 \ne v$.
\beqa
A_\mu = \frac{g_R}{\sqrt{2}}W_{R\mu}^+ T_R^+ + \frac{g_R}{\sqrt{2}}W_{R\mu}^- T_R^- + g_R Z_{R\mu} T_R^3, \nn
\eeqa
where $T_R^\pm = T_R^1 \pm i T_R^2$ and generally, $g_R\ne g_L=g$. One may also denote respectively $W_R^\pm,Z_R$ as $W^{\prime\pm},Z^\prime$. While in the effective Lagrangian description, one may integrate out these heavy particles field.

For mater field $\Phi$ which transforms as $\Phi\to {\mathfrak h}(\Pi,g) \Phi$, then the covariant derivative is
\beqa
\nabla_\mu \Phi = (\partial_\mu + ie_\mu^a T^a )\Phi.
\eeqa
Therefore, ordinary derivative $\partial_\mu$ can be promoted to be covariant derivative 
\beqa
\nabla_\mu \equiv \partial_\mu + i e_\mu , 
\eeqa
which is gauge invariant under ${\mathcal H}$. 

One may also introduce the vector resonances $\rho^{\pm}$. The strength of the external gauge fields transforms under the local symmetry ${\mathcal H}$, $F_{\mu\nu}^{\pm}(\Pi) \to {\mathfrak h}(\Pi,g)F_{\mu\nu}^{\pm}(\Pi) {\mathfrak h}^{-1}(\Pi,g)$. $F_{\mu\nu}^\pm\equiv \rho_{\mu\nu}^{\pm}$, which takes even and odd ${\mathcal R}$-parity respectively, i.e., $F_{\mu\nu}^+\equiv \rho_{\mu\nu}^{+a}T^a\in \text{adj}({\mathcal H})$, $F_{\mu\nu}^{-}\equiv\rho_{\mu\nu}^{-\hat{a}}T^{\hat{a}}\in {\mathcal G}/{\mathcal H}$ and transform under ${\mathcal G}$. 
Then kinetic and mass term of the gauge field are
\beqa
{\mathcal L}_F = \frac{1}{2} \text{Tr} [\nabla^\mu F_{\mu\sigma}^\pm \nabla_\nu F^{\pm\nu \sigma }] + \frac{1}{4}m_\pm^2 \text{Tr} [ F_{\mu\nu}^\pm F^{\pm\mu\nu}], \qquad \quad
\eeqa
where
\beqa
\nabla_\mu A_\nu = \partial_\mu A_\nu + i [e_\mu, A_\nu].
\eeqa
One may also introduce the next order effective Lagrangian, by coupling the external gauge symmetry with the building blocks as
\beqa
{\mathcal L} &=& c_{g_+} \text{Tr} \left(F_{\mu\nu}^+ [d_\mu, d_\nu] \right) + c_{f_+} \text{Tr} \left( F_{\mu\nu}^{+} f^{+\mu\nu} \right) \nn\\
&+& c_{f_-} \text{Tr} \left( F_{\mu\nu}^{-} f^{-\mu\nu} \right),
\eeqa
which is relevant to the $S$ parameter.

\subsection{Non-linear chiral Lagrangian}

It is worthy of noticing that the GBs field $\Pi$ in Eq.(\ref{Omega}) enters in the effective Lagrangian only through its derivatives. The derivative of $\Omega$ field can be expanded through the $1$-form $\omega_\mu$ as
\beqa
\omega_\mu &=& \sum_{n=1}^{\infty} \frac{(-i)^{n-1}}{n!}[\Pi,[\Pi,\ldots,[\Pi,\partial_\mu \Pi]\ldots]] \nn\\
&=& \partial_\mu \Pi - \frac{i}{2} [\Pi,\partial_\mu \Pi] - \frac{1}{6}[\Pi, [\Pi,\partial_\mu\Pi]] \nn\\
&& + \frac{i}{24}[\Pi,[\Pi, [\Pi,\partial_\mu\Pi]]] + \cdots,
\eeqa
where $ [\Pi,\partial_\mu\Pi] = \Pi \overleftrightarrow{\partial}_\mu \Pi $.
In this case, the leading order expanding the filed $d_\mu$ and $e_\mu$ are
\beqa
d_\mu &=& \partial_\mu \Pi  - \frac{1}{6}[\Pi, [\Pi,\partial_\mu\Pi]]  + {\mathcal O}(\Pi^5), \nn\\
e_\mu &=&  - \frac{i}{2} [\Pi,\partial_\mu \Pi]  + \frac{i}{24}[\Pi,[\Pi, [\Pi,\partial_\mu\Pi]]] + {\mathcal O}(\Pi^6). \qquad \label{Eq:d_mu-E_mu}
\eeqa
When gauge interactions are turned on, i.e., $\partial_\mu \to D_\mu= \partial_\mu + i A_\mu^a T^a$, the Higgs will be interacting with gauge bosons too. Then, the covariant derivative of the $\Pi$ field is
\beqa
D_\mu \Pi^{\hat{a}} = (\partial_\mu + i A_\mu^a T^a )\Pi^{\hat{a}} = \partial_\mu \Pi^{\hat{a}} + i A_\mu^a (T^a)^{\hat{a}}_{~\hat{b}}\Pi^{\hat{b}}, \qquad
\eeqa
where for simplicity, one may restrict to the case in which $A_\mu$ belongs to ${\mathcal H}^0\subset {\mathcal H}$ is given by Eq.(\ref{Eq:Amu_H0}). 
\beqa
\omega_\mu & \equiv & -i \Omega^\dag D_\mu \Omega  = -i \Omega^\dag \partial_\mu \Omega  + \Omega^\dag A_\mu \Omega \nn\\
& = & \overline{d}_\mu + \overline{e}_\mu  + A_\mu  . 
\label{Eq:omegaB_mu}
\eeqa
Thus, for gauged version, one can redefine $e_\mu$ in a way as
\beqa
\bar{e}_\mu = e_\mu - A_\mu, \quad \bar{d}_\mu = d_\mu.
\eeqa
and the gauge field strength is
\beqa
\overline{e}_{\mu\nu} = - F_{\mu\nu} + e_{\mu\nu} .
\eeqa
Under the automorphism symmetry ${\mathcal R}$ in Eq.(\ref{R_automorphism}), we have
\beqa
 {\mathcal R}(d_\mu^{\hat{a}}) \to - d_\mu^{\hat{a}}, \quad {\mathcal R}(e_\mu^a) \to e_\mu^a.
\eeqa



By using the building blocks $d_\mu^{\hat{a}}$, the lowest order Lagrangian can be written as
\beqa
{\mathcal L}_{dd}^{(2)} &=& \frac{f^2}{2} d_\mu^{\hat{a}} d^{\hat{a}\mu}  
= \frac{f^2}{2}\left(\partial_\mu \Pi \partial^\mu \Pi+ \frac{1}{3}(\Pi \overleftrightarrow{\partial}^\mu \Pi) (\Pi\overleftrightarrow{\partial}_\mu \Pi) + \ldots\right)\nn\\
&=& \frac{1}{2}(\partial_\mu \pi)^2 + \frac{1}{3f^2}(\pi \overleftrightarrow{\partial}^\mu \pi)^{\hat{a}} (\pi\overleftrightarrow{\partial}_\mu \pi)^{\hat{a}} + {\mathcal O}(\pi^6) ,\qquad \quad \label{Eq:Leff-dd}
\eeqa
where $\Pi\overleftrightarrow{\partial}_\mu \Pi = [\Pi,\partial_\mu\Pi] $ and in the last equality we make the notation as $\Pi \equiv \pi/f$ and $\pi= h \Xi/2$. In this case, phenomenologically, $\pi$ has extra derivative self-interactions, and interactions with higgs $h$. In the presence of the external gauge field $A_\mu$, the gauge interactions are introduced with $\partial_\mu \to D_\mu$, the second term involves interactions of $2$-derivative invariant operators with four Higgs doublet.

The next leading order Lagrangian with derivative up to  the $4$-derivative for the gauge field $e_{\mu\nu}$ can be written as
\beqa
{\mathcal L}_{ee}^{(4)} &=&  \text{Tr}[e_{\mu\nu}e^{\mu\nu}] = e_{\mu\nu}^a e^{a\mu\nu} = (\partial_\mu e_\nu - \partial_\nu e_\mu +i[e_\mu, e_\nu])^2 \nn\\
&=& -\frac{1}{4} (\partial_\mu ([\Pi,\partial_\nu \Pi]) - \partial_\nu ([\Pi,\partial_\mu \Pi]) +i[[\Pi,\partial_\mu \Pi], [\Pi,\partial_\nu \Pi]])^2 \nn\\
&=& -\frac{1}{4} (2 [\partial_\mu \Pi,\partial_\nu \Pi] +i[[\Pi,\partial_\mu \Pi], [\Pi,\partial_\nu \Pi]])^2 \nn\\
& = & - \frac{1}{f^4}[\partial_\mu \pi, \partial_\nu \pi]^2 + {\mathcal O}(\pi^6)  \nn\\
&=& - (f^{abe}f^{cde})\frac{1}{f^4}\partial_\mu \pi^a \partial_\nu \pi^b \partial^\mu \pi^c \partial^\nu \pi^d + \cdots .
\eeqa
The interactions are suppressed by $f^{-4}$, while when the gauge interactions are turned on, the external gauge field $A_\mu$ is present, the first term involves interactions of $4$-derivative invariant operators with four Higgs doublet, which are enhanced with $g^4$.

Another next leading order Lagrangian with mixing between $e_\mu$ and $d_\mu$ fields are
\beqa
{\mathcal L}_{de}^{(4)} &=&  \text{Tr}[ [d_\mu, d_\nu] e^{\mu\nu} ] =   f^{abc} d_\mu^{\hat{a}} d_{\nu}^{\hat{b}} e^{c\mu\nu} \nn\\
&=& -\frac{i}{2} f^{abc} (\partial_\mu \Pi^a + \ldots)(\partial_\nu \Pi^b + \ldots)( 2[\partial^\mu\Pi,\partial^\nu \Pi]^c  + \ldots) \nn\\
&=& -\frac{i}{f^4}f^{abc} f^{dec}\partial_\mu \pi^a \partial_\nu \pi^a \partial^\mu \pi^d \partial^\nu \pi^e + \cdots .
\eeqa
Thus, these lowest order Lagrangian relating to the gauge field will appear at least at order higher than $p^4/f^4$. When the gauge interactions are turned on, one just make the replacement for $e_{\mu\nu} \to e_{\mu\nu} - F_{\mu\nu}$ while keep $d_\mu$ the same, then the interactions involving two Nambu-Goldstone boson and two gauge fields will be appear.

\subsection{Explicit expression of CCWZ}

\subsubsection{$d_\mu$ field}

In the NMCHM, the explicit expression of $d_\mu$ fields in the coset sector $SO(6)/SO(5)$ are
\beqa
d_\mu^{1,2} &=& \frac{g W^{1,2}_\mu }{\sqrt{2}} \cos \left(\frac{\psi }{f}\right) \sin \left(\frac{\phi }{f}\right), \nn \\
d_\mu^{3} &=& \frac{ g W^3_\mu-  g'  B_\mu }{\sqrt{2}}\cos \left(\frac{\psi }{f}\right) \sin \left(\frac{\phi }{f}\right) , \nn \\
d_\mu^{4} &=& \frac{\sqrt{2}}{f} \left[\partial_\mu\phi  \cos \left(\frac{\psi }{f}\right)-\partial_\mu\psi  \sin \left(\frac{\psi }{f}\right) \sin \left(\frac{\phi }{f}\right)\right], \nn \\
d_\mu^{5} &=& \frac{\sqrt{2} }{f}\left[\partial_\mu\phi  \sin\left(\frac{\psi }{f}\right)+\partial_\mu\psi  \cos \left(\frac{\psi }{f}\right) \sin \left(\frac{\phi }{f}\right)\right]. \qquad \quad \label{d_phi-psi}
\eeqa

\subsubsection{$d_{\mu,\nu}$ field}

From the $d_\mu$ fields above, one can obtain their anti-commutators as
\beqa
d_{\mu,\nu}^{a} \equiv [d_\mu, d_\nu]^a.
\eeqa
The explicit expression of $d_{\mu,\nu}^a$ in $(\textbf{1},\textbf{3})+(\textbf{3},\textbf{1})$ are given in 
\beqa
i (d_{\mu,\nu}^{L,1} + d_{\mu,\nu}^{R,1}) &=& \frac{1}{2} g \cos ^2\left(\frac{\psi }{f}\right) \sin ^2\left(\frac{\phi }{f}\right) \left(W^2_{\nu } Z_{\mu }-W^2_{\mu } Z_{\nu }\right), \nn\\
i (d_{\mu,\nu}^{L,1} - d_{\mu,\nu}^{R,1}) &=& \frac{1}{f}g \cos^2 \left(\frac{\psi }{f}\right) \sin \left(\frac{\phi }{f}\right) \bigg(  \left(\phi _{\mu } W^1_{\nu }-\phi _{\nu } W^1_{\mu }\right)  \nn\\
&+& \tan \left(\frac{\psi }{f}\right) \sin \left(\frac{\phi }{f}\right) \left(\psi _{\nu } W^1_{\mu }-\psi _{\mu } W^1_{\nu }\right)   \bigg), \nn\\   
i (d_{\mu,\nu}^{L,2} + d_{\mu,\nu}^{R,2}) &=& -\frac{1}{2} g \cos ^2\left(\frac{\psi }{f}\right) \sin ^2\left(\frac{\phi }{f}\right) \left(W^1_{\nu } Z_{\mu }-W^1_{\mu } Z_{\nu }\right) , \nn\\
i (d_{\mu,\nu}^{L,2} - d_{\mu,\nu}^{R,2}) &=& \frac{1}{f} g \cos^2 \left(\frac{\psi }{f}\right) \sin \left(\frac{\phi }{f}\right) \bigg( \left(\phi _{\mu } W^2_{\nu }-\phi _{\nu } W^2_{\mu }\right) \nn\\
&& + \tan \left(\frac{\psi }{f}\right) \sin \left(\frac{\phi }{f}\right) \left(\psi _{\nu } W^2_{\mu }-\psi _{\mu } W^2_{\nu }\right) \bigg) , \nn\\
i (d_{\mu,\nu}^{L,3} + d_{\mu,\nu}^{R,3}) &=&  \frac{1}{2} g^2 \cos ^2\left(\frac{\psi }{f}\right) \sin ^2\left(\frac{\phi }{f}\right) \left(W^2_{\mu } W^1_{\nu }-W^1_{\mu } W^2_{\nu }\right), \nn\\
i (d_{\mu,\nu}^{L,3} - d_{\mu,\nu}^{R,3}) &=& \frac{1}{f}\cos \left(\frac{\psi }{f}\right) \sin \left(\frac{\phi }{f}\right) \bigg(  \left(\phi _{\nu } Z_{\mu }-\phi _{\mu } Z_{\nu }\right) \nn\\
&& + \tan \left(\frac{\psi }{f}\right) \sin \left(\frac{\phi }{f}\right) \left(\psi _{\mu } Z_{\nu }-\psi _{\nu } Z_{\mu }\right) \bigg). \qquad
\eeqa
From above, it is obvious that chiral sector is relatively suppressed with a order $1/f$ comparing to the vector sector.   \qquad

For those in $(\textbf{2},\textbf{2})$ due to $T^\alpha$, we have
\beqa
i  d_{\mu,\nu}^{1} &=& \frac{1}{\sqrt{2} f} g \cos \left(\frac{\psi }{f}\right) \sin \left(\frac{\phi }{f}\right) \bigg(\sin \left(\frac{\psi }{f}\right) \left(\phi _{\mu } W_{\nu }-\phi _{\nu } W_{\mu
   }\right) \nn\\
   && +\cos \left(\frac{\psi }{f}\right) \sin \left(\frac{\phi }{f}\right) \left(\psi _{\mu } W_{\nu }-\psi _{\nu } W_{\mu }\right)\bigg), \nn\\
i  d_{\mu,\nu}^{2} &=& \frac{1}{\sqrt{2} f} g \cos \left(\frac{\psi }{f}\right) \sin \left(\frac{\phi }{f}\right) \bigg(\sin \left(\frac{\psi }{f}\right) \left(\phi _{\mu } W^2_{\nu }-\phi _{\nu }
   W^2_{\mu }\right)\nn\\
   &&+\cos \left(\frac{\psi }{f}\right) \sin \left(\frac{\phi }{f}\right) \left(\psi _{\mu } W^2_{\nu }-\psi _{\nu } W^2_{\mu}\right)\bigg) , \nn\\
i  d_{\mu,\nu}^{3} &=& \frac{1}{\sqrt{2} f}\cos \left(\frac{\psi }{f}\right) \sin \left(\frac{\phi }{f}\right) \bigg(\sin \left(\frac{\psi }{f}\right) \left(\phi _{\mu } Z_{\nu }-\phi _{\nu } Z_{\mu
   }\right) \nn\\
   && +\cos \left(\frac{\psi }{f}\right) \sin \left(\frac{\phi }{f}\right) \left(\psi _{\mu } Z_{\nu }-\psi _{\nu } Z_{\mu }\right)\bigg) \nn\\
i  d_{\mu,\nu}^{4} &=& \frac{\sqrt{2} }{f^2}\sin \left(\frac{\phi }{f}\right) \left(\psi _{\mu } \phi _{\nu }-\phi _{\mu } \psi _{\nu }\right) .
\eeqa
Comparing to the leading order, the first three $d_{\mu,\nu}^{1,2,3}$ is suppressed by $1/f$, while the fourth $d_{\mu,\nu}^{4}$ is suppressed by $1/f^2$.

\subsubsection{$e_\mu$ field}

The $e_\mu$ fields in the unbroken $SO(5)$ sector, has $10$ components, which can be decomposed into $\textbf{10}=(\textbf{3},1)+(1,\textbf{3})+(\textbf{2},\textbf{2})\in SU(2)_L\times SU(2)_R \simeq SO(4)$ components. The $(\textbf{3},1)+(1,\textbf{3})$ due to the generators $T_{L,R}^{a}$ gives
\beqa
 e_{L\mu}^{1,2}  &=&  g W_{L\mu}^{1,2} -g W_{L\mu}^{1,2} \cos ^2\left(\frac{\psi }{f}\right) \sin ^2\left(\frac{\phi }{2 f}\right), \nn\\
 e_{L\mu}^3  &=&  g W_{L\mu}^{3} - \cos ^2\left(\frac{\psi }{f}\right) \sin ^2\left(\frac{\phi }{2 f}\right) \left(g W_{L\mu}^3 - g' B_\mu \right) ; \nn\\
 e_{R\mu}^{1,2}  &=& g W_{L\mu}^{1,2} \cos ^2\left(\frac{\psi }{f}\right) \sin ^2\left(\frac{\phi }{2 f}\right), \nn\\
 e_{R\mu}^3 &=&  \cos ^2\left(\frac{\psi }{f}\right) \sin ^2\left(\frac{\phi }{2 f}\right) \left( g W_{L\mu}^3 - g' B_\mu \right). 
\eeqa
where hence and forth, we have made the abbreviation of the subscript $L$ for $W$ bosons field, i.e., $W_\mu = W_{L\mu}$.
The $(\textbf{2},\textbf{2})$ due to $T^{\alpha}$ gives
\beqa
 e_{\mu}^{1,2} &=& -\frac{g W_{\mu}^{1,2} }{\sqrt{2}}\sin \left(\frac{2 \psi }{f}\right) \sin ^2\left(\frac{\phi }{2 f}\right), \nn\\
 e_{\mu}^{3} &=& - \frac{1}{\sqrt{2}}\left(  g W_\mu^3 - g' B_\mu \right) \sin \left(\frac{2 \psi }{f}\right) \sin ^2\left(\frac{\phi }{2 f}\right) , \nn \\
 e_{\mu}^4 &=& \frac{\sqrt{2} }{f}\partial_\mu\psi  \left[1 - \cos \left(\frac{\phi }{f}\right) \right]. \qquad
\eeqa

\subsubsection{$f_{\mu\nu}^{\pm}$ field}

In the Omega representation, the building blocks of gauge field are not $F_{\mu\nu}$, but $f_{\mu\nu}^{\pm}$ as we will give the explicit expressions in the following, 

According to the definitions in Eq.(\ref{fpm}), we have the gauge field $f_{\mu\nu}^+=(f_{\mu\nu}^{L,a},f_{\mu\nu}^{R,a},f_{\mu\nu}^{+,\alpha})$ in the unbroken sector as
\beqa
&& f^{L,1,2}_{\mu \nu } = g W^{1,2}_{\mu \nu } \left[1-\cos ^2\left(\frac{\psi }{f}\right) \sin ^2\left(\frac{\phi }{2 f}\right)\right], \nn\\
&& f^{L,3}_{\mu \nu } = g W^3_{\mu \nu } - \cos ^2\left(\frac{\psi }{f}\right) \sin ^2\left(\frac{\phi }{2 f}\right) \left(g W^3_{\mu \nu } - g' B_{\mu \nu } \right) , \nn\\
&& f^{R,1,2}_{\mu \nu } = g W^{1,2}_{\mu \nu } \cos ^2\left(\frac{\psi }{f}\right)\sin ^2\left(\frac{\phi }{2 f}\right), \nn\\
&& f^{R,3}_{\mu \nu } = g' B_{\mu \nu } + \cos ^2\left(\frac{\psi }{f}\right) \sin ^2\left(\frac{\phi }{2 f}\right) \left(g W^3_{\mu \nu }- g' B_{\mu \nu } \right) ,  \qquad \quad
\eeqa
which leads to an equivalent linear independent combination as
\beqa
&& f^{L,1,2}_{\mu \nu } + f^{R,1,2}_{\mu \nu } = g W^{1,2}_{\mu \nu } , \nn\\
&& f^{L,1,2}_{\mu \nu } - f^{R,1,2}_{\mu \nu } = g W^{1,2}_{\mu \nu }\left[ 1 - 2 \cos ^2\left(\frac{\psi }{f}\right)\sin ^2\left(\frac{\phi }{2 f}\right)\right] ,  \nn\\
&& f^{L,3}_{\mu \nu } +  f^{R,3}_{\mu \nu } = g W^3_{\mu \nu } + g' B_{\mu \nu } \equiv A_{\mu\nu} , \nn\\
&& f^{L,3}_{\mu \nu } - f^{R,3}_{\mu \nu } = Z_{\mu\nu} \left[ 1  - 2\cos ^2\left(\frac{\psi }{f}\right) \sin ^2\left(\frac{\phi }{2 f}\right) \right] ,  \qquad \quad
\eeqa
where $Z_{\mu\nu}\equiv g W^3_{\mu \nu } - g' B_{\mu \nu } $. It's obvious that the vector $f^{L,3}_{\mu \nu } +  f^{R,3}_{\mu \nu }$ is nothing but the gauge stress tensor for photon fields in the SM.
The components in the $(\textbf{2},\textbf{2})$ are
\beqa
&& f^{+,1}_{\mu \nu } = -\frac{g }{\sqrt{2}}W^1_{\mu \nu } \sin \left(\frac{2 \psi }{f}\right) \sin ^2\left(\frac{\phi }{2 f}\right), \nn\\
&& f^{+,2 }_{\mu \nu } = -\frac{g}{\sqrt{2}}W^2_{\mu \nu } \sin \left(\frac{2 \psi }{f}\right) \sin ^2\left(\frac{\phi }{2 f}\right), \nn\\
&& f^{+,3 }_{\mu \nu } = - \frac{1}{\sqrt{2}}  \left(g W^3_{\mu \nu } - g'  B_{\mu \nu } \right)\sin \left(\frac{2 \psi }{f}\right) \sin ^2\left(\frac{\phi }{2 f}\right) , \nn\\
&& f^{+,4 }_{\mu \nu } = 0, 
\eeqa

For the gauge field $f_{\mu\nu}^-$ in the coset $SO(6)/SO(5)$ sector,
\beqa
&& f_{\mu\nu}^{-, 1} = \frac{g }{\sqrt{2}} W^1_{\mu \nu } \cos \left(\frac{\psi }{f}\right) \sin \left(\frac{\phi }{f}\right) , \nn\\
&& f_{\mu\nu}^{-, 2} = \frac{g }{\sqrt{2}} W^2_{\mu \nu } \cos \left(\frac{\psi }{f}\right) \sin \left(\frac{\phi }{f}\right) , \nn\\
&& f_{\mu\nu}^{-, 3} = \frac{1}{\sqrt{2}} (g W^3_{\mu \nu }-B_{\mu \nu } g') \cos \left(\frac{\psi }{f}\right) \sin \left(\frac{\phi}{f}\right)  , \nn\\
&& f_{\mu\nu}^{-, 4} = 0, \quad f_{\mu\nu}^{- 5} = 0,
\eeqa
where $W^a_{\mu\nu}$ and $B_{\mu\nu}$ are gauge field stress tensor for $SU(2)_L$ and $U(1)_L$ symmetry
\beqa
&& W^a_{\mu\nu} = \partial_\mu W_\nu^a - \partial_\nu W_\mu^b - g \epsilon^{abc} W_\mu^b W_\nu^c, \nn\\
&& B_{\mu\nu} =  \partial_\mu B_\nu - \partial_\nu B_\mu.
\eeqa
since we have defined $F_{\mu\nu}=\partial_\mu A_\nu - \partial_\nu A_\mu + i [A_\mu, A_\nu]$.

\subsubsection{$e_{\mu\nu}$ field}

In NMCHM,  the gauge field stress tensor in Eq.(\ref{emn}) becomes
\beqa
e_{\mu \nu}  
 =  e_{\mu \nu}^{L}+e_{\mu \nu}^{R} + e_{\mu\nu}^{(\alpha)}
 \eeqa
where $e_{\mu \nu}^{L/R} = e_{\mu \nu}^{L/R, a} T_{a}^{L/R}$ and $e_{\mu\nu}^{(\alpha)}= e_{\mu \nu}^{{(\alpha)}\alpha} T_{\alpha}$.

The $e_{\mu\nu}$ can be computed through separating into two sector as
\beqa
e_{\mu\nu} 
= 2 \partial_{[\mu} e_{\nu]}, - i e_{\mu,\nu},
\eeqa
where the Abelian sector of the fields are
\beqa
\partial_\mu e_\nu - \partial_\nu e_\mu \equiv 2 \partial_{[\mu} e_{\nu]}, \label{emn-Abelian}
\eeqa
while the non-Abelian sector are
\beqa
e_{\mu,\nu} \equiv [e_\mu, e_\nu].  \label{emn-nonAbelian}
\eeqa


The explicitly expression in the NMCHM, turns out to be
\beqa
i (e_{\mu,\nu}^{L,1}+ e_{\mu,\nu}^{R,1} ) &=& 2 g \cos ^2\left(\frac{\psi }{f}\right) \sin ^4\left(\frac{\phi }{2 f}\right) \left(W^2_{\nu } Z_{\mu }-W^2_{\mu } Z_{\nu }\right), \nn\\
i (e_{\mu,\nu}^{L,1}- e_{\mu,\nu}^{R,1} ) &=& \frac{2 g }{f}\sin \left(\frac{2 \psi }{f}\right) \sin ^4\left(\frac{\phi }{2 f}\right) \left(\psi _{\nu } W^1_{\mu }-\psi _{\mu } W^1_{\nu }\right), \nn\\
i (e_{\mu,\nu}^{L,2}+ e_{\mu,\nu}^{R,2} ) &=& 2 g \cos ^2\left(\frac{\psi }{f}\right) \sin ^4\left(\frac{\phi }{2 f}\right) \left(W^1_{\mu } Z_{\nu }-W^1_{\nu } Z_{\mu }\right), \nn\\
i (e_{\mu,\nu}^{L,2}- e_{\mu,\nu}^{R,2} ) &=& \frac{2 g }{f}\sin \left(\frac{2 \psi }{f}\right) \sin ^4\left(\frac{\phi }{2 f}\right) \left(\psi _{\nu } W^2_{\mu }-\psi _{\mu } W^2_{\nu }\right), \nn\\
i (e_{\mu,\nu}^{L,3}+ e_{\mu,\nu}^{R,3} ) &=& 2 g^2 \cos ^2\left(\frac{\psi }{f}\right) \sin ^4\left(\frac{\phi }{2 f}\right) \left(W^2_{\mu } W^1_{\nu }-W^1_{\mu } W^2_{\nu }\right), \nn\\
i (e_{\mu,\nu}^{L,3}- e_{\mu,\nu}^{R,3} ) &=& \frac{2 }{f} \sin \left(\frac{2 \psi }{f}\right) \sin ^4\left(\frac{\phi }{2 f}\right) \left(\psi _{\nu } Z_{\mu }-\psi _{\mu } Z_{\nu }\right). \qquad
\eeqa
The situation is similar to what has happened in $d_{\mu,\nu}$, the vector contribution is larger than the chiral contribution.
For $e_{\mu,\nu}$ in $(\textbf{2},\textbf{2})$, they are
\beqa
ie_{\mu,\nu}^1 &=& \frac{2 \sqrt{2}}{f} g \cos ^2\left(\frac{\psi }{f}\right) \sin ^4\left(\frac{\phi }{2 f}\right) \left(\psi _{\mu } W^1_{\nu }-\psi _{\nu } W^1_{\mu }\right), \nn\\
ie_{\mu,\nu}^2 &=& \frac{2 \sqrt{2} }{f}g \cos ^2\left(\frac{\psi }{f}\right) \sin ^4\left(\frac{\phi }{2 f}\right) \left(\psi _{\mu } W^2_{\nu }-\psi _{\nu } W^2_{\mu }\right), \nn\\
ie_{\mu,\nu}^3 &=& \frac{2 \sqrt{2} }{f}\cos ^2\left(\frac{\psi }{f}\right) \sin ^4\left(\frac{\phi }{2 f}\right) \left(\psi _{\mu } Z_{\nu }-\psi _{\nu } Z_{\mu }\right), \nn\\
e_{\mu,\nu}^4 &=& 0.
\eeqa

We have also calculate all of the commutative part of $e_{\mu\nu}$ in Eq.(\ref{emn-Abelian}),
\beqa
\partial_{[\mu} e_{\nu]}^{V ,1,2,3}  &=& 0 , \nn\\
\partial_{[\mu} e_{\nu]}^{A ,1,2} &=&  g \cos ^2\left(\frac{\psi }{f}\right) \sin ^2\left(\frac{\phi }{2 f}\right) \left( \partial_\nu W^{1,2}_\mu - \partial_\mu W^{1,2}_\nu\right) \nn\\
&+&\frac{1}{2f}g\bigg( \cos ^2\left(\frac{\psi }{f}\right) \sin\left(\frac{\phi }{f}\right) \left(\phi _{\nu } W^{1,2}_{\mu }-\phi _{\mu } W^{1,2}_{\nu }\right) \nn\\
&+& 2 \sin\left(\frac{2 \psi }{f}\right) \sin ^2\left(\frac{\phi }{2 f}\right) \left(\psi _{\mu } W^{1,2}_{\nu }-\psi _{\nu } W^{1,2}_{\mu }\right) \bigg), \nn \\
\partial_{[\mu} e_{\nu]}^{A ,3}  &=&  \cos ^2\left(\frac{\psi }{f}\right) \sin ^2\left(\frac{\phi }{2 f}\right) \left( \partial_\nu Z_\mu - \partial_\mu Z_\nu \right)\nn\\
&+&\frac{1}{2f}\bigg( \cos ^2\left(\frac{\psi }{f}\right) \sin \left(\frac{\phi
   }{f}\right) \left(\phi _{\nu } Z_{\mu }-\phi _{\mu } Z_{\nu }\right) \nn\\
&+& 2 \sin \left(\frac{2 \psi}{f}\right) \sin ^2\left(\frac{\phi }{2 f}\right) \left(\psi _{\mu } Z_{\nu }-\psi _{\nu } Z_{\mu }\right) \bigg), \qquad \quad
\eeqa
where we have made the recombination
$\partial_{[\mu} e_{\nu]}^{V/A ,a}  \equiv \partial_{[\mu} e_{\nu]}^{L ,a} \pm \partial_{[\mu} e_{\nu]}^{R a}$ 
with $a=1,2,3$

For the $(\textbf{2},\textbf{2})$, one has
\beqa
 \partial_{[\mu} e_{\nu]}^{1,2} &=& \frac{g }{2\sqrt{2}}\sin \left(\frac{2 \psi }{f}\right) \sin ^2\left(\frac{\phi }{2 f}\right)
   \left( \partial_\nu W^{1,2}_\mu - \partial_\mu W^{1,2}_\nu \right)  \nn\\
   && + \frac{1}{2\sqrt{2}f}g\bigg( \frac{1}{2}  \sin \left(\frac{2 \psi }{f}\right) \sin \left(\frac{\phi
   }{f}\right) \left(\phi _{\nu } W^{1,2}_{\mu }-\phi _{\mu } W^{1,2}_{\nu }\right)\nn\\
  && +2  \cos \left(\frac{2 \psi }{f}\right) \sin ^2\left(\frac{\phi }{2 f}\right) \left(\psi _{\nu }
   W^{1,2}_{\mu }-\psi _{\mu } W^{1,2}_{\nu }\right) \bigg), \nn\\
  \partial_{[\mu} e_{\nu]}^{3} &=& \frac{1}{2\sqrt{2}}\sin \left(\frac{2 \psi }{f}\right) \sin ^2\left(\frac{\phi }{2 f}\right)
   \left(\partial_\nu Z_\mu - \partial_\mu Z_\nu \right) \nn\\
  && + \frac{1}{2\sqrt{2}f}\bigg( \frac{1}{2} \sin \left(\frac{2 \psi }{f}\right) \sin \left(\frac{\phi }{f}\right) \left(\phi
   _{\nu } Z_{\mu }-\phi _{\mu } Z_{\nu }\right) \nn\\
  &&+2 \cos \left(\frac{2 \psi }{f}\right) \sin ^2\left(\frac{\phi }{2 f}\right) \left(\psi _{\nu } Z_{\mu }-\psi _{\mu }
   Z_{\nu }\right) \bigg), \nn\\
  \partial_{[\mu} e_{\nu]}^{4} &=&  \frac{1}{\sqrt{2} f^2}\sin \left(\frac{\phi }{f}\right) \left(\phi _{\mu } \psi _{\nu }-\psi _{\mu } \phi _{\nu }\right).
\eeqa

According to Eq.(\ref{fmn_d_emn}), the field strength $e_{\mu\nu}$ are also related to the unbroken sector $f_{\mu\nu}^+$ and $d_\mu$ field through an identity
\beqa
e_{\mu\nu} = f_{\mu\nu}^+ - i [d_\mu, d_\nu] = f_{\mu\nu}^+ - i d_{\mu,\nu}.
\eeqa
Thus, they are not linear independent of $f_{\mu\nu}^{+}$. Thus, we can either use $e_{\mu\nu}$ or $f_{\mu\nu}^+$ as the building block.
The most easies but non-trivial check to determine the sign is through $\alpha=4$ components, which leads to
\beqa
e_{\mu\nu}^4 =  \partial_\mu e_\nu^4 - \partial_\nu e_\mu^4 = 2 \partial_{[\mu} e_{\nu]}^{4} = - i d_{\mu,\nu}^4,
\eeqa
since $f_{\mu\nu}^{+,4}=0$. 

\section{EW chiral Lagrangian and Operators}
\label{app:LCEFT}

\subsection{Building blocks of low energy EW chiral Lagrangian}


The (pseudo-)scalar $\mathbf{T} $ and vector chiral fields $\mathbf{V}_{\mu} $ 
are defined as~\cite{Appelquist:1980vg,Alonso:2014wta,Buchalla:2014eca}
\beqa
\mathbf{T}  \equiv \mathbf{U}  \sigma^{3} \mathbf{U}^{\dagger} , \quad \mathbf{V}_{\mu} \equiv \left(\mathbf{D}_{\mu} \mathbf{U} \right) \mathbf{U}^{\dagger} , 
\label{T-Vm}
\eeqa
where $\mathbf{U}$ is the three Goldstone bosons (GBs) in the coset $SU(2)_L\times SU(2)_R/SU(2)_C$, are parameterized as the longitudinal components of the SM gauge bosons by a non-linear $\sigma$-model as a dimensionless unitary matrix $\mathbf{U} $ at low energies as
\beqa
\mathbf{U}  = \exp{\left(i \frac{1}{v} \sigma^a \varphi^a   \right)}, \label{U_v}
\eeqa 
where $v$ is the scale associated to the SM GBs, and $\sigma^a$ are the usual Pauli matrices. The dimensionless unitary SM GB matrix transforms as a bi-doublet under the global $SU(2)_L\times SU(2)_R$ symmetry as
\beqa
\mathbf{U}  \rightarrow \mathfrak{g}_L \mathbf{U}  \mathfrak{g}_R^\dag.
\eeqa

After EWSB, the global $SU(2)_L\times SU(2)_R$ symmetry is spontaneously broken down to the diagonal $SU(2)_C$ in terms of custodial symmetry, and explicitly broken by gauging the $U(1)_Y$ hypercharge and by the fermion mass splittings. 

The covariant derivative are\footnote{The bold of the characters implies that we have adopt abbreviations for $SU(2)_L$ generators.}
\beqa
\mathbf{D}_{\mu} \mathbf{U}  \equiv & \partial_{\mu} \mathbf{U} + i g \mathbf{W}_{\mu}   \mathbf{U} - i g^{\prime} B_{\mu}  \mathbf{U}  { \sigma_{3}}/{2}, 
\eeqa
where $\mathbf{W}_{\mu}  \equiv W_{\mu}^{a}  \sigma_{a} / 2$ with $W_\mu^a $ and $B_\mu$ denote the $SU(2)_L$ and $U(1)_Y$ gauge bosons, respectively, and $g,g'$ are the corresponding gauge coupling. Both $\mathbf{T} $ and $\mathbf{V}_{\mu} $ transform in the adjoint symmetry representation of $SU(2)_L$ as 
\beqa
\mathbf{T}  \rightarrow \mathfrak{g}_L \mathbf{T}  \mathfrak{g}_L^\dag, \quad \mathbf{V}_{\mu}   \rightarrow \mathfrak{g}_L  \mathbf{V}_{\mu}  \mathfrak{g}_L^\dag ,
\eeqa
while the chiral scalar field $\mathbf{T}$ breaks explicitly the $SU(2)_R$ symmetry nor is invariant under $SU(2)_C$. Thus, it can be considered as a custodial symmetry breaking term. 
Thus, the covariant derivative $\mathbf{D}_{\mu}$ denotes that in the adjoint representation of SU$(2)_L$, i.e., when acting upon ${\mathbf V}_\mu$, is given by
\beqa
\mathbf{D}_{\mu} \mathbf{V}_{\nu} \equiv \partial_{\mu} \mathbf{V}_{\nu}+i g\left[\mathbf{W}_{\mu}, \mathbf{V}_{\nu}\right], \label{DV}
\eeqa
and satisfies a useful identity $({\mathbf D}_\mu \mathbf{U})^\dag=-\mathbf{U}^{-1}({\mathbf D}_\mu \mathbf{U})\mathbf{U}^\dag$.
In this case, one obtains frequently useful equalities as~\cite{Appelquist:1980vg}
\beqa
&& {\mathbf{V}_{\mu \nu} \equiv \mathbf{D}_{\mu} \mathbf{V}_{\nu}-\mathbf{D}_{\nu} \mathbf{V}_{\mu}=i g \mathbf{W}_{\mu \nu}-i \frac{g^{\prime}}{2} B_{\mu \nu} \mathbf{T}+\left[\mathbf{V}_{\mu}, \mathbf{V}_{\nu}\right]} , \nn\\
&& \mathbf{D}_{\mu} \mathbf{T}=\left[\mathbf{V}_{\mu}, \mathbf{T}\right] , \quad {\left[\mathbf{D}_{\mu}, \mathbf{D}_{\nu}\right] \mathcal{O}=i g\left[\mathbf{W}_{\mu \nu}, \mathcal{O}\right]},  
\eeqa
where $\mathcal{O}$ is a generic operator covariant under $SU(2)_L$ and invariant under $U(1)_Y$. 

It's worthy of noticing that it is concise to use the building block $\mathbf{V}_\mu$, comparing to use the explicit expression of $\mathbf{U}$ as
\beqa
\text{Tr} \left(\mathbf{V}_\mu {\mathbf V}^\mu\right) =  \text{Tr}  \left[\mathbf{U}^\dag \left(\mathbf{D}_{\mu} \mathbf{U} \right) \mathbf{U}^{\dagger}\left(\mathbf{D}_{\mu} \mathbf{U} \right) \right] , 
\label{Vm-Lm}
\eeqa
where in the last equality, we have used the Hermitian condition for Lagrangian up to kinetic terms.

\subsection{Low energy EW chiral Lagrangian}

\subsubsection{Higgs singlet}

The physical Higgs $h$ is an iso-singlet of the SM gauge symmetry with vacuum expectation value (vev) at EW scale $v\approx 246\gev$. In the low energy effective Lagrangian, there are four pure Higgs operators. One is that with two derivatives, and the other three are those with four derivatives as~\cite{Brivio:2013pma}
\beqa
\begin{array}{lll}
\mathcal{L}_{H} &=& \frac{1}{2}\left(\partial_{\mu} h\right)\left(\partial^{\mu} h\right) \\
\mathcal{L}_{\square H} &=& \frac{1}{v^{2}}\left(\square h\right)^{2} \\
\mathcal{L}_{\Delta H} &=& \frac{1}{v^{3}}\left(\partial_{\mu} h\right)\left(\partial^{\mu} h\right) \square h \\
\mathcal{L}_{D H} &=& \frac{1}{v^{4}}\left[\left(\partial_{\mu} h\right)\left(\partial^{\mu} h\right)\right]^{2}, \\
\end{array} \label{L_H}
\eeqa
where $\square=\partial_{\mu} \partial^{\mu} $.

\subsubsection{ CP-even case}

\textbf{Operators with two derivatives} are~\cite{Longhitano:1980iz}
\beqa
\begin{array}{lll}
 \mathcal{L}_{C} &=& - \frac{(v+h)^2}{4}\text{Tr}(\bf{V}^\mu \bf{V}_\mu) , \\
 \mathcal{L}_{T} &=& \frac{(v+h)^{2}}{4}\operatorname{Tr}\left( \bf{T} \bf{V}_{\mu}\right) \operatorname{Tr}\left( \bf{T} \bf{V}^{\mu}\right) ,  \\
\end{array} \label{L_C-T}
\eeqa
where $C$ and $T$ indicates the custodial preserving and custodial breaking, respectively.


\textbf{Operators withfour derivatives} one has\footnote{These 26 $p^4$ operators in the EW chiral Lagrangian is redundant if the fermion sector is included~\cite{Buchalla:2013rka}.

Thus, we only need to focus on the NMCHM setup with only composite vector boson states included.
}~\cite{Appelquist:1980vg,Longhitano:1980tm,Appelquist:1993ka,Alonso:2012px,Alonso:2012px,Brivio:2013pma,Alonso:2014wta,Brivio:2016fzo}
\beqa
\begin{array}{lll}
\mathcal{L}_{B} &=&-\frac{g'^2}{4} B_{\mu \nu} B^{\mu \nu}    \\ 
\mathcal{L}_{W} &=& -\frac{g^2}{2} \operatorname{Tr}\left(\mathbf{W}_{\mu \nu} \mathbf{W}^{\mu \nu}\right)    \\ 
\mathcal{L}_{1} &=&g g^{\prime} B_{\mu \nu} \operatorname{Tr}\left(\mathbf{T} \mathbf{W}^{\mu \nu}\right)     \\  
\mathcal{L}_{2} &=&i g^{\prime} B_{\mu \nu} \operatorname{Tr}\left(\mathbf{T}\left[\mathbf{V}^{\mu}, \mathbf{V}^{\nu}\right]\right)    \\ 
\mathcal{L}_{3} &=&i g \operatorname{Tr}\left(\mathbf{W}_{\mu \nu}\left[\mathbf{V}^{\mu}, \mathbf{V}^{\nu}\right]\right)     \\
\mathcal{L}_{4} &=&i g^{\prime} B_{\mu \nu} \operatorname{Tr}\left(\mathbf{T} \mathbf{V}^{\mu}\right) \partial^{\nu}(h / v)     \\ 
\mathcal{L}_{5} &=&i g \operatorname{Tr}\left(\mathbf{W}_{\mu \nu} \mathbf{V}^{\mu}\right) \partial^{\nu}(h / v)    \\ 
\mathcal{L}_{6} &=&\left[\operatorname{Tr}\left(\mathbf{V}_{\mu} \mathbf{V}^{\mu}\right)\right]^{2}    \\ 
\mathcal{L}_{7} &=&\operatorname{Tr}\left(\mathbf{V}_{\mu} \mathbf{V}^{\mu}\right) \partial_{\nu} \partial^{\nu}(h / v)   \\
\mathcal{L}_{8} &=&\operatorname{Tr}\left(\mathbf{V}_{\mu} \mathbf{V}_{\nu}\right) \partial^{\mu}(h / v )\partial^{\nu}(h / v)  \\ 
\mathcal{L}_{9} &=&\operatorname{Tr}[\left(\mathbf{D}_{\mu} \mathbf{V}^{\mu}\right)^{2}]     \\
\mathcal{L}_{10} &=&\operatorname{Tr}\left(\mathbf{V}_{\nu} \mathbf{D}_{\mu} \mathbf{V}^{\mu}\right) \partial^{\nu}(h / v)    \\ 
\mathcal{L}_{11} &=&[\operatorname{Tr}\left(\mathbf{V}_{\mu} \mathbf{V}_{\nu}\right)]^{2}  \\
\mathcal{L}_{12}&=&g^{2}\left[\operatorname{Tr}\left(\mathbf{T} \mathbf{W}_{\mu \nu}\right)\right]^{2}   \\ 
\mathcal{L}_{13}&=&i g \operatorname{Tr}\left(\mathbf{T} \mathbf{W}_{\mu \nu}\right) \operatorname{Tr}\left(\mathbf{T}\left[\mathbf{V}^{\mu}, \mathbf{V}^{\nu}\right]\right)   \\
\mathcal{L}_{14}&=&g \epsilon_{\mu \nu \rho \lambda} \operatorname{Tr}\left(\mathbf{T} \mathbf{V}^{\mu}\right) \operatorname{Tr}\left(\mathbf{V}^{\nu} \mathbf{W}^{\rho \lambda}\right)  \\
\mathcal{L}_{15}&=& \operatorname{Tr}\left(\mathbf{T} \mathbf{D}_{\mu} \mathbf{V}^{\mu}\right) \operatorname{Tr}\left(\mathbf{T} \mathbf{D}_{\nu} \mathbf{V}^{\nu}\right)  \\
\mathcal{L}_{16}&=&\operatorname{Tr}\left(\left[\mathbf{T}, \mathbf{V}_{\nu}\right] \mathbf{D}_{\mu} \mathbf{V}^{\mu}\right) \operatorname{Tr}\left(\mathbf{T} \mathbf{V}^{\nu}\right)   \\
\mathcal{L}_{17} &=&i g \operatorname{Tr}\left(\mathbf{T} \mathbf{W}_{\mu \nu}\right) \operatorname{Tr}\left(\mathbf{T} \mathbf{V}^{\mu}\right) \partial^{\nu}(h / v)   \\
\mathcal{L}_{18} &=&\operatorname{Tr}\left(\mathbf{T}\left[\mathbf{V}_{\mu}, \mathbf{V}_{\nu}\right]\right) \operatorname{Tr}\left(\mathbf{T} \mathbf{V}^{\mu}\right) \partial^{\nu}(h / v)  \\
\mathcal{L}_{19} &=&\operatorname{Tr}\left(\mathbf{T} \mathbf{D}_{\mu} \mathbf{V}^{\mu}\right) \operatorname{Tr}\left(\mathbf{T} \mathbf{V}_{\nu}\right) \partial^{\nu}(h / v)  \\
 \mathcal{L}_{20} &=&\operatorname{Tr}\left(\mathbf{V}_{\mu} \mathbf{V}^{\mu}\right) \partial_{\nu}(h / v) \partial^{\nu}(h / v)  \\
 \mathcal{L}_{21} &=&\left[\operatorname{Tr}\left(\mathbf{T} \mathbf{V}_{\mu}\right)\right]^{2} \partial_{\nu}(h / v) \partial^{\nu}(h / v)   \\
 \mathcal{L}_{22} &=&\operatorname{Tr}\left(\mathbf{T} \mathbf{V}_{\mu}\right) \operatorname{Tr}\left(\mathbf{T} \mathbf{V}_{\nu}\right) \partial^{\mu}(h / v)  \partial^{\nu}(h / v)   \\
 \mathcal{L}_{23} &=&\operatorname{Tr}\left(\mathbf{V}_{\mu} \mathbf{V}^{\mu}\right)\left[\operatorname{Tr}\left(\mathbf{T} \mathbf{V}^{\nu}\right)\right]^{2}  \\
 \mathcal{L}_{24} &=&\operatorname{Tr}\left(\mathbf{V}_{\mu} \mathbf{V}_{\nu}\right) \operatorname{Tr}\left(\mathbf{T} \mathbf{V}^{\mu}\right) \operatorname{Tr}\left(\mathbf{T} \mathbf{V}^{\nu}\right)   \\
 \mathcal{L}_{25} &=&\left[\operatorname{Tr}\left(\mathbf{T} \mathbf{V}_{\mu}\right)\right] \left[\operatorname{Tr}\left(\mathbf{T} \mathbf{V}^{\mu}\right)\right] \partial_{\nu} \partial^{\nu}(h / v)   \\
  \mathcal{L}_{26} &=&\left[\operatorname{Tr}\left(\mathbf{T} \mathbf{V}_{\mu}\right) \operatorname{Tr}\left(\mathbf{T} \mathbf{V}_{\nu}\right)\right]^{2},  \\
\end{array} \label{L_CP-even}
\eeqa
where the first $13$ Lagrangian $\mathcal{L}_{B,W,1,\ldots,13}$ corresponds to the custodial preserving ones, while the residue corresponds to the custodial breaking ones, which describes tree-level effects of custodial breaking sources beyond the SM ones. Since the gauging of the SM symmetry breaks explicitly the custodial symmetries. Consequently, these custodial symmetry breaking operators are generated due to the quantum corrections induced by the SM interactions.
The covariant derivative of $\mathbf{V}_\mu$ are defined as~\cite{Appelquist:1980vg}
\beqa
{\mathbf D}_\mu {\mathbf V}_\nu \equiv \partial_\mu {\mathbf V}_\nu + i g [W_\mu, \mathbf{V}_\nu].
\eeqa
In the absence of a light CP-even Higgs-like scalar singlet $h$ in the low energy spectrum, the $12$ operators containing derivatives of Higgs are absent. Thus, there are a complete $18$ (independent) CP-even operators those preserving $SU(2)_L\times U(1)_Y$ symmetry. 

Among the operators, two $SU(2)_C$ custodial symmetry preserving and three custodial violating operators ${\mathcal L}_{6,11;23,24,26}$ exhibit quartic vector-boson interactions, which leads to new anomalous quartic couplings~\cite{Belyaev:1998ih} such as $Z_\mu Z_\nu Z^\mu Z^\nu$, $W_\mu^+ W_\nu^- Z^\mu Z^\nu$ and $W_\mu^+ W_\nu^- W^{+\mu} W^{-\nu} $.


The CP-even low energy effective Lagrangian in Eq.(\ref{L_CP-even}) can be expressed more explicitly in the unitary gauge as
\beqa
\begin{array}{lll} 
\mathcal{L}_{C}  &=& \frac{1}{8}v^2 \left[ g^2[ (W_\mu^1)^2 + (W_\mu^2)^2  ] + (g W_\mu^3 - g' B_\mu)^2 \right]  , \\
\mathcal{L}_{T}  &=& -\frac{1}{4}v^2(g W^{3}_{\mu} - g' B_\mu)^2 , \\
\mathcal{L}_{B}  &=& - \frac{g'^2}{4}B_{\mu\nu}B^{\mu\nu}  ,  \\
\mathcal{L}_{W} &=&  - \frac{g^2}{4}W_{\mu\nu}^aW^{a \mu\nu}   ,  \\
\mathcal{L}_{1}  &=& gg' B_{\mu\nu} W^{3\mu\nu}   ,  \\
{\mathcal L}_{2} &=& g^2 g' B^{\mu\nu}(W^{1}_{\mu} W^{2}_{\nu}- W^{2}_{\mu} W^{1}_{\nu})  , \\
{\mathcal L}_{4} &=& g' B^{\mu\nu}(g'B_\mu - g W^{3}_{\mu})\partial_\nu h /v    , \\ 
{\mathcal L}_{3} &=& \frac{g^2}{2}[g W^{3 \mu\nu}(W_{\mu}^1 W_{\nu}^2 - W_{\mu}^2 W_{\nu}^1   )  \\
&& + (g W_\nu^3 - g' B_\nu)(W^{1\mu\nu}W_\mu^2 - W_\mu^1 W^{2\mu\nu} )  \\
&& - (g W_\mu^3 - g' B_\mu)(W^{1\mu\nu} W_\nu^2 - W_\nu^1 W^{2\mu\nu} ) ]  , \\
{\mathcal L}_{5} &=& - \frac{g}{2} [ g(W^{1}_{\mu}W^{1\mu\nu}+W^{2}_{\mu}W^{2\mu\nu})  \\
&& + (gW_\mu^3 - g' B_\mu ) W^{3\mu\nu} ] \partial_\nu h /v  , \\
{\mathcal L}_{6} &=& \frac{1}{4}[ g^2(W^1_\mu)^2+g^2(W^2_\mu)^2 + (g W^3_\mu - g' B_\mu)^2]^2 , \\
{\mathcal L}_{7} &=& - \frac{1}{2v} [g^2((W_\mu^1)^2+(W_\mu^2)^2)+(gW_\mu^3-g' B_\mu)^2] \Box h , \\
{\mathcal L}_{8} &=& -\frac{1}{2v^2}[g^2( W_\mu^1 W_\nu^1 + W_\mu^2 W_\nu^2 )  \\
&& + (g W_\mu^3 - g' B_\mu) (g W_\nu^3 - g' B_\nu) ]\partial^\mu h \partial^\nu h  , \\
{\mathcal L}_{9} &=& -\frac{1}{2}[ g^2 ( \partial_\mu W^{1\mu} + g'  B_\mu W^{2\mu} )^2  \\
&& + g^2 ( \partial_\mu W^{2\mu}  - g' B_\mu W^{1\mu}  )^2  \\
&& + (g \partial_\mu W^{3\mu} - g' \partial_\mu B^\mu)^2 ] , \\
{\mathcal L}_{10} &=& -\frac{1}{2v}[ g^2(W_{\mu}^1 \partial_\nu W^{1\nu}+W_{\mu}^2 \partial_\nu W^{2\nu})  \\
&& + (g W_\mu^3 - g' B_\mu)\partial^\nu(g W_\nu^3 - g' B_\nu^3)] \partial^\mu h ,  \\
{\mathcal L}_{11} &=& \frac{1}{4}[g^2( W_\mu^1 W_\nu^1 + W_\mu^2 W_\nu^2 )  \\
&& + (g W_\mu^3 - g' B_\mu) (g W_\nu^3 - g' B_\nu) ]^2  , \\
{\mathcal L}_{12} & =& g^2 W^{3\mu\nu}W^3_{\mu\nu}   , \\
{\mathcal L}_{13} & =& g^3W^{3\mu\nu}(W^1_\mu W^2_\nu-W^1_\nu W^2_\mu)  , \\
{\mathcal L}_{14} & =& g(gW^{3\mu}-g' B^\mu)[g(W^{1\nu}\widetilde{W}^1_{\mu\nu}+W^{2\nu}\widetilde{W}^2_{\mu\nu}) \\
&& + (g W^{3\nu}-g' B^\nu)\widetilde{W}^3_{\mu\nu}] , \\
{\mathcal L}_{15} & =& - (g \partial_\mu W^{3\mu} - g' \partial_\mu B^\mu)^2 ,\\
{\mathcal L}_{16} & =& -g^2(gW^{3\nu}-g' B^\nu)[ (W_\nu^2 \partial_\mu W^{1\mu} - W_\nu^1 \partial_\mu W^{2\mu} )  \\
&& + g' B^\mu(W^1_\mu W^1_\nu + W^2_\mu W^2_\nu) ] , \\
{\mathcal L}_{17} & =& - \frac{1}{v}g(g W^3_\mu - g' B_\mu) W^{3\mu\nu}\partial_\nu h , \\
{\mathcal L}_{18} & =& \frac{1}{v}g^2(g W^{3\mu} - g' B^\mu)(W^1_\mu W^2_\nu - W^2_\mu W^1_\nu)\partial^\nu h , \\
{\mathcal L}_{19} & =& -\frac{1}{v}(g W^{3\nu}-g' B^\nu) (g\partial_\mu W^{3\mu} - g' \partial_\mu B^\mu) \partial_\nu h  ,\\
{\mathcal L}_{20} &=& -\frac{1}{2v^2}[g^2( (W_\mu^1)^2+(W_\mu^2)^2) + (g W_\mu^3 - g' B_\mu)^2] (\partial_\nu h)^2  ,\\
{\mathcal L}_{21} & =& -\frac{1}{v^2} (g W^{3\mu} - g' B^\mu)^2 \partial_\nu h \partial^\nu h  ,\\
{\mathcal L}_{22} & =& -\frac{1}{v^2} (g W^{3\mu} - g' B^\mu)(g W^{3\nu} - g' B^\nu)\partial_\mu h \partial_\nu h  ,\\
{\mathcal L}_{23} & =& \frac{1}{2}(gW^{3\nu}-g' B^\nu)^2 [ g^2 (W^{1\mu}W^1_{\mu}+W^{2\nu}W^2_{\nu})  \\
&& + (gW^{3\mu} - g' B^\mu)^2  ] , \\
{\mathcal L}_{24} & =& \frac{1}{2}(gW^{3\mu}-g' B^\mu)(gW^{3\nu}-g' B^\nu)\times  \\
&& [ g^2 (W^1_\mu W^1_\nu + W^2_\mu W^2_\nu )   +  (gW^{3}_{\mu}-g' B_\mu) (gW^{3}_{\nu}-g' B_\nu) ] , \\
{\mathcal L}_{25} & =& - \frac{1}{v}(g W^{3\mu} - g' B^\mu)^2 \Box h ,  \\
{\mathcal L}_{26} & =& (gW^{3\mu} - g' B^\mu)^2(gW^{3\nu} - g' B^\nu)^2,     \\
\end{array} \label{LCEFT_CP-even}
\eeqa
where $\widetilde{W}^{a\mu\nu}=\epsilon^{\mu\nu\rho\sigma}W^a_{\rho\sigma}/2$.
Note that in the strong coupling limit, i.e., $f\to\infty$ or $\xi\to 0$, ${\mathcal L}_{4}$ just recovers that in the low energy effective Lagrangian, while  ${\mathcal L}_{4}^{(s)}$ just decoupled. It is worthy of noticing that for NMCHM with symmetry breaking pattern as $SO(6)/SO(5)$,  the custodial violating operator $\mathcal{L}_{T}=-2\mathcal{L}_{C}+\cdots$, is not independent of $\mathcal{L}_{C}$.

\subsubsection{CP-odd case}

\textbf{Operators with two derivatives}
\beqa
\begin{array}{lll}
\mathcal{L}_{\widetilde{C}} = 0  \nn\\
\mathcal{L}_{\widetilde{T}} = i \frac{(v+h)^{2}}{4}\left[\operatorname{Tr}\left( \bf{T} \mathbf{D}^{\mu} \bf{V}_{\mu}\right)\right]^{2} ,  \\
\end{array} \label{L_Ct-Tt}
\eeqa
where $\widetilde{C}$ and $\widetilde{T}$ indicates the custodial preserving and custodial breaking, respectively.

\textbf{Operators with four derivatives} one has~\cite{Gavela:2014vra,Hierro:2015nna,Brivio:2016fzo}
\beqa
\begin{array}{lll}
\mathcal{L}_{\widetilde{B}}  &=&-\frac{g^{\prime 2}}{4} \widetilde{B}_{\mu \nu} B^{\mu \nu}  \\ 
\mathcal{L}_{\widetilde{W}}  &=&-\frac{g^{2}}{2} \operatorname{Tr}\left(\widetilde{\mathbf{W}}_{\mu \nu} \mathbf{W}^{\mu \nu}\right)    \\
\mathcal{L}_{\widetilde{1}}  &=&2 g g^{\prime} {B}_{\mu \nu}\operatorname{Tr}\left(\mathbf{T} \widetilde{\mathbf{W}}^{\mu \nu}\right)   \\ 
\mathcal{L}_{\widetilde{2}}  &=&2 i g^{\prime} \widetilde{B}_{\mu \nu}\operatorname{Tr}\left(\mathbf{T} \mathbf{V}^{\mu}\right) \partial^{\nu}(h / v)   \\ 
\mathcal{L}_{\widetilde{3}}  &=&2 i g \operatorname{Tr}\left(\widetilde{\mathbf{W}}^{\mu \nu} \mathbf{V}_{\mu}\right) \partial_{\nu}(h / v)     \\
\mathcal{L}_{\widetilde{4}}  &=&g \operatorname{Tr}\left(\mathbf{W}^{\mu \nu} \mathbf{V}_{\mu}\right) \operatorname{Tr}\left(\mathbf{T} \mathbf{V}_{\nu}\right)    \\
\mathcal{L}_{\widetilde{5}}  &=&i \operatorname{Tr}\left(\mathbf{V}^{\mu} \mathbf{V}^{\nu}\right) \operatorname{Tr}\left(\mathbf{T} \mathbf{V}_{\mu}\right) \partial_{\nu}(h / v)    \\
\mathcal{L}_{\widetilde{6}}  &=&i \operatorname{Tr}\left(\mathbf{V}^{\mu} \mathbf{V}_{\mu}\right) \operatorname{Tr}\left(\mathbf{T} \mathbf{V}^{\nu}\right) \partial_{\nu}(h / v)     \\
\mathcal{L}_{\widetilde{7}}  &=&g \operatorname{Tr}\left(\mathbf{T}\left[\mathbf{W}^{\mu \nu} , \mathbf{V}_{\mu}\right]\right) \partial_{\nu}(h / v)    \\
\mathcal{L}_{\widetilde{8}}  &=&2 g^{2} \operatorname{Tr}\left(\mathbf{T} \widetilde{\mathbf{W}}^{\mu \nu}\right) \operatorname{Tr}\left(\mathbf{T} \mathbf{W}_{\mu \nu}\right)    \\
\mathcal{L}_{\widetilde{9}}  &=&2 i g \operatorname{Tr}\left(\mathbf{T} \widetilde{\mathbf{W}}^{\mu \nu}\right) \operatorname{Tr}\left(\mathbf{T} \mathbf{V}_{\mu}\right) \partial_{\nu}(h / v)   \\ 
\mathcal{L}_{\widetilde{10}}  &=&i \operatorname{Tr}\left(\mathbf{V}^{\mu} \mathbf{D}^{\nu} \mathbf{V}_{\nu}\right) \operatorname{Tr}\left(\mathbf{T} \mathbf{V}_{\mu}\right)   \\ 
\mathcal{L}_{\widetilde{11}}  &=&i \operatorname{Tr}\left(\mathbf{T} \mathbf{D}^{\mu} \mathbf{V}_{\mu}\right) \operatorname{Tr}\left(\mathbf{V}^{\nu} \mathbf{V}_{\nu}\right)   \\
\mathcal{L}_{\widetilde{12}}  &=&i \operatorname{Tr}\left(\left[\mathbf{V}^{\mu}, \mathbf{T}\right] \mathbf{D}^{\nu} \mathbf{V}_{\nu}\right) \partial_{\mu}(h / v)   \\
\mathcal{L}_{\widetilde{13}}  &=&i \operatorname{Tr}\left(\mathbf{T} \mathbf{D}^{\mu} \mathbf{V}_{\mu}\right) \partial^{\nu} \partial_{\nu}(h / v)  \\
\mathcal{L}_{\widetilde{14}}  &=&i \operatorname{Tr}\left(\mathbf{T} \mathbf{D}^{\mu} \mathbf{V}_{\mu}\right) \partial^{\nu}(h / v) \partial_{\nu}(h / v)  \\
\mathcal{L}_{\widetilde{15}}  &=&i \operatorname{Tr}\left(\mathbf{T} \mathbf{V}^{\mu}\right)\left(\operatorname{Tr}\left(\mathbf{T} \mathbf{V}^{\nu}\right)\right)^{2} \partial_{\mu}(h / v)  \\
\mathcal{L}_{\widetilde{16}}  &=&i \operatorname{Tr}\left(\mathbf{T} \mathbf{D}^{\mu} \mathbf{V}_{\mu}\right)\left(\operatorname{Tr}\left(\mathbf{T} \mathbf{V}^{\nu}\right)\right)^{2}, \\
\end{array} \label{L_CP-odd}
\eeqa
where the first five Lagrangian $\mathcal{L}_{\widetilde{B},\widetilde{W},\widetilde{1},\widetilde{2},\widetilde{3}}$ corresponds to the custodial preserving ones, while the residue operators to (tree-level) custodial breaking ones.  In the custodial breaking class, the presence of the scalar chiral field $\mathbf{T} $, which implies that the custodial symmetry is violating. The dual tensor are defined by
$\widetilde{B}_{\mu \nu} \equiv \epsilon_{\mu \nu \rho \sigma} B^{\rho \sigma}$ ,  and $\widetilde{\mathbf{W}}_{\mu \nu} \equiv \epsilon_{\mu \nu \rho \sigma} \mathbf{W}^{\rho \sigma}$.
The covariant derivative of $\mathbf{V}$ is defined in Eq.(\ref{DV}).  In the absence of a light Higgs-like, i.e., CP-odd scalar singlet $h$ in the low energy spectrum, the $10$ operators containing derivatives of Higgs are absent. Thus, there are a complete $9$ (independent) CP-even operators those preserving $SU(2)_L\times U(1)_Y$ symmetry. 


The CP-odd low energy effective Lagrangian in Eq.(\ref{L_CP-odd}) can be expressed more explicitly in the unitary gauge as
\beqa
\begin{array}{lll} 
 \mathcal{L}_{\widetilde{C}} &=& 0,  \\
 \mathcal{L}_{\widetilde{T}} &=& - \frac{i}{4}v^2 (g\partial_\mu W^{3\mu} - g' \partial_\mu B^\mu)^2,  \\
 \mathcal{L}_{\widetilde{W}} &=& - \frac{1}{4}g'^2 \widetilde{W}_{\mu\nu}^a W^{a\mu\nu},  \\
 \mathcal{L}_{\widetilde{B}} &=& - \frac{1}{4}g'^2 \widetilde{B}_{\mu\nu}B^{\mu\nu} ,  \\
 \mathcal{L}_{\widetilde{1}} &=& 2 gg' B_{\mu\nu}\widetilde{W}^{3 \mu\nu} ,  \\
 \mathcal{L}_{\widetilde{2}} &=& -\frac{2}{v}g' \widetilde{B}^{\mu\nu}( g W_\mu^3 - g' B_\mu )\partial_\nu h ,  \\
 \mathcal{L}_{\widetilde{3}} &=& -\frac{1}{v}[ g^2(W_\mu^1 \widetilde{W}^{1\mu\nu} + W_\mu^2 \widetilde{W}^{2\mu\nu}  )  \\
&&+ g \widetilde{W}^{3 \mu\nu}(gW_\mu^3 - g' B_\mu) ]\partial_\nu h , \\
 \mathcal{L}_{\widetilde{4}} &=& -\frac{1}{2}[ g^2 (W_\mu^1 W^{1\mu\nu} + W_\mu^2 W^{2\mu\nu}  )  \\
 && + g(g W_\mu^3  - g' B_\mu)W^{3\mu\nu}) ](g W_\nu^3 - g' B_\nu) ,  \\
 \mathcal{L}_{\widetilde{5}} &=& \frac{1}{2v} (g W_\mu^3 - g' B_\mu)[ g^2(W^{1\mu}W^{1\nu}+W^{2\mu}W^{2\nu})  \\
 && + (g W^{3\mu}-g' B^\mu)(g W^{3\nu}-g' B^\nu)   ] \partial_\nu h,  \\
 \mathcal{L}_{\widetilde{6}} &=& \frac{1}{2v}(g W^{3\nu}-g' B^\nu)[ (gW_\mu^1 + g W_\mu^2)^2  \\
 && + (g W_\mu^3-g' B_\mu)^2 ]\partial_\nu h,  \\
 \mathcal{L}_{\widetilde{7}} &=& \frac{1}{v}g^2(W_\mu^1 W^{2\mu\nu} - W_\mu^2 W^{1 \mu\nu}  )\partial_\nu h,  \\
 \mathcal{L}_{\widetilde{8}} &=& 2g^2 W_{\mu\nu}^3 \widetilde{W}^{3\mu\nu},  \\
 \mathcal{L}_{\widetilde{9}} &=& - \frac{2g}{v}(gW_\mu^3 - g' B_\mu) \widetilde{W}^{3\mu\nu} \partial_\nu h,  \\
 \mathcal{L}_{\widetilde{10}} &=& \frac{1}{2}(g W_\mu^3 - g' B_\mu)[ g^2(W^{1\mu} \partial_\nu W^{1\nu} + W^{2\mu} \partial_\nu W^{2\nu} )  \\
 && + (g W^{3\mu}-g' B^\mu)(g\partial_\nu W^{3\nu}-g' \partial_\nu B^\nu)  \\
 && + g^2 g' B^\nu (W^{1\mu} W_\nu^2 - W^{2\mu} W_\nu^1)   ] ,  \\
 {\mathcal L}_{\widetilde{11}} &=& \frac{1}{2}[g^2(W_{\nu}^1W^{1\nu} +W_{\nu}^2W^{2\nu}) + (gW_\nu^3 - g' B_\nu)^2 ] \\
 && \times (g\partial_\mu W^{3\mu} - g' \partial_\mu B^\mu),  \\
 {\mathcal L}_{\widetilde{12}} &=& \frac{g^2}{v}   [ g' B_\nu ( W^{1\mu}W^{1\nu} + W^{2\mu}W^{2\nu}  )  \\
 && + ( W^{2\mu} \partial_\nu W^{1\nu} - W^{1\mu} \partial_\nu W^{2\nu}  )   ] \partial_\mu h ,  \\
 {\mathcal L}_{\widetilde{13}} &=& - \frac{1}{v}(g\partial_\mu W^{3\mu} - g' \partial_\mu B^\mu) \Box h ,  \\
 {\mathcal L}_{\widetilde{14}} &=& - \frac{1}{v^2}(g\partial_\mu W^{3\mu} - g' \partial_\mu B^\mu) (\partial_\nu h)^2 ,  \\
 {\mathcal L}_{\widetilde{15}} &=& \frac{1}{v} (gW^{3\nu} - g' B^\nu)^2 (g W^{3\mu}-g' B^\mu) \partial_\mu h,  \\
 {\mathcal L}_{\widetilde{16}} &=& (gW^{3\nu} - g' B^\nu)^2(g\partial_\mu W^{3\mu} - g' \partial_\mu B^\mu).  \\
\end{array}  \label{LCEFT_CP-odd}
\eeqa

\end{CJK}



\end{document}